\documentclass[a4paper,11pt]{article}
\pdfoutput=1
\usepackage{jheppub}

\usepackage[T1]{fontenc}
\usepackage{amsfonts}
\usepackage{gensymb}
\usepackage{amsmath}
\usepackage{amssymb}
\usepackage{multirow}
\usepackage{graphicx}
\usepackage{subfig}
\usepackage{comment}
\usepackage{tensor}

\usepackage{scrextend}

\usepackage{tikz-cd}

\usepackage{mathtools}
\usepackage{array,ragged2e}

\usepackage{multicol}
\usepackage{blindtext}
\usepackage{floatrow}
\newfloatcommand{capbtabbox}{table}[]
\usepackage{xfrac}

\DeclareMathOperator*{\IM}{Im}
\DeclareMathOperator*{\res}{res}

\DeclareMathOperator{\diag}{diag}
\DeclareMathOperator{\disc}{disc}

\newcommand{\dd}{\text{d}}

\newcommand{\Low}{{\mathrm{Low}}}
\newcommand{\High}{{\mathrm{High}}}

\numberwithin{equation}{section}
\usepackage{graphicx}

\begin{document}

\title{Positivity Bounds on Massive Vectors}

\author[a]{Francesco Bertucci,}
\emailAdd{francesco.bertucci@phd.unipi.it}
\affiliation[a]{Department of Physics, University of Pisa and INFN, Largo Pontecorvo 3, 56127 Pisa, Italy 
}
\author[a,b]{Johan Henriksson,}
\emailAdd{johan.henriksson@ipht.fr}
\affiliation[b]{Universit\'e Paris--Saclay, CEA, Institut de Physique Th\'eorique, 91191, Gif-sur-Yvette, France}
\author[a,c]{Brian McPeak,}
\emailAdd{brian.mcpeak@mcgill.ca}
\affiliation[c]{McGill University, 3600 rue University
Montréal, QC
Canada H3A 2T8}

\author[d]{Sara Ricossa,}
\emailAdd{sara.ricossa@unige.ch}
\affiliation[d]{Départment de Physique Théorique, Université de Genève,
24 quai Ernest-Ansermet, 1211 Genève 4, Switzerland}

\author[d]{Francesco Riva,}
\emailAdd{francesco.riva@unige.ch}

\author[a]{Alessandro Vichi}
\emailAdd{alessandro.vichi@df.unipi.it}

\abstract{
In this paper, we explore  positivity bounds for the effective field theory~(EFT) of a single weakly coupled massive vector field. The presence of both mass and spin makes the crossing properties of the amplitudes vastly  complicated -- we address this by parametrizing the amplitudes as products of a polarization matrix and a vector of appropriately chosen functions with  simpler crossing properties. 
The resulting 
framework involves sum rules and null constraints that allows us to constrain any combination of low-energy observables, such as  EFT amplitudes.   
By varying the value of the vector mass over the cutoff scale, some of our bounds asymptote to the bounds obtained in the context of photons and massless scalars. This work paves the way for future applications to \emph{e.g.} non-abelian massive vectors, glueballs and theories with spin larger than one.
}

\maketitle

\section{Introduction}

Effective field theory (EFT) provides a powerful framework for understanding the dynamics of particles at low energies, particularly when the complete underlying theory is unknown. EFTs are quite flexible and can describe a wide array of different situations and particle content. Among these, spin one particles hold a special place due to their role as force carriers in the Standard Model of particle physics. These include massless spin one fields---gluons and photons---and the massive  $W$ and $Z$ bosons that mediate the weak force. The latter display a remarkably intricate structure due to their interaction with the Higgs boson and consequent symmetry breaking. Elucidating this structure was one of the crowning achievements of 20\textsuperscript{th} century theoretical physics. Spinning massive particles also play a prominent role in the IR description of QCD: a number of the excited states including glueballs and many of the hadrons are massive and have spin. Understanding the structure of resonances and their interactions is a longstanding dream.

A basic question is: how flexible is this structure? Are the particle content and interaction strengths in the Standard Model ``necessary,'' or does our world simply represent a point in a vast landscape of possible theories? Effective field theory alone is too permissive to answer this kind of question. However with additional physical input---in this case, the requirement that the EFT admits a UV completion with an $S$-matrix that is unitary and analytic---we can say quite a lot more. The simplest such constraints came by using dispersion relations to relate certain EFT coefficients to the scattering cross section~\mbox{\cite{Pham:1985cr, Pennington:1994kc,  Ananthanarayan:1994hf, Comellas:1995hq, Dita:1998mh, Adams:2006sv}}. The positivity of the cross section then implies positivity of the EFT coefficient that appears in the forward-limit expansion of elastic amplitudes. Eventually a precise numerical framework, inspired in part by the conformal bootstrap \cite{Rattazzi:2008pe}, was developed for optimizing the bounds obtained from dispersion relations \cite{ Bellazzini:2020cot, Tolley:2020gtv,Caron-Huot:2020cmc,  Sinha:2020win,Arkani-Hamed:2020blm}, resulting in closed allowed regions for ratios of coefficients, leading to the quest to understand the ``extremal effective theories'' that lie on the boundaries. Overall, the approach of using dispersion relations to bound EFT coefficients has proved to be extremely broad and useful, leading to a large number of studies~\cite{Manohar:2008tc, Mateu:2008gv, Nicolis:2009qm, Baumann:2015nta, Bellazzini:2015cra, Bellazzini:2016xrt, Cheung:2016yqr, Bonifacio:2016wcb,  Cheung:2016wjt, deRham:2017avq, Bellazzini:2017fep, deRham:2017zjm, deRham:2017imi, Hinterbichler:2017qyt, Bonifacio:2017nnt, Bellazzini:2017bkb, Bonifacio:2018vzv, deRham:2018qqo, Zhang:2018shp, Bellazzini:2018paj, Bellazzini:2019xts, Melville:2019wyy, deRham:2019ctd, Alberte:2019xfh, Alberte:2019zhd, Bi:2019phv, Remmen:2019cyz, Ye:2019oxx, Herrero-Valea:2019hde, Zhang:2020jyn, Trott:2020ebl,Zhang:2021eeo, Wang:2020jxr, Li:2021lpe, Du:2021byy, Davighi:2021osh, Chowdhury:2021ynh, Henriksson:2021ymi, Caron-Huot:2021enk, Caron-Huot:2022jli, Caron-Huot:2022ugt, Bern:2021ppb, Henriksson:2022oeu, Fernandez:2022kzi, Albert:2023jtd,Bellazzini:2023nqj, Mcpeak:2023wmq}.\footnote{A complementary approach involving propagation on classical backgrounds was envisioned in \cite{Adams:2006sv} and developed in \cite{Cheung:2014ega}. Related recent work includes \cite{CarrilloGonzalez:2022fwg, CarrilloGonzalez:2023cbf, CarrilloGonzalez:2023rmc} and \cite{Cremonini:2023epg} Another related approach investigated recently is the set of EFT coefficients which are consistent with the Classical Regge Growth conjecture, see \cite{Chowdhury:2019kaq, Chowdhury:2023fwb}.}

In this paper we initiate a systematic and complete study of positivity bounds for EFTs containing non-scalar massive particles. As a first step in this direction we consider a massive vector, neutral under any flavor symmetry. A concrete realization of this scenario is the EFT of the lightest spin-1 glueball in QCD, where all heavier states have been integrated out.\footnote{In QCD, vector glueballs are not the lightest states -- scalars and spin-2 glueballs are observed to have smaller mass. Indeed, lattice calculations, QCD sum rules, flux tube, and constituent glue models agree that the lightest glueballs have quantum numbers $J^{PC} = 0^{++}$ and $2^{++}$. Lattice calculations predict for the ground state ($0^{++}$) a mass around 1600--1700 MeV with an uncertainty of about 100 MeV, while the first excited state ($2^{++}$) has a mass of about 2300 MeV. Heavier glueballs with quantum numbers $0^{-+}, 2^{-+}, 1^{+-}, \ldots$ are predicted above 2500 MeV and the lowest exotic ones (non-$q\bar{q}$)  are expected above 4000 MeV \cite{Lucini:2001ej,Lucini:2004my,Lucini:2010nv,Athenodorou:2021qvs}.}
This simple scenario allows us to identify and address general problems  related to the non-zero mass of  external spinning particles, such as the proliferation of helicity amplitudes, all mixed via crossing symmetry. This paves the road 
 to a generalization to more  complicated setups---such as scattering massive higher-spin particles in QCD at large~$N$.

We focus on  the most general possible weakly coupled EFT which only includes a massive vector $A_\mu$ and no other light states below the cutoff. We do not require the theory to be weakly coupled above the cutoff. 
Previous work in this direction has  focused on  a subset of theories with ghost free equations of motion (which we discuss in  appendix \ref{app:Proca-theory}). In particular Refs.~\cite{Bonifacio:2016wcb,deRham:2018qqo}  studied elastic amplitudes in this context in the strict forward limit, 
while Ref.~\cite{deRham:2022sdl}, extends this  beyond the forward limit by allowing for one $t$-derivative to act on the amplitude.
 See also \cite{Chowdhury:2023fwb} for a classification of the four-point interactions in massive vector EFTs which do not violate the classical Regge growth conjecture, which requires that the amplitude grows no faster than $s^2$ at large (but finite) energies.

Our work extends this by systematically constructing the most general massive spin-1 EFT and classifying the complete set of positivity sum rules. Finally we obtain the most stringent bounds from full crossing symmetry of the amplitude and compare them with known UV completions.

\vspace{0.5cm}

The article is structured as follows. In Section~\ref{sec:setup} we introduce the general parametrization of the scattering amplitude for massive vectors, both in the low energy EFT regime and in the UV regime.  Next, we discuss the consequences of crossing symmetry and unitarity of the $S$-matrix. We also review how to derive dispersion relations in the present framework.

In Section~\ref{sec:general} we tackle the main differences compared to the massless scalar case. Firstly, the presence of a non-zero mass makes it hard to express individual EFT coefficients in terms of a dispersive sum rule. This because if we define EFT coefficients by expanding an amplitude $f(s,t)$ around $s,t\sim0$, we will also encounter the expansion of $f(s,u)$ around $u\sim 4m^2$, which contains infinitely many EFT coefficients. To overcome this problem we identify a set of observables, closely related to the arcs of \cite{Bellazzini:2020cot}, and we discuss their interpretation in the massless limit. A second problem concerns the optimal choice of subtractions to be taken inside dispersion relations. A conservative approach would consist in assuming that all scattering amplitudes are bounded by $s^2$ at large energies and always take two subtractions. As we will see in this paper, and as has been pointed out in other contexts, such as \cite{Caron-Huot:2022jli} for gravity, scattering states with spin lead to \textit{super-convergence} -- the polarizations add extra powers of $s$ to the amplitude so the actual functions parameterizing the interactions have a softer Regge behavior. As recently discussed in \cite{Mcpeak:2023wmq}, the EFT constraints are highly sensitive to the number of subtractions, so it is crucial to extract all of the valid ``low-lying'' null constraints. At the end of the section we spell out how to construct such combinations. We also discuss how to obtain null constraints.

We  present our bounds in section~\ref{sec:Bounds}. We first identify a set of observables that are positive definite. These are integrals over elastic amplitudes $A^{\lambda_1 \lambda_2 \to \lambda_1 \lambda_2}$ and they will serve as normalizations for the other observables, which may be non-positive. One novelty here is that, depending on the observable we are studying, it is possible to use different normalization conditions. We show how to obtain analytic bounds in the simplified forward limit $t=0$. Finally we discuss numerical bounds obtained using the full power of the EFT bootstrap, by working to higher order in the small-$t$ expansion and including a large number of null constraints. 
Furthermore, we compare our results to the UV completions that arise from integrating out scalars and vectors at tree-level, and find that each of these lies on the edge of the plots, meaning that they are extremal.

\section{Massive vector scattering amplitudes}
\label{sec:setup}

\subsection{Parametrization of the amplitude}

The first task is to work out a proper parametrization of the amplitudes. In general this is given by \textit{structures}, involving contractions of the polarization tensors with each other and with external momenta, and \textit{functions} of the momenta. Under the assumption of weak coupling  these functions are simple polynomials of the Mandelstam variables within the IR EFT\footnote{\label{ftnt:loops}Here we work in the tree-level limit in which the EFT non-analyticities (such as logarithms coming from loops) are ignored; as discussed in Refs.~\cite{Bellazzini:2020cot,Bellazzini:2021oaj} this is a reasonable assumption in the scattering of massive particles. Moreover  the EFT of a single massive vector admits no poles, see   section \ref{sec:nothreepoint}.  }. So the amplitudes will take the form:
\begin{equation}\label{eq:AsumEF}
	A^I(s,t) =\sum_J \tensor E{^I_J} (s,t) F_J (s,t),
\end{equation}
where $E{^I_J}$ is a $17 \times 17$  matrix of structures, and $F_J$ is a $17$-vector of functions.  The $J$-index is a structure index, which we discuss in the next paragraph.
The $I$-index, instead, labels the outside spins, for which we use the helicity basis. An alternative choice would be the transversity basis, discussed in this context in Ref.~\cite{deRham:2017zjm}, which diagonalizes crossing symmetry. The helicity basis has the advantage that its $m\to 0$ limit manifestly matches the amplitudes of massless spin-0 and spin-1 theories, according to the equivalence theorem. This allows a more transparent comparison with previous literature.

\subsubsection{Structures}

The amplitudes $A^I = A^I(p_i,\epsilon_i)$ for vector scattering are functions of the momenta and structures, with a linear dependence on each $\epsilon_i$. The polarizations can be contracted with each other, or with external momenta. We denote these contractions by
\begin{align}
    (\epsilon_i \epsilon_j) \equiv (\epsilon(p_i))^\mu  (\epsilon(p_j))_\mu \, , \qquad (\epsilon_i p_j) \equiv (\epsilon(p_i))^\mu (p_j)_\mu \, .
\end{align} 
Each \textit{structure} is a unique way of contracting the polarizations. There is a large number of these -- 43, if we use momentum conservation to remove $k_4$, and work in general dimensions. 

There are three discrete symmetries---parity, time reversal and boson exchange---which require that
\begin{align}
    \mathcal{P}: \quad  A^{\lambda_1 \lambda_2 \lambda_3 \lambda_4}(p_1, p_2, p_3, p_4) \ &=(-1)^{\lambda_1 +\lambda_2 +\lambda_3 +\lambda_4} A^{-\lambda_1 -\lambda_2 -\lambda_3 -\lambda_4}(p_4, p_3, p_2, p_1)  \, , \\
    \mathcal{T}: \quad  A^{\lambda_1 \lambda_2 \lambda_3 \lambda_4}(p_1, p_2, p_3, p_4) \ &=(-1)^{-\lambda_1 -\lambda_2 -\lambda_3 +\lambda_4} \ A^{\lambda_3 \lambda_4 \lambda_1 \lambda_2}(p_4, p_3, p_2, p_1)  \, ,\\
    \mathcal{B}:  \quad  A^{\lambda_1 \lambda_2 \lambda_3 \lambda_4}(p_1, p_2, p_3, p_4) \ &= \ A^{\lambda_2 \lambda_1 \lambda_4 \lambda_3}(p_2, p_1, p_4, p_3) \, .
\end{align}
Only $\mathcal{P}$,$\mathcal{T}$, and $\mathcal{B}$-invariant combinations of structures can appear in the amplitude, which reduces the number of independent structures to 19. These provide a valid basis for the amplitude in general dimensions; however, in 4d there exist two additional constraints which reduce the number of independent structures to 17. The 19 structures, along with the 4d constraints, are given in appendix~\ref{app:structures}. The result of this analysis is a basis of 17 structures:
\begin{align}\label{eq:epslist}
\begin{split}
     {e}_1 \quad &= \quad (\epsilon_1\epsilon_2) \  (\epsilon_3 \epsilon_4) \, , \\
    {e}_2 \quad &= \quad (\epsilon_1 \epsilon_3) \ (\epsilon_2 \epsilon_4) \, , \\
    {e}_3 \quad &= \quad (\epsilon_1 \epsilon_4) \ (\epsilon_2 \epsilon_3) \, , \\ \\
    {e}_4  \quad &= \quad (\epsilon_1 \epsilon_{2}) \  (\epsilon_3 p_{4}) \ (\epsilon_4 p_{3}) + (\epsilon_3 \epsilon_{4}) \  (\epsilon_1 p_{2}) \ (\epsilon_2 p_{1}) \, , \\
    {e}_5  \quad &= \quad (\epsilon_1 \epsilon_{3}) \  (\epsilon_2 p_{4}) \ (\epsilon_4 p_{2}) + (\epsilon_2 \epsilon_{4}) \  (\epsilon_1 p_{3}) \ (\epsilon_3 p_{1}) \, , \\
    {e}_{6}  \quad &= \quad (\epsilon_1 \epsilon_{4}) \  (\epsilon_2 p_{3}) \ (\epsilon_3 p_{2}) + (\epsilon_2 \epsilon_{3}) \  (\epsilon_1 p_{4}) \ (\epsilon_4 p_{1}) \, , \\
    {e}_7  \quad &= \quad (\epsilon_1 \epsilon_{2}) \  (\epsilon_3 p_{1}) \ (\epsilon_4 p_{2}) + (\epsilon_3 \epsilon_{4}) \  (\epsilon_1 p_{3}) \ (\epsilon_2 p_{4}) \, , \\
    {e}_8  \quad &= \quad (\epsilon_1 \epsilon_{2}) \  (\epsilon_3 p_{2}) \ (\epsilon_4 p_{1}) + (\epsilon_3 \epsilon_{4}) \  (\epsilon_2 p_{3}) \ (\epsilon_1 p_{4}) \, , \\
    {e}_9  \quad &= \quad (\epsilon_1 \epsilon_{3}) \  (\epsilon_2 p_{1}) \ (\epsilon_4 p_{3}) + (\epsilon_2 \epsilon_{4}) \  (\epsilon_1 p_{2}) \ (\epsilon_3 p_{4}) \, , \\
    {e}_{10}  \quad &= \quad (\epsilon_1 \epsilon_{3} ) \  (\epsilon_2 p_{3}) \ (\epsilon_4 p_{1}) + (\epsilon_2 \epsilon_{4}) \  (\epsilon_1 p_{4}) \ (\epsilon_3 p_{2} ) \, , \\
    {e}_{11}  \quad &= \quad (\epsilon_1 \epsilon_{4}) \  (\epsilon_2 p_{1}) \ (\epsilon_3 p_{4}) + (\epsilon_2 \epsilon_{3}) \  (\epsilon_1 p_{2}) \ (\epsilon_4 p_{3}) \, , \\
    {e}_{12}  \quad &= \quad (\epsilon_1 \epsilon_{4} )\  (\epsilon_2 p_{4}) \ (\epsilon_3 p_{1}) + (\epsilon_2 \epsilon_{3}) \  (\epsilon_1 p_{3}) \ (\epsilon_4 p_{2}) \, , \\ \\
    {e}_{13}  \quad &= \quad (\epsilon_1 p_3) \ (\epsilon_2 p_4) \ (\epsilon_3 p_2) \ (\epsilon_4 p_1) + (\epsilon_1 p_4) \ (\epsilon_2 p_3) \ (\epsilon_3 p_1) \ (\epsilon_4 p_2) \, , \\ 
    {e}_{14}  \quad &= \quad (\epsilon_1 p_2) \ (\epsilon_2 p_3) \ (\epsilon_3 p_4) \ (\epsilon_4 p_1 )+ (\epsilon_1 p_4) \ (\epsilon_2 p_1) \ (\epsilon_3 p_2) \ (\epsilon_4 p_3) \, , \\ 
    {e}_{15}  \quad &= \quad (\epsilon_1 p_2) \ (\epsilon_2 p_4) \ (\epsilon_3 p_1) \ (\epsilon_4 p_3) + (\epsilon_1 p_3) \ (\epsilon_2 p_1) \ (\epsilon_3 p_4) \ (\epsilon_4 p_2 )\, , \\ 
    {e}_{16}  \quad &= \quad (\epsilon_1 p_3) \ (\epsilon_2 p_4) \ (\epsilon_3 p_1) \ (\epsilon_4 p_2) \, , \\ 
    {e}_{17}  \quad &= \quad (\epsilon_1 p_4) \ (\epsilon_2 p_3) \ (\epsilon_3 p_2) \ (\epsilon_4 p_1) \, .  \\ 
\end{split}
\end{align}
Of course any choice of 17 linearly independent combinations of these structures may serve as a basis; however, this choice has the nice property that none of the functions multiplying these structures have poles in their low energy expansions.

Each structure depends on the external polarizations, so we shall denote them by 
\begin{align}
   E_J^{\lambda_1 \lambda_2 \lambda_3 \lambda_4} = e_J(\epsilon_1^{\lambda_1},\epsilon_2^{\lambda_2},\epsilon_3^{\lambda_3},\epsilon_4^{\lambda_4},p_1,p_2,p_3,p_4)\, ,
\end{align}
with $J = 1,  \ldots , 17$ and where $\lambda_i$ is the polarization ($1$, $0$, or $-1$) of the $i^{\text{th}}$ particles, using all-ingoing conventions. See appendix~\ref{app:conventions} for a full explanation of our conventions for the momenta and polarizations. 

\subsubsection{Functions}

Each structure $F_J$ multiplies a function in the amplitude. Since many of the structures are related by crossing symmetry, the functions must be related too. This reduces the set of 17 functions $F_J$ to only five functions, which we will call $f$, $g$, $\tilde g$, $h$, and $\tilde h$. These are related to the $F_J$ by
\begin{align}\label{eq:functions}
    &F_1(s,t,u) \ = \ f(s|t,u),  &F_2(s,t,u) \ = \ f(t|s,u), & & F_3(s,t,u) \ = \ f(u|s,t), \nonumber \\
    &F_4(s,t,u) \ = \ g(s|t,u),  &F_5(s,t,u) \ = \ g(t|s,u), & & F_6(s,t,u) \ = \ g(u|s,t), \nonumber \\
    &F_7(s,t,u) \ = \ \tilde g(s,t,u),  &F_8(s,t,u) \ = \ \tilde g(s,u,t), & & F_9(s,t,u) \ = \ \tilde g(t,s,u), \nonumber \\
    &F_{10}(s,t,u) \ = \ \tilde g(t,u,s),  &F_{11}(s,t,u) \ = \ \tilde g(u,s,t), & & F_{12}(s,t,u) \ = \ \tilde g(u,t,s), \nonumber 
    \\
    &F_{13}(s,t,u) \ = \ h(s|t,u)+\tilde h(s,t,u)+\tilde h(s,u,t), \hspace{-4cm}\nonumber \\
    &F_{14}(s,t,u) \ = \ h(t|u,s)+\tilde h(s,u,t), \hspace{-1cm} \nonumber \\
    &F_{15}(s,t,u) \ = \ h(u|s,t)+\tilde h(s,t,u), \hspace{-1cm} \nonumber \\
    &F_{16}(s,t,u) \ = \ 2 \tilde{h}(t,s,u)  \nonumber \,,\\
    &F_{17}(s,t,u) \ = \ 2 \tilde{h}(u,s,t)\, .  
\end{align}
Here $f$, $g$, and $h$ denote functions multiplying the 2-momentum, 4-momentum, and 6-momentum structures, respectively. The result is that $f(s|t,u)$, $g(s|t,u)$, and $h(s|t,u)$ have a partial crossing symmetry in the last two arguments ($t \leftrightarrow u$), $\tilde{g}(s,t,u)$ has no symmetry, and $\tilde{h}(s,t,u)$ is $s \leftrightarrow t$ antisymmetric and satisfies the cyclic identity $\tilde h(s,t,u)+\tilde h(t,u,s)+\tilde h(u,s,t)=0$. For the derivation of this particular set of functions, see appendix~\ref{app:functions}.

Using their symmetries, we can determine the most general low-energy expansion of these functions:
\begin{align}\label{eq:ffun}
    f(s|t,u) \ & = \ f_{0,0} + f_{1,0} (t + u) + f_{2,0} (t + u)^2 + f_{0,1} t u +   \ldots,  \\
    \label{eq:gfun}
    g(s|t,u) \ & = \ g_{0,0} + g_{1,0} (t + u) + g_{2,0} (t + u)^2 + g_{0,1} t u + \ldots,  \\
    \label{eq:tildegfun}
    \tilde g(s,t,u) \ & = \ \tilde{g}_{0,0} + \tilde{g}_{1,0} s + \tilde{g}_{0,1} t + \tilde{g}_{2,0} s^2 + \tilde{g}_{1,1} st + \ldots,  \\
    \label{eq:hfun}
    h(s|t,u) \ & = \ h_{0,0} + h_{1,0} (t + u) + h_{2,0} (t + u)^2 + h_{0,1} t u + \ldots, \\
    \label{eq:tildehfun}
    \tilde h(s,t,u) \ & = \ (s-t) \left( \tilde{h}_{0,0} + \tilde{h}_{1,0} (s+t) + \tilde{h}_{2,0} (s+t)^2 + \tilde{h}_{0,1} st +   \ldots \right).
\end{align}

\subsubsection{Amplitudes}

The structures and the amplitudes both depend on the helicity, so our amplitude looks like
\begin{equation}\label{eq:AequalEf}
	A^{\lambda_1\lambda_2\lambda_3\lambda_4}(s,t) =\sum_J E^{\lambda_1\lambda_2\lambda_3\lambda_4}_J F_J(s,t),
\end{equation}
It will convenient to choose a canonical order for the amplitudes by defining
\begin{align}\label{eq:helamplorder}
	\nonumber
	A^1 &\equiv A^{++++}, &\quad A^2 &\equiv A^{+++0}, &\quad A^3 &\equiv A^{+++-}, &\quad A^4 &\equiv A^{000+}, \\
	\nonumber
	A^5 &\equiv A^{0000}, &\quad A^6 &\equiv A^{++00}, &\quad A^7 &\equiv A^{+0+0}, &\quad A^8 &\equiv A^{0++0}, \\
	A^9 &\equiv A^{++0-}, &\quad A^{10} &\equiv A^{+0+-}, &\quad A^{11} &\equiv A^{0++-}, &\quad A^{12} &\equiv A^{++--}, \\
	\nonumber
	A^{13} &\equiv A^{+-+-}, &\quad A^{14} &\equiv A^{-++-}, &\quad A^{15} &\equiv A^{+00-}, &\quad A^{16} &\equiv A^{+0-0}, \\
	\nonumber
	A^{17} &\equiv A^{+-00}.
\end{align}
which, as promised, allows us to rewrite \eqref{eq:AequalEf} in matrix form as in equation \eqref{eq:AsumEF},
where the upper index $I$ denotes the particular choice of helicities, using the ordering defined by~\eqref{eq:helamplorder}. As mentioned earlier, there are 17 unique choices for helicity---the rest are determined by discrete symmetries. As a result, $\tensor E{^I_J}$ is a $17\times17$ matrix, with entries
\begin{align}
\begin{split}
        & \tensor E{^1_1}=E^{++++}_{1}=(\epsilon^+_1\cdot\epsilon^+_2)(\epsilon^+_3\cdot\epsilon^+_4)=1, \\
    & \tensor E{^1_2}=E^{++++}_{2}=(\epsilon^+_1\cdot\epsilon^+_3)(\epsilon^+_2\cdot\epsilon^+_4)=\frac{t^2}{(s-4m^2)^2} \, , \qquad \text{etc.}\\
\end{split}
\end{align}
Equation~\eqref{eq:AsumEF} can be easily inverted:
\begin{equation}
	F_J(s,t) \ = \ \sum_I \tensor {(E^{-1})}{_J^I}(s,t) A^I(s,t).
\end{equation}
where
\begin{equation}
    \tensor{(E^{-1})}{_1^1}=\frac12, \quad \tensor{(E^{-1})}{_1^2}=0,\quad \tensor{(E^{-1})}{_1^3}=\frac s{m^2}\, , \qquad \text{etc.}
\end{equation}

\paragraph{Removing kinematic branch cuts}
When the sum of external polarizations is odd, the amplitude $A(s,t)$ is proportional to an overall factor $\sqrt{stu} =\sqrt{-t s (s+t-4m^2)}$. Hence, in these cases there is a kinematic branch cut in the $s$-plane over the entire $\mathrm{Re}(s)>0$ semi-axis. We can easily cure this branch cut introducing
\begin{equation}\label{eq:Atildedef}
	\tilde{A}^I(s,t) =\begin{cases}
		\frac{m s}{\sqrt{-t s (s+t-4m^2)}}A^I(s,t) \quad \mathrm{if} \quad I=2,4,9,10,11, \\
		A^I(s,t) \quad \mathrm{otherwise.}
	\end{cases}
\end{equation}
The extra factor of $ms$ is chosen to preserve the dimensionality and Regge boundedness of the amplitude.

With this definition
\begin{equation}\label{eq:AFTildeDef}
	\tilde A^I(s,t) =\sum_J \tensor{\tilde E}{^I_J} (s,t) F_J (s,t),
 \qquad
	F_J(s,t) \ = \ \sum_I \tensor {(\tilde E^{-1})}{_J^I}(s,t) \tilde A^I(s,t),
\end{equation}
where we defined a new matrix $\tilde E(s,t)=\diag(1,\frac{ms}{\sqrt{stu}},1,\frac{ms}{\sqrt{stu}},1,1,\ldots)E(s,t)$.

\subsubsection{No IR poles in the amplitude}
\label{sec:nothreepoint}

The functions we used to parameterize our IR amplitudes are simply polynomials, and in particular have no poles. 
This is a general consequence of rotational invariance and Bose symmetry, as we now explain.

Poles are associated with non-vanishing on-shell
 3-point amplitudes, which appear as residues in the factorisation of 4-point amplitudes.\footnote{On-shellness for 3-point amplitudes can only be satisfied when one or more of the momenta are complex, yet they imply a pole at real 4-point kinematics.}
We can build a list of all possible tensor structures that can appear in these 3-point amplitudes,  allowing for parity-odd
vectors, which means the amplitude can be built from all the $\epsilon_i$, some $p_i$, and
possibly a Levi-Civita tensor $\varepsilon$.
Momentum conservation and on-shellness imply,\begin{equation}
   p_i\cdot\epsilon_j = -p_k\cdot\epsilon_j,
   \label{eq:3ptconservation}
\end{equation}
where $(i,j,k)$ is any permutation of $\{1,2,3\}$ -- crossing symmetry is implemented by summing over all such permutations.

Moreover, there are no kinematic (Mandelstam) invariants  at 3-points, as
 can be seen  explicitly through momentum conservation:
$ \left(p_1+p_2+p_3\right)^\mu = 0$,
which implies,
$  2(p_1\cdot p_2) =  2(p_1\cdot p_3) = 2(p_2\cdot p_3) = m^2$. Therefore, the 3-point amplitude must be a linear combination of,
\begin{multicols}{2}
\begin{enumerate}
    \item $(\epsilon_i\cdot\epsilon_j)(p_i\cdot\epsilon_k)$,
    \item $(\epsilon_i\cdot p_j)(\epsilon_j\cdot p_i)(\epsilon_k\cdot p_i)$,
    \item $\varepsilon^{\mu\nu\rho\sigma}\epsilon_\mu^i\epsilon_\nu^j\epsilon_\rho^kp_\sigma^i$,
    \item $(\epsilon_j\cdot\epsilon_k)\varepsilon^{\mu\nu\rho\sigma}p_\mu^ip_\nu^jp_\rho^k\epsilon_\sigma^i$,
    \item $(p_i\cdot\epsilon_j)\varepsilon^{\mu\nu\rho\sigma}\epsilon_\mu^i\epsilon_\nu^kp_\rho^ip_\sigma^k$,
    \item $(p_i\cdot\epsilon_j)\varepsilon^{\mu\nu\rho\sigma}\epsilon_\mu^i\epsilon_\nu^kp_\rho^ip_\sigma^j$\,.
\end{enumerate}
\end{multicols}

It is simple enough to check that the crossing symmetric sums of each of these terms vanish. For example, for term 1, we have,
\begin{equation}
    (\epsilon_i\cdot\epsilon_j)(p_i\cdot\epsilon_k) \xrightarrow[i\leftrightarrow j]{}
    (\epsilon_i\cdot\epsilon_j)(p_j\cdot\epsilon_k) = -(\epsilon_i\cdot\epsilon_j)(p_i\cdot\epsilon_k).
    \label{eq:3ptcancellation}
\end{equation}
We obtain similar relations for terms 2, 3, 4, and 5, by crossing $i\leftrightarrow j$, $j\leftrightarrow k$, $j\leftrightarrow k$,
and $i\leftrightarrow k$, respectively.
Term 6, by crossing $i\leftrightarrow k$, becomes,
\begin{equation}
    (p_i\cdot\epsilon_j)\varepsilon^{\mu\nu\rho\sigma}\epsilon_\mu^i\epsilon_\nu^kp_\rho^ip_\sigma^j
    \xrightarrow[i\leftrightarrow k]{}
    (p_i\cdot\epsilon_j)\varepsilon^{\mu\nu\rho\sigma}\epsilon_\mu^i\epsilon_\nu^kp_\rho^kp_\sigma^j,
\end{equation}
but the sum of these two terms vanishes by momentum conservation and antisymmetry of the Levi-Civita tensor:
\begin{equation}
    (p_i\cdot\epsilon_j)\varepsilon^{\mu\nu\rho\sigma}\epsilon_\mu^i\epsilon_\nu^k\left(p_\rho^i+p_\rho^k\right)p_\sigma^j
    = - (p_i\cdot\epsilon_j)\varepsilon^{\mu\nu\rho\sigma}\epsilon_\mu^i\epsilon_\nu^kp_\rho^jp_\sigma^j = 0.
\end{equation}

These cancellations do not occur in theories with colour (the polarization vectors then carry a colour index, which is not symmetrized over). This is in fact related to  color-kinematics duality \cite{Bern:2010ue}: non-trivial kinematics requires non-trivial colour structure.

\subsection{Dispersion relations}

We want to derive bounds on the \emph{low-energy} amplitudes using unitarity of the \emph{high-energy} amplitudes. These are related by dispersion relations, which are derived with the following standard assumptions: 
\begin{itemize}
	\item The theory is weakly coupled at least up to the scale $M$; below this scale the amplitude is well described by the tree-level EFT amplitude (see footnote \ref{ftnt:loops}).
	\item Regge boundedness: for fixed $t<0$ the amplitude falls off faster than $s^2$ for large values of $s$, \emph{i.e.}
	\begin{equation}
		\lim_{|s|\to\infty} \frac{A(s,t)}{s^2} =0.
	\end{equation}
	\item Analyticity: for fixed $t<0$, $A(s,t)$ is analytic in the upper half plane $\mathrm{Im}(s)>0$.
 \item Real analyticity: $A^*(s, t) = A(s^*, t^*)$ for parity-respecting theories. This allows us to define $A(s,t)$ in the lower-half $s$-plane by $A^*(s,t)$ in the upper-half $s$-plane, when $t$ is real. 
\end{itemize}

\subsubsection{Contour integral}

Regge-boundedness implies that the following contour integrals vanish
\begin{equation}
\label{eq:mastersumrule}
    \mathcal I_{k}(t)=\oint_\infty\frac{ds}{2\pi i }\frac{\tilde A(s,t)}{s^{k+1}}=0, \qquad k=2,3,\ldots,
\end{equation}
where the contour is a circle at infinity in the complex $s$ plane. Analyticity implies that the contour can be deformed to encircle the low-energy region $|s|\lesssim M^2$, where it produces low-energy poles calculable from the EFT, plus integrals along high-energy branch cuts where the physics is unknown but unitary:
\begin{itemize}
	\item Branch cuts: $(-\infty,-M^2-t+4m^2)$ and $(M^2,\infty)$.
	\item Poles: $s=0$, \text{other potential poles of $\tilde{A}$}.
\end{itemize}
We call the equation $ \mathcal I_{k}(t)=0$ a sum rule.

In our conventions, all helicity amplitudes have a pole at $s=4m^2$. The contour can be deformed to produce the equation $\mathcal I_k^I(t)=0$, taking the form
\begin{equation}\label{eq:matrixeq}
	\begin{split}
& 
\left( \res_{s=0}+\res_{s=4m^2} \right)\left[\frac{\tilde A^I_{\Low}(s,t)}{s^{1+k}}\right] 
\\&
\quad = \int_{-\infty}^{-M^2-t+4m^2} \frac{\mathrm{d}s}{\pi}\,\disc\left[ \frac{\tilde A^I_{\High}(s,t)}{s^{k+1}} \right]
  +
  \int_{M^2}^{\infty} \frac{\mathrm{d}s}{\pi}\, \disc\left[ \frac{\tilde A^I_{\High}(s,t)}{s^{k+1}}  \right].
	\end{split}
\end{equation}
As stated above, for parity respecting theories, we have real analyticity, which implies
\begin{align}
    \disc[A(s,t)] = \lim_{\epsilon \to 0} \frac{1}{2i} \left[ A(s+ i \epsilon, t) - A(s - i \epsilon, t) \right] = \IM A(s, t)
\end{align}
so the discontinuity of the amplitude is equal to the imaginary part.

\subsubsection{Crossing the cuts}
\label{sec:crossing}
The two integrals in \eqref{eq:matrixeq} are related by a crossing transformation that exchanges $s\leftrightarrow u=4m^2-s-t$. Given the symmetry properties of the functions \eqref{eq:functions}, we can express
\begin{equation}
    \label{eq:functioncrossing}
	F_I(4m^2-s-t,t)=\sum_J C^{su}_{IJ} F_J(s,t),
\end{equation}
where $C^{su}$ is a block diagonal matrix with blocks
\begin{align}
C^{su}_{[1\text{---}3]}&=C^{su}_{[4\text{---}6]}=\left(\begin{smallmatrix}
    0 & 0 & 1 \\
 0 & 1 & 0 \\
 1 & 0 & 0
\end{smallmatrix}\right), & C^{su}_{[7\text{---}12]}&=\left(\begin{smallmatrix}
0 & 0 & 0 & 0 & 0 & 1 \\
 0 & 0 & 0 & 0 & 1 & 0 \\
 0 & 0 & 0 & 1 & 0 & 0 \\
 0 & 0 & 1 & 0 & 0 & 0 \\
 0 & 1 & 0 & 0 & 0 & 0 \\
 1 & 0 & 0 & 0 & 0 & 0 
\end{smallmatrix}\right), & C^{su}_{[13\text{---}17]}&=\left(\begin{smallmatrix}
 0 & 0 & 1 & 0 & 1 \\
 0 & 1 & 0 & 0 & 1 \\
 1 & 0 & 0 & 0 & 1 \\
 0 & 0 & 0 & 1 & -1 \\
 0 & 0 & 0 & 0 & -1 
\end{smallmatrix}\right).
\label{eq:Csu}
\end{align}
Therefore we can implement crossing for the amplitude as
\begin{equation}\label{eq:matrixcrossing}
	\begin{split}
		\tilde A^I(u,t) &=\sum_{J} \tensor{\tilde E}{^I_J}(u,t)F_J(u,t) =
  \sum_{J,K} \tensor{\tilde E}{^I_J}(u,t)C^{su}_{JK}F_K(s,t) \\
    &=\sum_{J,K,L}\tensor{\tilde E}{^I_J}(u,t)C^{su}_{JK}\,\tensor {(\tilde E^{-1})}{_K^L}(s,t) \tilde A^L(s,t).
	\end{split}
\end{equation}
where $u=4m^2-s-t$. 

Since we always work with the amplitudes $\tilde A$, this way of formulating crossing is free from any additional signs that may arise due to analytic continuations across the $\sqrt{stu}$ branch cuts. See \textit{e.g.} \cite{Hebbar:2020ukp} for an alternative formulation which correctly deals with the continuation around the branch cut.

With the change of variables $s \to u(s)=4m^2-s-t$ \eqref{eq:matrixeq} becomes
\begin{equation}\label{eq:disprelmatrix}
	\begin{split}
		&\left( \res_{s=0}+\res_{s=4m^2} \right)
  \left[\frac{\tilde A^I_{\Low}(s,t)}{s^{1+k}}\right] =\int_{M^2}^{\infty} \frac{\mathrm{d}s}{\pi}\, \disc
  \Bigg[\bigg( \frac{\tilde A^I_{\High}(s,t)}{s^{1+k}} \\
		& 
  \qquad \qquad
  -\sum_J \big(\tilde E(u(s),t)C^{su}\tilde E^{-1}(s,t) \big)^{IJ} 
  \frac{\tilde A^J_{\High}(s,t)}{(4m^2-s-t)^{1+k}} \bigg)\Bigg].
	\end{split}
\end{equation}

For crossing $s \leftrightarrow t$, we similarly have
\begin{equation}
    \label{eq:stcrossing}
	F_I(t,s)=\sum_J C^{st}_{IJ} F_J(s,t),
\end{equation}
where 
\begin{align}
C^{st}_{[1\text{---}3]}&=C^{st}_{[4\text{---}6]}=\left(\begin{smallmatrix}
    0 & 1 & 0 \\
 1 & 0 & 0 \\
 0 & 0 & 1
\end{smallmatrix}\right), & C^{st}_{[7\text{---}12]}&=\left(\begin{smallmatrix}
0 & 0 & 1 & 0 & 0 & 0 \\
 0 & 0 & 0 & 1 & 0 & 0 \\
 1 & 0 & 0 & 0 & 0 & 0 \\
 0 & 1 & 0 & 0 & 0 & 0 \\
 0 & 0 & 0 & 0 & 0 & 1 \\
 0 & 0 & 0 & 0 & 1 & 0 
\end{smallmatrix}\right), & C^{st}_{[13\text{---}17]}&=\left(\begin{smallmatrix}
 0 & 1 & 0 & 1 & 0 \\
 1 & 0 & 0 & 1 & 0 \\
 0 & 0 & 1 & 1 & 0 \\
 0 & 0 & 0 & -1 & 0 \\
 0 & 0 & 0 & -1 & 1 
\end{smallmatrix}\right).
\label{eq:Cst}
\end{align}

\subsection{Unitarity and discrete symmetries}

The high-energy form of the amplitude is allowed to be anything which is consistent with unitarity. As such, we parametrize it with partial waves. In 4d, this takes the form
\begin{equation}\label{eq:PWD}
    A^{\lambda_1\lambda_2\lambda_3\lambda_4}(s,t) =\sum_\ell 16\pi (2\ell+1)\sqrt{\frac{s}{s-4m^2}} d^{(\ell)}_{\lambda_1-\lambda_2,\lambda_4-\lambda_3}(\theta) A^{\lambda_1\lambda_2\lambda_3\lambda_4}_\ell(s) \, ,
\end{equation}
where the scattering angle $\theta$ is given by 
\begin{align} 
\cos \theta = 1 + \frac{2t}{s - 4m^2} \, .
\end{align}
The dynamical information is contained in the partial wave density $A_\ell$. The Wigner $d$ matrices $d^{(\ell)}_{a,b}$, represent the kinematic part. We review their definition in appendix~\ref{app:wigner}. For the case of zero external helicities, they reduce to the more familiar Legendre polynomials~\cite{Correia:2020xtr,Buric:2023ykg}.

We also define
\begin{equation}\label{eq:rhodef}
    \rho_\ell^{\lambda_1\lambda_2\lambda_3\lambda_4}(s) \equiv \disc\big[ A^{\lambda_1\lambda_2\lambda_3\lambda_4}_\ell(s) \big],
\end{equation}
which is the discontinuity of the partial wave density on the right-hand cut. Let us analyze the former in detail:
\begin{equation}
    \int_{M^2}^\infty \frac{ds}{\pi s} \disc\left[\frac{  A^{\lambda_1\lambda_2\lambda_3\lambda_4}(s,t) }{s^k}\right] =\sum_{\ell}16(2\ell+1)\int_{M^2}^\infty \frac{ds}s \sqrt{\frac{s}{s-4m^2}}\, \rho_{\ell}^{\lambda_1\lambda_2\lambda_3\lambda_4}(s)\,  \frac{d^{\ell}_{\lambda_{12},\lambda_{43}}(\theta)}{s^k}.
\end{equation}
The right-hand side of this expression consists of three ingredients:
\begin{itemize}
    \item A sum and integral with an overall \textbf{positive measure}: $\sum_{\ell}\int ds \,\mu(s)$, with
    \begin{equation}
        \mu(s)=\frac1s \sqrt{\frac s{s-4m^2}}.
        \label{eq:positiveMeasure}
    \end{equation}
    \item A \textbf{kinematic part}: $d_{\lambda_{12},\lambda_{43}}^\ell(\theta)/s^k$. This part carries no dynamical information and it can always be written in explicit form. Any additional factors put on the left-hand side, for instance the denominator $s^{k}$ in the master sum rule \eqref{eq:mastersumrule}, will always be included in this remaining kinematic part. In general, we introduce the notation $V^{\lambda_1\lambda_2\lambda_3\lambda_4}(\ell,s,t)$ for such a kinematic part. 
    \item A \textbf{dynamical part}: $16(2\ell+1)\rho_{\ell}(s)^{\lambda_1\lambda_2\lambda_3\lambda_4}(s)$. For scalar amplitudes or general forward amplitudes, it is manifestly positive $\rho_\ell(s)\geq 0$. More generally, including the case of helicity amplitudes of this paper, unitarity implies that a matrix of $\rho$'s is positive semi-definite. These densities are not fixed, and we remain ignorant of them---we merely find regions of low-energy coefficients compatible with their positivity.
\end{itemize}
The purpose of the rest of this section is to rewrite the dynamical part in a way that makes positivity manifest and is suitable for numerical implementation.
In practice, this will lead to the following modifications: the positive measure will remain invariant, but the dynamical part and the kinematic part will take a matrix form.

\paragraph{Generalized optical theorem}

The unitarity of the $S$-matrix implies
\begin{equation}
	S^\dagger S =1 \quad \Longrightarrow \quad 2\IM[T] =T^\dagger T.
\end{equation}
Contracting the right-hand side with external states of definite helicities $\lambda_i$ and spin $\ell$, and recalling that $\disc[A] = \IM[A]$, gives
\begin{equation}\label{eq:complrel}
\begin{split}
	\disc[A^{\lambda_1 \lambda_2 \lambda_3 \lambda_4}_\ell(s)] \ &= \ \frac12 \langle s,\ell,\lambda_3,\lambda_4 |T^\dagger T|s,\ell,\lambda_1,\lambda_2 \rangle \\ &= \ \frac12 \sum_X \langle s,\ell,\lambda_3,\lambda_4 |T^\dagger|X,\ell\rangle\langle X,\ell|T|s,\ell,\lambda_1,\lambda_2 \rangle,
\end{split}
\end{equation}
where we used completeness and inserted a set of intermediate states with spin $\ell$ labeled by $X$.

If we define
\begin{equation}
	c_{\ell,X}^{\lambda_i \lambda_j}(s) \equiv \sqrt{8(2\ell+1)} \langle X,\ell|T|s,\ell,\lambda_i,\lambda_j \rangle
\end{equation}
and use \eqref{eq:rhodef} we get 
\begin{equation}
	16(2\ell+1)\rho_\ell^{\lambda_1\lambda_2\lambda_3\lambda_4}(s) =\sum_X 
 \Big( c_{\ell,X}^{-\lambda_3 -\lambda_4}(s) 
 \Big)^*c_{\ell,X}^{\lambda_1 \lambda_2}(s) ,
\end{equation}
\emph{i.e.} by viewing pairs of polarizations as matrix indices, the spectral density is a positive definite Hermitian matrix.

\paragraph{Positivity of the dynamical part}

Now we will rewrite the dynamical part to make positivity manifest. The optical theorem means that we will consider expressions of the form
\begin{equation}
\rho^{\lambda_1\lambda_2\lambda_3\lambda_4}_\ell(s)V^{\lambda_1\lambda_2\lambda_3\lambda_4}(\ell,s)=\frac12 \sum_X \langle -\lambda_3-\lambda_4|T^\dagger|X\rangle V^{\lambda_1\lambda_2\lambda_3\lambda_4}(\ell,s)\langle X|T|\lambda_1\lambda_2\rangle, 
\end{equation}
where we have suppressed $s$ and $\ell$ in the external states, and where $V^{\lambda_1\lambda_2\lambda_3\lambda_4}(\ell,s)$ is representing the kinematic part of the sum rule. We could write this as\footnote{The minus signs in the outgoing polarizations originate from our choice of the all-ingoing convention.}
\begin{equation}
16(2\ell+1)\rho^{\lambda_1\lambda_2\lambda_3\lambda_4}_\ell(s)V^{\lambda_1\lambda_2\lambda_3\lambda_4}(\ell,s)=\sum_X \left(c_{X,\ell}^{-\lambda_3-\lambda_4}(s)\right)^* V^{\lambda_1\lambda_2\lambda_3\lambda_4}(\ell,s) c_{X,\ell}^{\lambda_1\lambda_2}(s),
\end{equation}
where $c_{X,\ell}^{\lambda_i\lambda_j}(s)=\sqrt{8(2\ell+1)}\langle X|T|\lambda_1\lambda_2\rangle$.

We will now write $c^{\lambda_1,\lambda_2}_{X,\ell}$ in terms of a fixed basis vector $\boldsymbol c_{X,\ell}(s)=(c^{++}_{X,\ell},c^{+0}_{X,\ell},\ldots )^T$ of length $9$ (this basis is explicitly shown below) giving,
\begin{align}
&16(2\ell+1)\rho^{\lambda_1\lambda_2\lambda_3\lambda_4}_\ell(s)V^{\lambda_1\lambda_2\lambda_3\lambda_4}(\ell,s)
\label{eq:matrixV}
= \sum_X \mathbf c_{X,\ell}(s)^\dagger \mathbf V^{\lambda_1\lambda_2\lambda_3\lambda_4}(\ell,s) \mathbf c_{X,\ell}(s)\,,
\end{align}
where we now have a matrix-valued kinematic part, and two vector-valued densities (half-amplitudes). 
Here the matrix $\mathbf V^{\lambda_1\lambda_2\lambda_3\lambda_4}(\ell,s) $ is $9\times 9$ in a basis of polarizations, and the only non-zero entry is the one corresponding to the arguments $\lambda_i$.\footnote{Specifically, the non-zero element sits at the row labeled by $(-\lambda_3,-\lambda_4)$ and the column labeled by $(\lambda_1,\lambda_2)$.} Below, we will reduce this further, by separating the sum over $X$ based on assumptions of spin and parity.

From \eqref{eq:matrixV}, we will be able to write all sum rules on the form
\begin{equation}
\label{eq:sumruleform}
g_i = \sum_\ell \int_{M^2}^\infty ds \mu(s)  \sum_X \mathbf c_{X,\ell}(s)^\dagger \mathbf V_{g_i}(\ell,s) \mathbf c_{X,\ell}(s).
\end{equation}
Here $g_i$ represents a specific low-energy coefficient (sum rules of this form were written as $g_i=\langle \mathbf V_{g_i}\rangle$ in related previous work \cite{Caron-Huot:2020cmc,Henriksson:2022oeu}). We have used the fact that using crossing, the left-hand cut can also be brought to the same form as the right-hand cut. Since crossing mixes the different amplitudes, the left-hand cut contribution from a given helicity amplitude will give rise to a matrix $\mathbf V$ where now many entries may be non-zero. 

The advantage of  rewriting the high-energy part in the form \eqref{eq:sumruleform} is that the dynamical part---the vector $\boldsymbol c_{X,\ell}(s)$---is the same for all sum rules. It is therefore possible to add linear combinations of sum rules in an algorithmic way. Specifically, we use the fact that for any real-valued symmetric matrix $\mathbf V$ and complex vector $\boldsymbol c$, the positive-semi-definiteness condition $\mathbf V \succcurlyeq 0$ implies $\boldsymbol c^\dagger\mathbf V\boldsymbol c\geq 0$.

\paragraph{Discrete symmetries}
Parity and boson exchange symmetries require that (we omit the $s$ dependence to avoid cluttering)
\begin{equation}\label{eq:constparityB}
\begin{split}
	c_{\ell,X}^{\lambda_i \lambda_j}& = P_X c_{\ell,X}^{-\lambda_j -\lambda_i}, \\
	c_{\ell,X}^{\lambda_i \lambda_j}& =(-1)^{\ell+\lambda_i+\lambda_j}c_{\ell,X}^{\lambda_j \lambda_i},
\end{split}
\end{equation}
where $P_X$ can be $+1$ or $-1$, and we used that the vector particle has spin 1.

Recall that for massive spin-1, $\lambda_i \in\{0,\pm1\}$. These equations can be easily solved once and for all. In what follows and throughout this article we will use $\pm$ instead of $\pm1$ in the superscripts.

Since there are four sectors, we will call the corresponding $c$'s using different letters. The only independent solutions of \eqref{eq:constparityB} are:\footnote{Terms that are not listed here vanish.}
\begin{description}
	\item[(ee)] $P_X=1$, $\ell$ even:
		\begin{equation}\label{eq:cee}
			a_{\ell,X}^{++}, \quad a_{\ell,X}^{+-}, \quad a_{\ell,X}^{+0}, \quad a_{\ell,X}^{00}.
		\end{equation}
The remaining entries are related to these by
\begin{align}
    a_{\ell,X}^{--} =a_{\ell,X}^{++}, & \quad & a_{\ell,X}^{-+}=a_{\ell,X}^{+-}, & \quad & a_{\ell,X}^{0+} =-a_{\ell,X}^{+0}, & \quad & a_{\ell,X}^{0-} =a_{\ell,X}^{+0}, & \quad & a_{\ell,X}^{-0} =-a_{\ell,X}^{+0}.
\end{align}
	\item[(oe)] $P_X=-1$, $\ell$ even:
	\begin{equation}\label{eq:coe}
		b_{\ell,X}^{++}, \quad b_{\ell,X}^{+0}.
	\end{equation}
 They are supplemented by
 \begin{align}
    b_{\ell,X}^{--} =-b_{\ell,X}^{++}, & \quad & b_{\ell,X}^{-0}=b_{\ell,X}^{+0}, & \quad & b_{\ell,X}^{0+} =-b_{\ell,X}^{+0}, & \quad & b_{\ell,X}^{0-} =-b_{\ell,X}^{+0}.
\end{align}
	\item[(eo)] $P_X=1$, $\ell$ odd:
	\begin{equation}\label{eq:ceo}
		c_{\ell,X}^{+-}, \quad c_{\ell,X}^{+0}.
	\end{equation}
 They are supplemented by
 \begin{align}
    c_{\ell,X}^{-+} =-c_{\ell,X}^{+-}, & \quad & c_{\ell,X}^{-0}=c_{\ell,X}^{+0}, & \quad & c_{\ell,X}^{0+} =c_{\ell,X}^{+0}, & \quad & c_{\ell,X}^{0-} =c_{\ell,X}^{+0}.
\end{align}
	\item[(oo)] $P_X=-1$, $\ell$ odd:
	\begin{equation}\label{eq:coo}
		d_{\ell,X}^{+0}.
	\end{equation}
It is supplemented by
 \begin{align}
     d_{\ell,X}^{-0} =-d_{\ell,X}^{+0}, & \quad & d_{\ell,X}^{0+} =d_{\ell,X}^{+0}, & \quad & d_{\ell,X}^{0-} =-d_{\ell,X}^{+0}.
 \end{align}
\end{description}

The above results are summarized in Table~\ref{tab:summary3pt}.
We can now substitute any factor with the corresponding element of the independent solution \eqref{eq:cee}, \eqref{eq:coe}, \eqref{eq:ceo}, \eqref{eq:coo}. Eventually, we obtain sum rules that can be written as
\begin{align}
  g_i  &= \int_{M^2}^\infty 
  ds\mu(s)
  \Bigg( 
  \sum_{\text{even }\ell,X} \mathbf{c}^{ee}_{\ell,X}(s)^\dagger\mathbf{V}^{ee}_{g_i}(\ell,s) \mathbf{c}^{ee}_{\ell,X}(s) 
  +
  \sum_{\text{even }\ell,X} \mathbf{c}^{oe}_{\ell,X}(s)^\dagger \mathbf{V}^{oe}_{g_i}(\ell,s) \mathbf{c}^{oe}_{\ell,X} (s)
  \nonumber
  \\
		&
 \qquad\qquad\qquad\qquad  +
  \sum_{\text{odd }\ell,X} \mathbf{c}^{eo}_{\ell,X} (s)^\dagger\mathbf{V}^{eo}_{g_i}(\ell,s) \mathbf{c}^{eo}_{\ell,X}(s)   
  +
  \sum_{\text{odd }\ell,X} \mathbf{c}^{oo}_{\ell,X} (s)^\dagger\mathbf{V}^{oo}_{g_i}(\ell,s) \mathbf{c}^{oo}_{\ell,X}(s) \Bigg),
  \label{eq:gisumrule}
\end{align}
where we defined the vectors (which depend on $s$)
\begin{equation}
	\mathbf{c}^{ee}_{\ell,X} \equiv\begin{pmatrix}
		 a_{\ell,X}^{++}   \\  a_{\ell,X}^{+-}   \\  a_{\ell,X}^{+0}   \\  a_{\ell,X}^{00}  
	\end{pmatrix}, \quad
	\mathbf{c}^{oe}_{\ell,X} \equiv\begin{pmatrix}
		 b_{\ell,X}^{++}   \\   b_{\ell,X}^{+0}  
	\end{pmatrix}, \quad
	\mathbf{c}^{eo}_{\ell,X} \equiv\begin{pmatrix}
		  c_{\ell,X}^{+-}   \\   c_{\ell,X}^{+0}  
	\end{pmatrix}, \quad
	\mathbf{c}^{oo}_{\ell,X} \equiv 
		  d_{\ell,X}^{+0}  .
\end{equation}
The expression \eqref{eq:gisumrule} contains four matrices $\mathbf{V}^{ee}_{g_i}(\ell,s)$, $\mathbf{V}^{oe}_{g_i}(\ell,s)$, $\mathbf{V}^{eo}_{g_i}(\ell,s)$ and $\mathbf{V}^{oo}_{g_i}(\ell,s)$ with dimension $4\times 4$, $2\times 2$, $2\times 2$ and $1\times 1$ respectively. Specifying a sum rule amounts to giving these four matrices.
In practice, the cases $\ell=0$ and $\ell=1$ need to treated separately, since in these cases not all $c^{\lambda_1\lambda_2}_{\ell,X}$ exist.\footnote{\label{footnote:couplings}Specifically, this gives rise to a length-2 vector $\mathbf{c}^{ee}_{0,X}=(a_{0,X}^{++}  ,  a_{0,X}^{00}  )^T$ in the $ee$ case, and scalars $\mathbf{c}^{oe}_{0,X}=b_{0,X}^{++}$, $\mathbf{c}^{eo}_{1,X}=c_{1,X}^{+0}$ and $\mathbf{c}^{oo}_{1,X}=d_{1,X}^{+0}$ in the other cases.}

We will compute these matrices for both the left and right cut and follow the usual logic: if a matrix $\mathbf{V}$ is symmetric, real and positive-semi-definite, then $\mathbf{c}^\dagger \mathbf{V} \mathbf{c} \geqslant0$ for any complex vector $\mathbf{c}$.

\begin{table}[]
    \centering
    \begin{tabular}{|c|c|c|c|}
    \hline
Sector & Spin &  Parity & Number of $c^{\lambda_1\lambda_2}_{\ell,X}$\\
\hline\hline
\textbf{(ee)} & $\ell=0$ & $+$ & 2\\
\hline
\textbf{(ee)} & $\ell\geq2,$ even & $+$ & 4\\
\hline
\hline
\textbf{(eo)} & $\ell=0$ & $-$ & 1\\
\hline
\textbf{(eo)} & $\ell\geq2,$ even  & $-$ & 2\\
\hline
\hline
\textbf{(oe)} & $\ell=1$ & $+$ & 1\\
\hline
\textbf{(oe)} & $\ell\geq3$, odd & $+$ & 2\\
\hline
\hline
\textbf{(oo)} & $\ell\geq1$, odd & $-$ & 1
\\\hline
    \end{tabular}
      \caption{Summary of independent three point vertices $c^{\lambda_1\lambda_2}_{\ell,X}$ for a state $X$ with spin $\ell$ and parity $P_X$.}
    \label{tab:summary3pt}
\end{table}

\section{EFT bootstrap}
\label{sec:general}

The method of using sum rules and null constraints (\emph{i.e.} sum rules which are zero on the low-energy side, see section \ref{sec:NCs}) to derive bounds via semi-definite optimisation was formalized in \cite{Caron-Huot:2020cmc} and further developed in \cite{Henriksson:2022oeu}. We refer to this setup as EFT bootstrap. For a bound involving two observables $g_i$, $g_j$, and $N$ null constraints, one considers the system
\begin{equation}
\label{eq:semidefiniteProblemSchematic}
    \begin{pmatrix}
        g_i \\ g_j \\ 0\\\vdots \\0
    \end{pmatrix}
    = \begin{pmatrix}
        \left\langle \mathbf V_{g_i}
        \right\rangle^{ee}+\ldots
        \\
        \left\langle \mathbf V_{g_j}
        \right\rangle^{ee}+\ldots
        \\
        \left\langle \mathbf V_{n_1}
        \right\rangle^{ee}+\ldots
        \\
        \vdots
        \\
        \left\langle \mathbf V_{n_N}
        \right\rangle^{ee}+\ldots
    \end{pmatrix}.
\end{equation}
Here the ``bracket notation'' represents the sum and integral defined in \eqref{eq:gisumrule}. Once such problem is formulated, it is possible to use semi-definite programming via SDPB  \cite{Simmons-Duffin:2015qma} to find the extremal values of the observables $g_i$, $g_j$.

In this section we will discuss how to compute the ingredients entering \eqref{eq:semidefiniteProblemSchematic} in the case of scattering of a generic massive vector particle. First, we will consider choices for the observables, and then discuss the systematic computation of null constraints.

The presence of a mass complicates things considerably. A main issue is that the \mbox{positivity} properties are most easily expressed in terms of the helicity amplitudes $A^{\lambda_1, \lambda_2, \lambda_3, \lambda_4}$, but these have complicated crossing properties: crossing particle $2$ and $3$, for instance, takes one out of the center-of-mass frame, so exchanging the momenta must be followed by a boost to restore the amplitude to that frame \cite{Hebbar:2020ukp}. However the boost acts in a very complicated way on the particles with spin, essentially mixing all of the amplitudes. So while for a photon, we have simple relations like $A^{++--}(s, t) = A^{+--+}(u, t)$, for massive vectors the ``crossed'' amplitude $ A^{+--+}(u, t)$ will in general be a linear combination of all 17 helicity amplitudes  $A^{\lambda_1, \lambda_2, \lambda_3, \lambda_4}(s,t)$.

The solution to this problem is that we will use the \textit{functions} $F_I(s,t)$, which have much simpler crossing properties (see equation~\ref{eq:functioncrossing}) for the null constraints.  On the other hand, we will choose observables to bound which are formed from the (helicity) amplitudes $A^I = E^I{}_J F_J$, where the positivity properties are more manifest.

\subsection{Observables}
\label{sec:observables}

An issue with defining observables is this: consider a function $f(s,t)$ which  is simply a power series in $s$ and $t$, 
\begin{align}
    f(s,t) = f_{0,1}(s + t) + f_{0,2}(s + t)^2 + f_{1,0} s t + \ldots .
\end{align}
Then the crossed version of the function:
\begin{align}
    f(u,t) = f_{0,1}(4m^2 - s) + f_{0,2}(4m^2 - s)^2 + f_{1,0} (4m^2 - s - t) t + \ldots  \,.
\end{align}
If we expand $f(u,t)$ in small $s$ and $t$, we no longer find a single EFT coefficient at each term in the expansion. Instead we find an infinite sum of coefficients with different powers of $m^2$. This problem may be avoided by taking some distance from the function coefficients, and defining observables directly at the level of the amplitude.

Starting from the dispersion relation \eqref{eq:matrixeq}, we define observables as\footnote{In section~\ref{sec:forward-limit-bounds} only, where we consider bounds in the forward limit, we will use   observables with manifestly crossing symmetric denominators $(s-2m^2)^{k+1}$
 to get simpler analytic expressions for these bounds.}
\begin{align}
    A^{\lambda_1\lambda_2\lambda_3\lambda_4}_{k,l} :=\left(\frac{\partial}{\partial t}\right)^l\left( \res_{s=0}+\res_{s=4m^2} \right)
  \left[\frac{\tilde A^{\lambda_1\lambda_2\lambda_3\lambda_4}_{\Low}(s,t)}{s^{k+1}}\right]  \bigg|_{t=0}
  \label{eq:arcs}
\end{align} 
For the case $l=0$, we just write
\begin{equation}
    A^{\lambda_1\lambda_2\lambda_3\lambda_4}_{k} :=A^{\lambda_1\lambda_2\lambda_3\lambda_4}_{k,0}.
    \label{eq:arcsl0}
\end{equation} 
The integrals described here are referred to as \textit{arcs} in the literature, see for instance \cite{Bellazzini:2020cot}. Arcs are  observables that, contrary to the EFT coefficients, are well defined also beyond the tree-level and even at strong coupling. For instance, when calculable IR loop effects are taken into account,  EFT coefficients run and are not observable, while arcs are still well-defined \cite{Bellazzini:2021oaj}. In the present case, we use them because their simpler relation to the helicity amplitudes leads to positivity properties that are not manifest in the EFT coefficients. 

\subsection{Null constraints}
\label{sec:NCs}

It is well understood that the bounds are significantly strengthened by including \emph{null constraints}, which are sum rules involving only high-energy data \cite{Tolley:2020gtv,Caron-Huot:2020cmc}. These are typically obtained by computing the same EFT coefficient using multiple sum rules -- the resulting high-energy parts are then required to be equal, and results in a constraint on the spectral densities. Thus these constraints can be thought of as arising from the sparseness of low-energy EFT coefficients. Let us see how to easily generate null constraints for our massive spinning setup.

\subsubsection{Regge bounds for reduced amplitudes}

\label{sec:regge-reduced}

To write a dispersive integral for the functions $F_J$, we need to understand their Regge behavior. Throughout this paper we assume that \begin{equation}
    |A^{\lambda_1\lambda_2\lambda_3\lambda_4}|< |s|^2
\end{equation}
as $|s|\to\infty$.
We refer to this equation as the Regge bound (and will drop the absolute values below). We will now use this to determine, in a systematic way, the Regge bound of each of the reduced amplitudes.
Following \cite{Caron-Huot:2022jli}, our analysis will lead to a \emph{general} Regge bound satisfied by all reduced amplitudes, and a finite set of \emph{stricter} bounds satisfied by specific combinations of amplitudes.

First we recall that by definition
\begin{equation}
   F_I(s,t) =  \sum_J \tensor {(E^{-1})}{_I^J}(s,t) A^J(s,t) = \sum_J \tensor {(\tilde E^{-1})}{_I^J}(s,t) \tilde A^J(s,t),
\end{equation}
where also the functions $\tilde A^I(s,t)$ are bounded by $s^2$:
\begin{equation}
\label{eq:Atildelimit}
    \lim_{|s|\to\infty} \frac{\tilde A^I(s,t)}{s^2}=0.
\end{equation}
What we are looking for is specific combinations of the functions which have particular Regge behavior. Consider a vector $c_I$, which may be a function of $s$ and $t$. Then we have
\begin{align}
    c_I(s,t) F_I(s,t) \ = \ c_I(s,t) (\tilde{E}^{-1})_I{}^J(s,t) \tilde{A}^J(s,t) \, .
\end{align}
Each element $\tilde{A}^J$ grows slower than $s^2$, and these are all independent in principle. So at large $s$
\begin{equation}\label{eq:ifaoif}
    c_I F_I < s^m \quad \leftrightarrow \quad c_I (\tilde{E}^{-1})_I{}^J \leq s^{m-2}\,,
\end{equation}
 which must hold for all values of $J$.

\paragraph{General bound from constant $c_I$}

The simplest constraint comes from assuming that $c$ is constant in $s$. Then we merely need to expand $\tilde{E}^{-1}$ at large $s$. From the explicit form of $\tilde E^{-1}$, we find that the leading power is $s$, meaning that
\begin{align}
    \tilde{E}^{-1}(s,t) \ = \ s e_1(t) + e_0(t) + \frac{1}{s} e_{-1}(t) + \ldots .
\end{align}
As a result, any generic $c_I$ will satisfy
\begin{align}
    c_I (\tilde{E}^{-1})_I{}^J \leq s^1 \, .
\end{align}
This implies that $m = 3$ in equation \eqref{eq:ifaoif}, so all $F_I$ must satisfy
\begin{align}
    F_I(s, t) < s^3
\end{align}
at large $s$. 

\paragraph{Improved bounds}

Many stronger bounds are possible because if we allow $c_I$ to depend on $s$ and $t$ then we can sometimes cancel the leading large-$s$ behavior of $c_I (\tilde{E}^{-1})_I{}^J$ to get improved behavior. Because we are expanding at large-$s$ and only encounter rational functions of $s$ and $t$, it is sensible to use a series expansion in powers of $1/s$. In principle, these could be arbitrarily large series, but in practice, we found no new lower-order null constraints going beyond order $1 / s^3$. So we first write the most general series
\begin{align}
    c_I(s,t) = \sum_{a, b < 4} c_I^{a,b} s^{-a} t^{-b}.
\end{align}
Then we compute the large-$s$ behavior of $c_I (\tilde{E}^{-1})_I{}^J$, and require that the leading entry is $s^{m-2}$. This implies that $c_I F_I < s^m$, so we have a combination with Regge behavior better than $s^m$. Since all $F$s have behavior better than $s^3$, we only need to do this for $m < 3$, and in practice we are able to enumerate all solutions.

\paragraph{Improvement in the crossed channel}

For the application we will need below, \textit{i.e.} deriving null constraints, we will actually need to determine combinations that have particular Regge behavior at large $s$ \textit{and} at large $t$. This means that we need to effectively do the procedure above in both channels. Therefore we start with 
\begin{align}
    c_I(s,t) = \sum_{a, b < 4} c_I^{a,b} s^{-a} t^{-b} 
\end{align}
and then we require that 
\begin{align}
    c_I(s, t) (\tilde{E}^{-1})_I{}^J (s,t) \sim s^{m-2}
\end{align}
at large $s$, but we also want
\begin{align}
    c_I(s,t) (\tilde{E}^{-1})_I{}^J (s,t) \sim t^{\ell-2}
\end{align}
at large $t$. Solving both of these constraints simultaneously leads to combinations of the form $c_I(s,t) F_I(s,t)$ which satisfy
\begin{align}
    \lim_{s \to \infty} c_I(s,t) F_I(s,t) \ &< \ s^m \, , \\
    \lim_{t \to \infty} c_I(s,t) F_I(s,t) \ &< \ t^\ell  \, .
\end{align}
This, finally, is the input that we shall need to derive null constraints. Below we shall report the numbers of functions with different improved Regge behavior, but first let us explain how the null constraints are obtained. 

\subsubsection{Systematics for null constraints}

There is a straightforward way to derive null constraints by writing down every possible sum rule up to a certain order---that is to consider the dispersion integral for each amplitude $A^I$, each power $k$ where the integrals converge, and for each power of $t$. Then one finds that there are more sum rules than low-energy coefficients, and many of the sum rules simply relate different integrals of the spectral densities, and thus are constraints on high-energy data only.

In our case, with a non-zero mass, the large number of coefficients and the fact that all helicities are mixed by crossing makes this method very challenging, so we will use a more efficient method. Here we shall follow the discussion of null constraints given in \cite{Albert:2023jtd}, which we will generalize to the massive case.

Consider a function $c_I F_I$ formed by contracting some (possibly $s$ and $t$ dependent) vector $c_I$ with $F_I(s,t)$. Suppose that $c_I F_I$ satisfies at fixed $t<0$
\begin{align}\label{eq:reggebehavior}
    \lim_{s \to \infty} \frac{c_I(s,t) F_I(s,t)}{s^m} \ = \ 0
\end{align}
for some integer $m$. Furthermore suppose that 
\begin{align}
    \lim_{t \to \infty} \frac{c_I(s,t) F_I(s,t)}{t^\ell}  = 0 \, ,
\end{align}
or in other words, that $c_I F_I$ grows slower than $t^\ell$ at large $t$. It is important to stress that $F_I(s,t)$ and $c_I(s,t) F_I(s,t)$ need not have any particular crossing symmetries at this point. Then we can use the Regge behavior we just derived to write
\begin{align}
    \oint_0 \frac{dt}{2 \pi i}\oint_\infty \frac{ds}{2 \pi i} \frac{1}{s t} \left(\frac{c_I(s,t) F_I(s,t)}{s^{m} t^\ell} -  \frac{c_I(t,s) F_I(t,s)}{t^{m} s^\ell} \right)  =  0 \, .
    \label{eq:gennull}
\end{align}
The polynomial nature of $F_I$ is such that these complex integrals automatically cancel the low-energy contribution, thus giving rise to a purely high-energy expression.

Let us make an example to see how this works in practice. Consider $F_{17}$. From the form of the crossing matrix we have
\begin{align}
    F_{17}(t,s) =-F_{16}(s,t) +F_{17}(s,t) \, .
\end{align}
Of course, both of these combinations grow slower than $s^3$, as we showed above, but they actually all have improved Regge behavior -- we find 
\begin{align}
    F_{17}(s,t)  \ &< \  s^{0} \, , \\
    -F_{16}(s,t) +F_{17}(s,t) \ &< \  s^{2}  \, .  
\end{align}
Therefore we can derive a null constraint by using 
\begin{align}
    \oint_0 \frac{dt}{2 \pi i}\oint_\infty \frac{ds}{2 \pi i} \frac{1}{s t} \left(\frac{ F_{17}(s,t)}{t^2} -  \frac{ F_{17}(t,s)}{ s^2} \right)  =  0 \, .
\end{align}
The null constraint that arises from this is
\footnotesize{
\begin{align}
    \footnotesize
    0 &=\sum_\ell 16(2\ell+1) \int_{M^2}^{\infty} ds\,\mu(s) \left[ -\frac{8 \ell (\ell+1) m^4 \left((\ell^2+\ell-2) m^2-s\right)}{s^2 (4 m^2-s)^5} \rho^1_\ell(s) \right.\\\nonumber
    &-\frac{4 \sqrt{2\ell(\ell+1)} m^3 \left(2 (\ell^4+2 \ell^3+5 \ell^2+4 \ell-12) m^2-3 (\ell^2+\ell+2) s\right)}{3 s^{3/2} \left(s-4 m^2\right)^5}\rho^2_\ell(s) \\\nonumber
    &-\frac{8 \sqrt{\ell(\ell-1)(\ell+1)(\ell+2)} (2 m^4-m^2 s) \left(4 (\ell^2+\ell-3) m^2-3 s\right)}{3 s^2 (4 m^2-s)^5}\rho^3_\ell(s)\\\nonumber
    &-\frac{8 \sqrt{2\ell(\ell+1)}m^3 \left(2 (\ell^4+2 \ell^3+5 \ell^2+4 \ell-12) m^2-3 (\ell^2+\ell+2) s\right)}{3 s^{3/2} (s-4 m^2)^5}\rho^4_\ell(s) \\\nonumber
    &-\frac{16 \ell (\ell+1) m^4 \left((\ell^2+\ell-2) m^2-s\right)}{s^2 (4 m^2-s)^5}\rho^5_\ell(s) -\frac{16 \ell (\ell+1) m^2 (2 m^2-s) \left((\ell^2+\ell-2) m^2-s\right)}{s^2 (4 m^2-s)^5}\rho^6_\ell(s) \\\nonumber
    &+\frac{16 \ell (\ell+1) m^4 \left(4 (\ell^2+\ell-2) m^2-3 s\right)}{3 s^2 (4 m^2-s)^5}\rho^7_\ell(s) +\frac{32 (\ell^2+\ell-1) m^4 s-32 (\ell^2+\ell-2)^2 m^6}{s^2 (4 m^2-s)^5}\rho^8_\ell(s) \\\nonumber
    &+\frac{4 \sqrt{2\ell(\ell+1)} m (3 m^2-s) \left(2 (\ell^4+2 \ell^3+5 \ell^2+4 \ell-12) m^2-3 (\ell^2+\ell+2) s\right)}{3 s^{3/2}
   (s-4 m^2)^5}\rho^9_\ell(s) \\\nonumber
   &-\frac{4 \sqrt{2} \ell (\ell+1) (\ell^2+\ell-2) \left((\ell^2+\ell+50) m^5 s-64 m^7-13 m^3 s^2+2 m s^3\right)}{3 \sqrt{(\ell-1)(\ell+2)} s^{5/2} (s-4
   m^2)^5}\rho^{10}_\ell(s) \\\nonumber
   &+4 \sqrt{2(\ell-1)(\ell+2)}\left(\frac{ 3 (\ell^2+\ell-10) m s^3+2 (\ell^4+2 \ell^3+\ell^2-156) m^5 s}{3 s^{5/2} (s-4 m^2)^5}
   \right.\\\nonumber&\qquad\qquad\qquad\qquad\qquad\left.+\frac{(-2 \ell^4-4 \ell^3+31 \ell^2+33 \ell-72) m^3
   s^2+384 m^7}{3 s^{5/2} (s-4 m^2)^5}\right)\rho^{11}_\ell(s) \\\nonumber
   &-\frac{4 \ell (\ell+1) (2 m^4-4 m^2 s+s^2) \left((\ell^2+\ell-2) m^2-s\right)}{s^2 (4 m^2-s)^5}\rho^{12}_\ell(s)\\\nonumber
   &+\frac{\ell (\ell+1) (\ell^2+\ell-2) (-28 m^6+24 m^4 s-6 m^2 s^2+s^3)}{3 s^2 (s-4 m^2)^5}\rho^{13}_\ell(s) \\\nonumber
   &+\left(\frac{-8 (\ell^2+\ell+20) m^4 s+4 (\ell^2+\ell-8) s^3+8 (\ell^4+2 \ell^3-9 \ell^2-10 \ell+60) m^6}{s^2 (s-4 m^2)^5}
   \right.\\\nonumber&\qquad\qquad\qquad\qquad\qquad-\left.\frac{4 (\ell^4+2 \ell^3-17 \ell^2-18 \ell+60) m^2
   s^2}{s^2 (s-4 m^2)^5}\right)\rho^{14}_\ell(s) \\\nonumber
   &-\frac{8 \ell (\ell+1) (2 m^4-m^2 s) \left(4 (\ell^2+\ell-2) m^2-3 s\right)}{3 s^2 (4 m^2-s)^5}\rho^{15}_\ell(s)\\\nonumber
   &-\frac{16 (2 m^4-m^2 s) \left((\ell^2+\ell-2)^2 m^2-(\ell^2+\ell-1) s\right)}{s^2 (s-4 m^2)^5}\rho^{16}_\ell(s) \\\nonumber
   &\left. -\frac{16 \sqrt{\ell(\ell-1)(\ell+1)(\ell+2)} m^4 \left(4 (\ell^2+\ell-3) m^2-3 s\right)}{3 s^2 (4 m^2-s)^5}\rho^{17}_\ell(s) \right].
\end{align}
}\normalsize

\paragraph{Null constraints used in bounds}

For numerical purposes, we need to use a finite number of null constraints. For massless theories, these can be nicely organized in inverse powers of the integrated variable $s$. By construction, our null constraints come with powers of $s$ and $s-4m^2$. We thus choose to organize them in inverse powers of $s^a (s-4m^2)^b$, and we will truncate by considering all null constraints with $a+b<N_{\mathrm{Max}}$. In table~\ref{tab:numberNullConstraints}, we give the number of null constraints for some low values of $N_{\mathrm{Max}}$.

\begin{table}[h]
    \centering
    \renewcommand*{\arraystretch}{1.25}
    \begin{tabular}{|l|lllllll|}
    \hline
      $N_{\mathrm{Max}}$   & $5$ & $6$ & $7$ & $8$ & $9$ & $10$ & $11$ \\\hline
      $\#\mathrm{NC}$   & $1$  & $11$  & $12$  & $44$  & $56$  & $106$ & $127$  
      \\\hline
    \end{tabular}
    \caption{Number of null constraints for a given $N_{\mathrm{Max}}$.}
    \label{tab:numberNullConstraints}
\end{table}

After experimenting with different orders, we chose $N_{\mathrm{Max}}=11$, which gives 127 null constraints. Spin convergence is achieved by summing $\ell$ up to $\ell_{\mathrm{Max}}=\#\mathrm{NC}+75=202$,\footnote{The sum in the partial wave decomposition is divided in even and odd spins. Thus effectively we will use $\ell_{\mathrm{Max}}/2$ even and odd spins.} plus sporadic higher values of $\ell$ and the asymptotic value at $\ell=\infty$.

\section{Bounds}
\label{sec:Bounds}

\subsection{Manifestly positive observables}
\label{sec:normalization}

Positivity  bounds are independent of the overall normalization of the scattering amplitude, and therefore apply only to \emph{ratios} of low-energy arcs or EFT coefficients -- they are projective bounds. In constructing such ratios, it is convenient to use manifestly positive quantities as denominators.

A set of manifestly positive quantities is given by the forward-limit expansion of elastic scattering amplitudes at the minimal (even) number of subtractions, which for us is 2,
\begin{equation}
    \oint \frac{ds}s\frac{A_{ab\to ab}(s,0)}{s^2} >0.
\end{equation}
This condition follows from the optical theorem using positivity of the total cross-section $\sigma^{\mathrm{tot}}_{ab\to\mathrm{anything}}(s)>0$ \cite{Adams:2006sv}. 
Contrary to for instance the case of a single massless scalar or photon, there is more than one such positive quantity with two subtractions. We have in principle four elastic amplitudes but 
two of them are related by crossing, leading thus to three positive quantities.

From the definitions \eqref{eq:arcsl0} and \eqref{eq:arcsForward} of the arcs, it is clear that a set of positive quantities is $A_{2}^{0000}$, $\left(A_{2}^{++--} + A_{2}^{-++-}\right)$, and $A_{2}^{+0-0}$.
The positivity of these quantities  can also be seen through the explicit sum rules,
\begin{align}
    A_2^{0000} &= \left\langle\left(
        \frac{1}{s^2} + \frac{s}{\left(s-4m^2\right)^3}
    \right)\rho_\ell^{0000}(s)\right\rangle,\label{eq:una}\\
    A_2^{++--}+A_2^{-++-} &= \left\langle
        \left(
        \frac{1}{s^2} + \frac{s}{\left(s-4m^2\right)^3}
    \right)
    \left(
    \rho_\ell^{-++-}(s)+
    \rho_\ell^{++--}(s)
    \right)\right\rangle,\\
    A_2^{+0-0} &= \left\langle\left(
        \frac{1}{s^2} + \frac{s}{\left(s-4m^2\right)^3}
    \right)\rho_\ell^{+0-0}(s)\right\rangle,\label{eq:terza}
\end{align}
where we define,
\begin{equation}
    \left\langle f_\ell(s) \right\rangle
    := \sum_\ell16(2\ell+1)\int_{M^2}^\infty ds\,\mu(s) f_\ell(s).
\end{equation}

\subsection*{Positive Denominators}

  Even if we consider an amplitude with fixed helicity structure, the sum rule will contain also its crossing symmetric counterpart, which is a mixture of all helicity amplitudes. Therefore, in the most general case, the bounds will project into a generic linear combination of the three positive amplitudes \eqref{eq:una}-\eqref{eq:terza}. For this reason, we employ the simple combination,
\begin{equation}\label{eq:canonicalNormalization}
    q=\frac14\left(
    \tfrac{1}{2} A_2^{0000}+ A_2^{++--}+A_2^{-++-}+ A_2^{+0-0}
    \right),
\end{equation}
as a universal denominator to our results.

This generic expectation breaks down in specific cases, in which a denominator involving a smaller subset of the three amplitudes suffices. Indeed, for observables with at most one derivative in $t$, the denominators do not need to include all three of these
quantities. This follows from the form of the crossing matrix which mixes amplitudes together in the dispersion relation \eqref{eq:disprelmatrix}: close to the forward limit, expanding in powers of $t$, it takes
the simple form~\cite{Bellazzini:2023nqj},
\begin{equation}
    {\big(E(u(s),t)CE^{-1}(s,t) \big)^{\lambda_1\lambda_2\lambda_3\lambda_4}}_{\lambda_3'\lambda_2'\lambda_1'\lambda_4'}
    = \mathcal{O}\left\{\left(-\frac{m^2t}{s^2}\right)^{\sum_i\frac{\left|\lambda_i-\lambda'_i\right|}{2}}\right\}.
    \label{eq:crossingMatrixExp}
\end{equation}

In the case of one $t$ derivative, we see that the crossing matrix will mix amplitudes that are at most related by two flips in
the polarizations. This means that the dispersion relation for a given observable, such as $A^{0000}_{2,1}$, does not
involve purely transverse amplitudes, such as $A^{++--}$, which has $\sum\left|\lambda_i-\lambda'_i\right|>2$. It is therefore enough to use $(A_2^{0000}+A_2^{+0-0})$ as a denominator to find bounds on $A_{2,1}^{0000}$.\footnote{When using the regularized crossing matrix built from $\tilde{E}$ instead of $E$, The matrix elements are all
the same, except when $\sum\left|\lambda_i-\lambda'_i\right|$ is odd, in which case we multiply by a factor of $\sim\sqrt{-t}$. The result is
that we must take the ceiling of the exponent in \eqref{eq:crossingMatrixExp}:
\begin{equation}
    {\big(\tilde{E}(u(s),t)C\tilde{E}^{-1}(s,t) \big)^{\lambda_1\lambda_2\lambda_3\lambda_4}}_{\lambda_3'\lambda_2'\lambda_1'\lambda_4'}
    = \mathcal{O}\left\{t^{\left\lceil\sum_i\frac{\left|\lambda_i-\lambda'_i\right|}{2}\right\rceil}\right\}.
    \label{eq:crossingMatrixExpTilde}
\end{equation}
}
This rule holds up to the fact that even though a given amplitude may not cross into one of the manifestly positive amplitudes, it may cross into an amplitude that is related to it by time reversal, parity, and boson exchange symmetry. For example, one would naively assume that
$A^{000+}_{2,1}$ does not need $A_2^{+0-0}$ in its denominator, since the corresponding matrix element in the (tilde) crossing matrix goes like $t^4$, but we have that $A^{000+} = A^{00-0}$, which crosses into $A^{+0-0}$ with a factor of $t^1$.

In the example of $A^{000+}_{2,1}$,
we applied parity and boson exchange on both the
in and out-going states. But we also need to account
for these symmetries on either of them separately. In the end,
this amounts to checking whether the dispersion
relations, when written in the form of \eqref{eq:gisumrule}, have non-zero matrix elements in the entries populated by each of the 4 manifestly positive amplitudes.

Applying these rules, we can classify all observables of the type $A^I_{k,1}$ in several categories, according to the number of
manifestly positive  amplitudes it is sufficient to include in the denominator in order to find a bound (see table \ref{tab:denominatorsForPols}).
We see that the observables nicely separate into three
categories, according to whether they have purely longitudinal, transverse, or mixed polarizations.

Already at order 2 in $t$, all the amplitudes mix into
all of the manifestly positive amplitudes, and we cannot conclude whether we may use less than all 4 terms in the denominator.

\begin{table}[]
    \centering
    \begin{tabular}{c|c}
        Amplitude index $I$ & Sufficient denominator for bounding $A^I_{k,1}$ \\
        \hline
        5 (longitudinal polarizations)& $A_2^{0000} + A_2^{+0-0}$\\
        1, 3, 12, 13, 14 (transverse polarizations)& $A_2^{++--}+A_2^{-++-} + A_2^{+0-0}$\\
        All others (mixed polarizations) & $A_2^{++--}+A_2^{-++-} + A_2^{0000} + A_2^{+0-0}$\\
    \end{tabular}
    \caption{Denominators sufficient to obtain
    a bound on the quantity $A^I_{k,1}$, $k\in\mathbb{N}$, if the bound exists. The sums on the right column could have different coefficients as long as they are all strictly positive. The bound
    may or may not still be there if one or more of the coefficients are zero. If there is no bound
    with all coefficients strictly positive, then
    the observable is unbounded.}
    \label{tab:denominatorsForPols}
\end{table}

\subsection{Forward-limit bounds}
\label{sec:forward-limit-bounds}

At $t \to 0$  (i.e. considering observables without taking any $t$ derivatives) the crossing matrix is diagonal. In this case, the constraints have a well-controlled dependence on $m$ and are much easier to obtain. Things can be made even simpler if we choose the subtractions to be crossing symmetric instead of taking all subtractions at $s=0$. This amounts to defining the observables as
\begin{align}
    \hat A^{\lambda_1\lambda_2\lambda_3\lambda_4}_{k} :=\left( \res_{s=2m^2}+\res_{s=4m^2} \right)
  \left[\frac{\tilde A^{\lambda_1\lambda_2\lambda_3\lambda_4}_{\Low}(s,0)}{\left(s-2m^2\right)^{k+1}}\right]  
  \label{eq:arcsForward}
\end{align}
rather than using \eqref{eq:arcs}.
  The arcs in \eqref{eq:arcsForward} satisfy the dispersion relation,
\begin{align}
  \hat A^I_k(M^2)=\sum_{\ell}16(2\ell+1)\int_{M^2}^\infty\frac{\dd s}{\pi\left(s-2m^2\right)^{k+1}} \Bigg[\rho_{\ell}^I (s) 
  +(-)^k\sum_J {X^I}_J 
  \rho_\ell^J(s)\Bigg] ,
  \label{eq:disprelforward}
\end{align}
where we defined the crossing matrix in the forward limit,
\begin{equation}
    {X^I}_J := {\big(E(u(s),t=0)CE^{-1}(s,t=0) \big)^I}_J\,.
\end{equation}
$X$ is special, in that it does not depend on $s$, and is equal to the crossing matrix in the massless limit.

If we perform a change of variables $\hat{s} = s - 2m^2$, and define
$\rho_\ell^I(\hat{s}+2m^2):=\hat{\rho}_\ell^I(\hat{s})$, the dispersion relations
\eqref{eq:disprelforward} are recast into equivalent dispersion relations
for a massless theory, with a modified cutoff scale $\hat{M}^2:=M^2-2m^2$:
\begin{align}
  \hat A^I_k(M^2)=\sum_{\ell}16(2\ell+1)\int_{M^2-2m^2}^\infty\frac{\dd\hat{s}}{\pi\hat{s}^{k+1}} \Bigg[\hat{\rho}_{\ell}^I (\hat{s}) 
  +(-)^k\sum_J {X^I}_J 
  \hat{\rho}_\ell^J(\hat{s})\Bigg] ,
\end{align}

\begin{figure}
    \centering
    \includegraphics[scale=0.76]{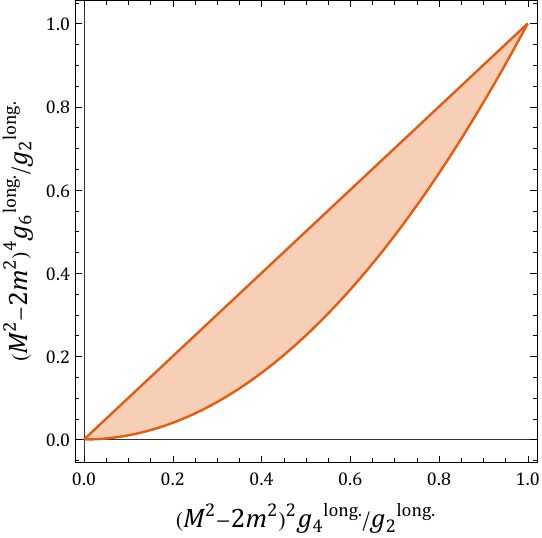}
    \qquad
    \includegraphics[scale=0.77]{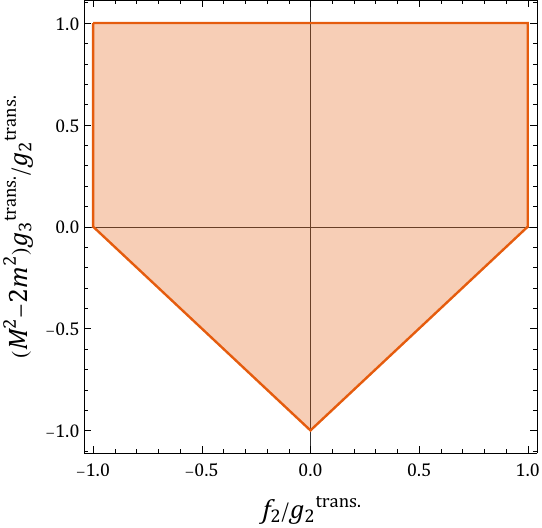}
    \caption{On the left: Bounds on the ratios $g^\text{long.}_6/g^\text{long.}_2$ and $g^\text{long.}_4/g^\text{long.}_2$ from the forward scattering of four longitudinal modes.
    On the right: Bounds on the ratios $f_2/g^\text{trans.}_2$ and $g^\text{trans.}_3/g^\text{trans.}_2$ from the forward scattering of four transverse modes.
    The bounds are exactly the same as a massless scalar/photon, with $M^2$ replaced by $M^2 - 2m^2$.
    In this particular case, they can be obtained analytically.}
    \label{fig:scalarg4g6}
\end{figure}

Since longitudinal and transverse modes do not mix through crossing,
we can immediately see that all the bounds involving only $A^{0000}$, or any of $A^{\pm\pm\pm\pm}$, with no derivative in $t$, are exactly the
same as those of a massless scalar, resp. massless photon, with the
cutoff scale $M^2$ replaced by $M^2-2m^2$. This applies also to 2d
bounds. See e.g. figure~\ref{fig:scalarg4g6}, where we use the notation of \cite{Caron-Huot:2021rmr} and \cite{Henriksson:2021ymi} in order to
define,
\begin{align}
    g^\text{long.}_{2k} &:= \hat A^{0000}_{2k} = \frac{1}{(2k)!}\partial_s^{2k}A^{0000}(s=2m^2,t=0),\\
    f_{2k} &:= 2^{-k}\hat A^{++++}_{2k} = \frac{1}{2^k(2k)!}\partial_s^{2k}A^{++++}(s=2m^2,t=0),\\
    g^\text{trans.}_k &:= \hat A^{++--}_k = \frac{1}{k!}\partial_s^kA^{++--}(s=2m^2,t=0).
\end{align}

The bounds shown can be obtained analytically by rephrasing the sum rules as a 1-dimensional moment problem \cite{Bellazzini:2020cot}.
We have that,
\begin{align}
    \left[\left(M^2-2m^2\right)\frac{g_4^\text{long.}}{g_2^\text{long.}}\right]^2 &<
    \left(M^2-2m^2\right)^2\frac{g_6^\text{long.}}{g_2^\text{long.}}
    < \frac{g_4^\text{long.}}{g_2^\text{long.}},\\
    \left|\frac{f_2}{g_2^\text{trans.}}\right|-1 &<
    \left(M^2-2m^2\right)\frac{g_3^\text{trans.}}{g_2^\text{trans.}}
    < 1.
\end{align}

Similarly, for observables with mixed helicities we define \footnote{There are two other
such elastic amplitudes devoid of singular terms $\sim\sqrt{stu}$, but it can be shown that they provide no bound in the
forward limit.}
\begin{align}
    g^\text{mixed}_{2k} := \hat A^{+0-0}_{2k} = \frac{1}{(2k)!}\partial_s^{2k}A^{+0-0}(s=2m^2,t=0)\,,
\end{align}
which are all positive and satisfy,

\begin{equation}
    0\leq\left(M^2-2m^2\right)^{k-l}\frac{g_{2k}^\text{mixed}}{g_{2l}^\text{mixed}}\leq 1 \quad\forall k\geq l\,.
\end{equation}

Beyond the forward limit (i.e. for observables involving derivatives), the analytic methods become less efficient (see however Refs.~\cite{Arkani-Hamed:2020blm,Chiang:2021ziz,Bellazzini:2021oaj}). We therefore  proceed with the numerical semi-definite optimisation techniques illustrated above.

\subsection{Non-forward bounds}

We now present the main results of this paper, namely the exclusion plots for a variety of observables as a function of the mass ratio $m/M$. We normalize the cutoff at $M=1$, so that $0\leqslant m \leqslant \frac12$. We choose the values $m=\{ \frac1{100},\frac1{10},\frac3{10}\}$ for illustration, unless otherwise specified.

\subsubsection{Scalar-like observables}
\begin{figure}
    \centering
    \includegraphics[width=0.85\textwidth]{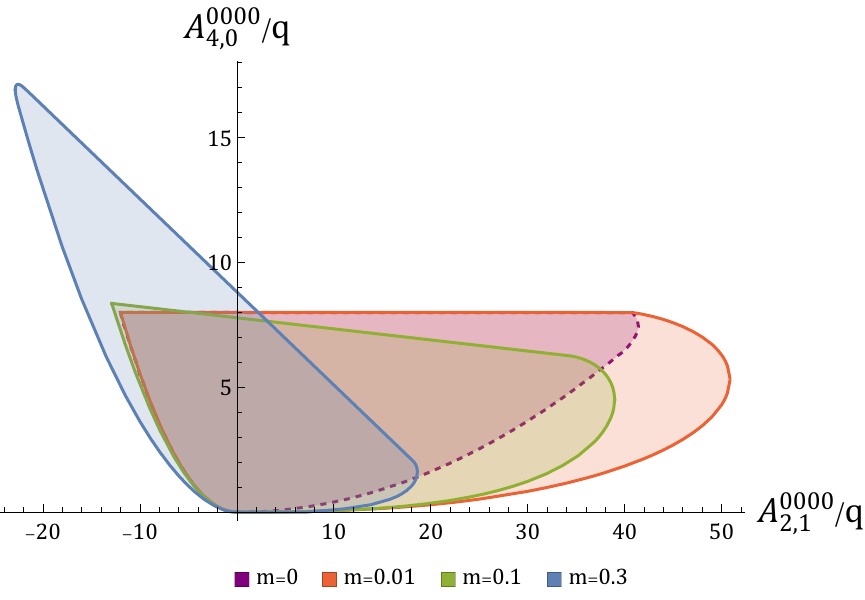}
    \caption{Bounds on scalar-like amplitudes as a function of the mass $m$ of the vector. The dashed region in purple shows the massless scalar plot extracted from \cite{Caron-Huot:2020cmc}.}
    \label{fig:scalarplot}
\end{figure}

In figure~\ref{fig:scalarplot} we show the bounds obtained with our numerical algorithm for two observables in the longitudinal sector, where all helicities are zero. We can see how the regions move as we deviate from the massless limit, while preserving the kink structure. For reference, we also show in purple the precise $m=0$ region  from \cite{Caron-Huot:2020cmc},  rescaled to agree with our conventions for the observables (in the language of \cite{Caron-Huot:2020cmc}, these are bounds in the plane $(g_3/g_2,\,g_4/g_2)$;  with respect of figure 8 of the same paper, our $x$-axis has a minus sign). We warn the reader that we cannot always make a direct comparison of our bounds with massless plots, because for any finite mass, the denominators in the two cases are different. That explains for instance why the region around the upper right kink of our plots moves to the left as we increase the mass, while the purple region seems to break this feature. 

\subsubsection{Photon-like observables}

\begin{figure}
    \centering
    \includegraphics[width=0.9\textwidth]{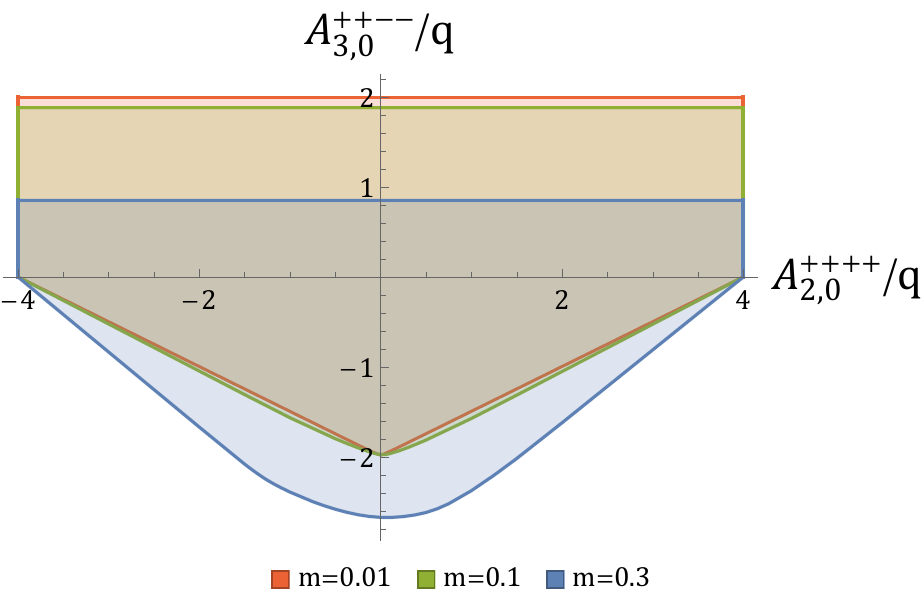}
    \caption{Bounds on photon-like amplitudes as a function of the mass $m$ of the vector. Compare with figure~12 (right) of \cite{Henriksson:2022oeu}.}
    \label{fig:chevron}
\end{figure}

In figure~\ref{fig:chevron} we show the bound on expressions $A_2^{++++}$ and $A_3^{++--}$, analogous to the bound in the plane $(M^2 g_3/g_2,\,f_2/g_2)$ in the case of photon scattering -- figure~12 (right) of \cite{Henriksson:2022oeu}.
Let us discuss this plot in detail.

It is interesting to note that the upper bound in figure~\ref{fig:chevron} becomes stronger as we increase the ratio $\kappa=\frac{m^2}{M^2}$, while the lower bound gets weaker. In fact, by explicitly looking at the sum rules, it is possible to derive an analytic expression for the upper bound:
\begin{equation}
    \frac{M^2g_3}{q}\leqslant \mathrm{min} \left\{
        \frac{2 - 24 \kappa + 96 \kappa^2 - 128 \kappa^3}{1 - 6 \kappa + 24 \kappa^2 - 32 \kappa^3} ,\ \frac{\zeta}{\kappa}
\right\}, 
\end{equation}
where $\zeta=0.0766602$ is a root of the polynomial $27 \zeta ^4-108 \zeta ^3+162 \zeta ^2-364 \zeta +27$, and the transition between the two expressions happens at $\kappa=0.0768738$ (root of $128 \kappa ^4-128 \kappa ^3+48 \kappa ^2-16 \kappa +1$). The lower bound in $g_3/q$, on the other hand, is dependent on the number of null constraints.

\begin{figure}
    \centering
    \includegraphics[width=0.9\textwidth]{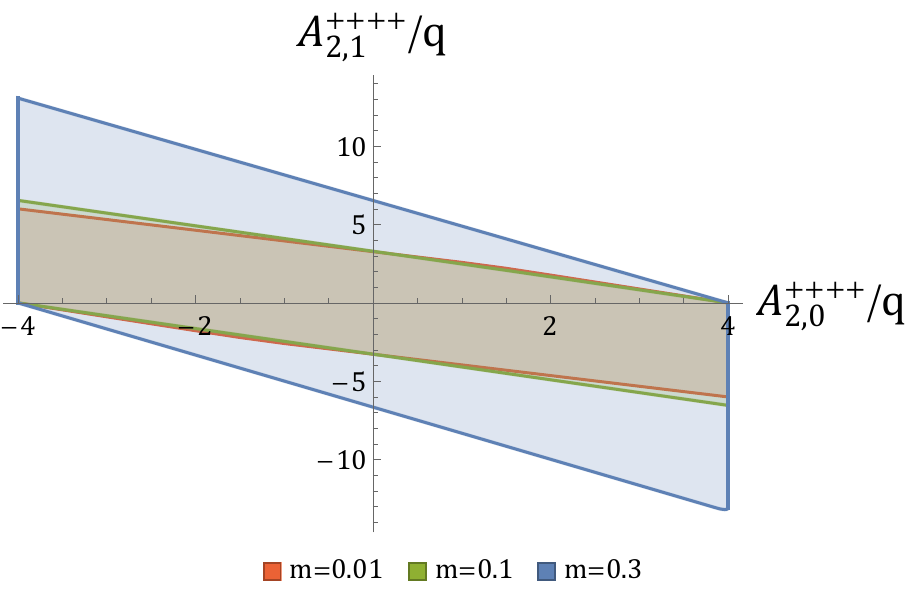}
    \caption{Bounds on photon-like amplitudes as a function of the mass $m$ of the vector. Compare with figure~12 (left) of \cite{Henriksson:2022oeu}.}
    \label{fig:photonplot2}
\end{figure}

Another photon-like plot is shown in figure~\ref{fig:photonplot2}. This is to be compared with the shape of figure~12 (left) of \cite{Henriksson:2022oeu}. In their language, this is the plane $(M^2 f_3/g_2,f_2/g_2)$. Our $y$-axis has a minus sign difference with respect to their plot.

The bottom right part of the plot for $m=3/10$ is special, in that the lower bound gently curves right before reaching the kink. We can understand this feature in terms of simple UV completions of the theory, see Section \ref{sec:UVcompl}.

In both examples, we can see that the main features of the photon exclusion plots are retained whenever we set $m=1/100$. This agrees with our expectation that the transverse sector of the theory reduces to a theory of interacting photons in the massless limit.

\clearpage

\subsubsection{Observables that vanish in the massless limit}

\begin{figure}
    \centering
    \includegraphics[scale=0.6]{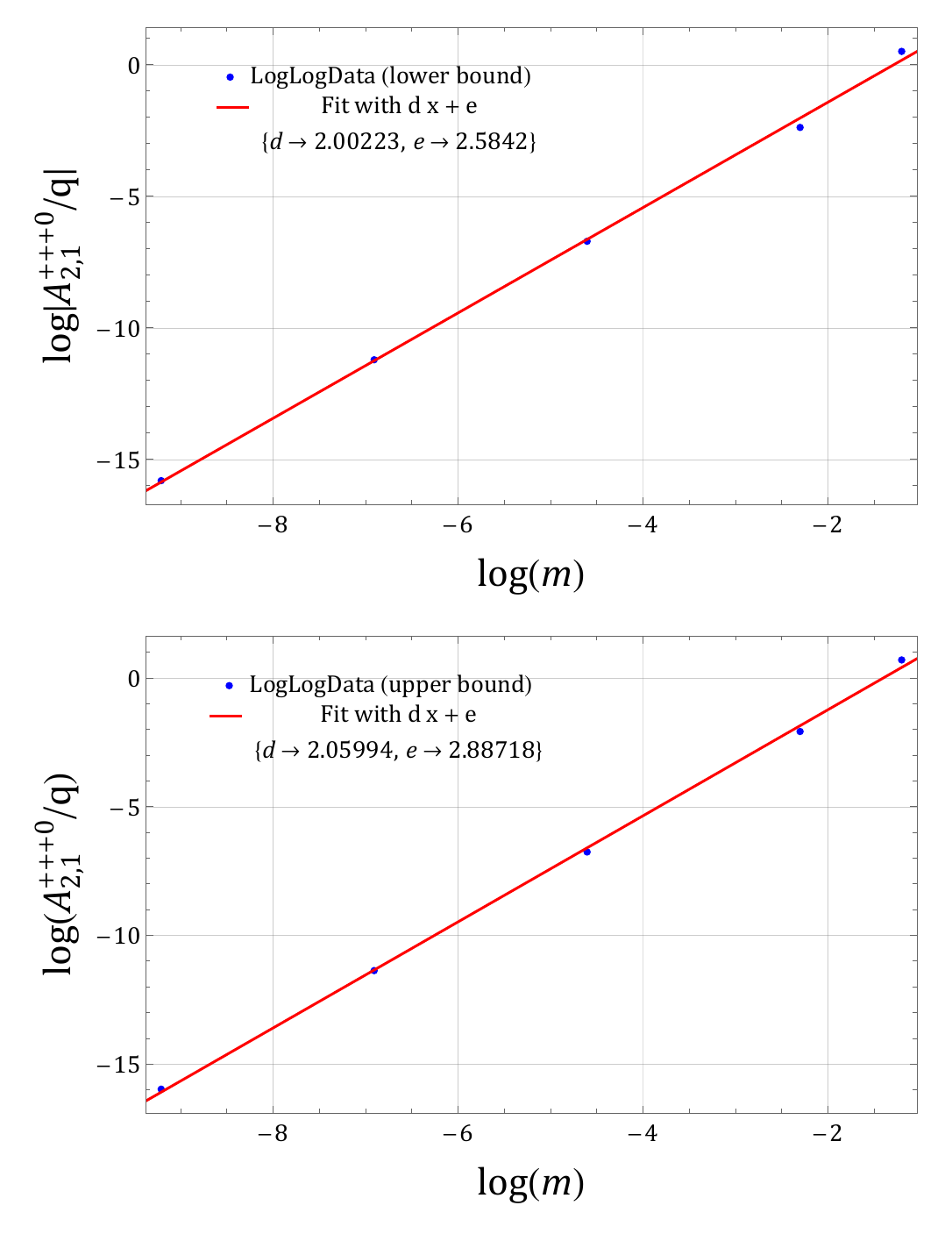}
    \caption{Log-log plot of the 1d bound on a particular mixed amplitude which vanishes as $m\to0$.}
    \label{fig:vanishingmasslessloglog}
\end{figure}

Some observables are special to the massive theory, and there is no counterpart in the massless case. In figure~\ref{fig:vanishingmasslessloglog} we show an example of such an object, in log-log scale. We study its lower and upper bounds as a function of $m=\{\frac1{10000},\frac1{1000},\frac1{100},\frac1{10},\frac3{10}\}$. A simple fit shows that these go to zero quadratically as $m\to0$.

\subsubsection{Mixed observables}
In the massless limit, a free theory of a massive spin-1 field reduces to a theory of photons and massless scalars, decoupled from each other. In the interacting theory, however, the massless limit still contains interactions between the scalar and photon sectors.

\begin{figure}
    \centering
    \includegraphics[width=0.9\textwidth]{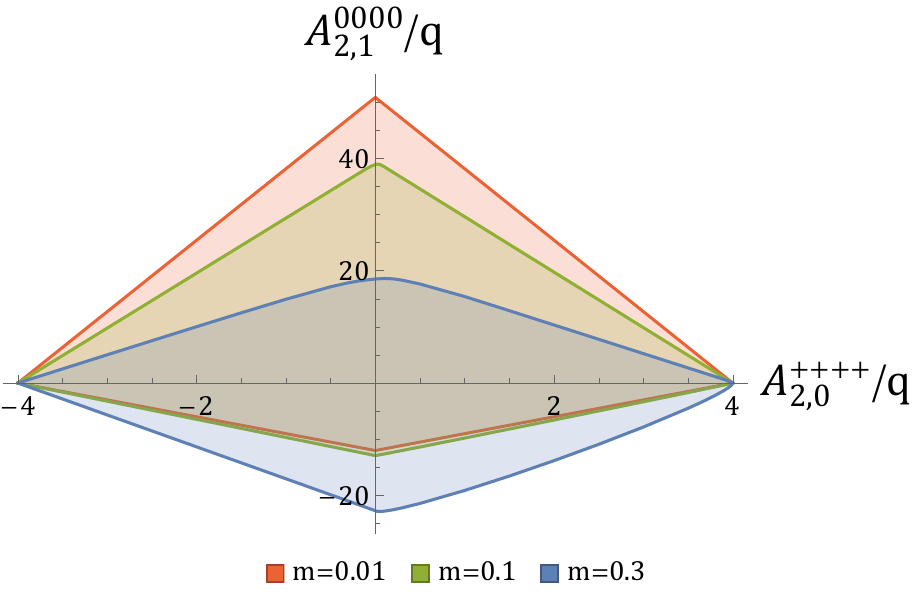}
    \caption{Bounds on scalar- and photon-like amplitudes as a function of the mass $m$ of the vector.}
    \label{fig:mixedplot1}
\end{figure}

Figure~\ref{fig:mixedplot1} focuses on the interplay between the longitudinal and transverse sectors, by considering one observable for each of them. As before, for large $m/M=3/10$ the lower bound near the right kink shows a curve which is barely visible without zooming in---we comment on this in Section \ref{sec:UVcompl}.

\begin{figure}
    \centering
    \includegraphics[width=0.95\textwidth]{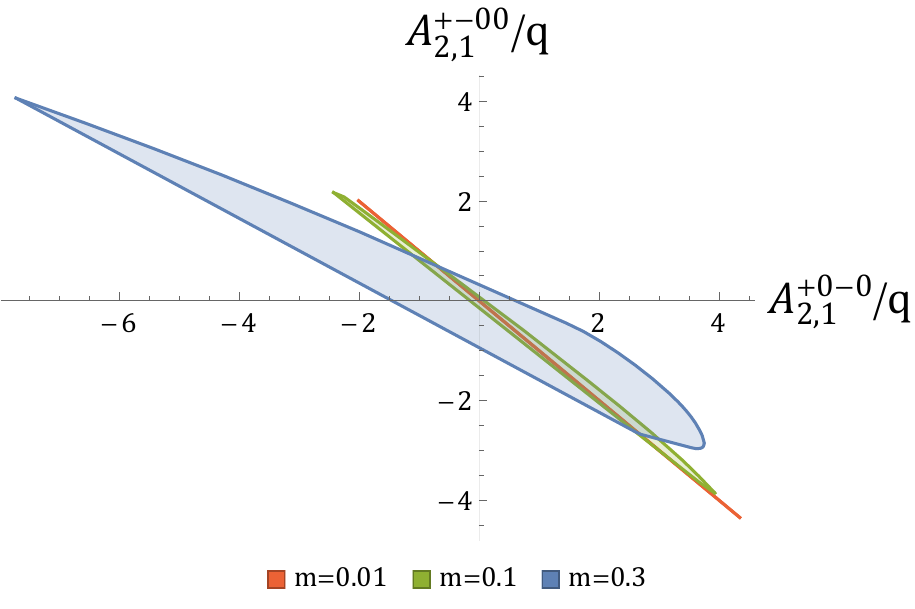}
    \caption{Bounds on mixed amplitudes amplitudes as a function of the mass $m$ of the vector.}
    \label{fig:mixedplot2}
\end{figure}

Figure~\ref{fig:mixedplot2} shows a different kind of observable. These are obtained from mixed amplitudes, whose helicities are both transverse and longitudinal. We can see a very distinct behavior as $m$ decreases: the allowed region shrinks and becomes a line. Further investigations with $m/M=10^{-5}$ suggest that the line converges to the $y=-x$ diagonal. This implies that, up to a sign, the two observables coincide as $m\to0$. 

This is a consequence of $s$-$t$ crossing in the massless limit. Indeed, the $s$-$t$ crossing matrix is given by,
\begin{equation}
    {\left(X^{st}\right)^I}_J = {\left(E(t,s)\cdot C^{st}\cdot E^{-1}(s,t)\right)^I}_J,
\end{equation}
where $C^{st}$ is defined in \eqref{eq:Cst}.  In the limit of $m\to0$, this matrix takes $A^{+-00}$ into $-A^{+0-0}$, and vice-versa, implying that the two amplitudes are the same (with opposite sign), and only a one-dimensional parameter space survives.

\subsubsection{The opposite limit: toward the cutoff}

We so far discussed many exclusion plots and observed as their shapes change as we vary the mass ratio $m/M$. We emphasized that the massless limit nicely agrees with the photon and massless scalar theories, which is a non-trivial result. Let us now briefly comment on the opposite limit, where the threshold gets close to the cutoff $4m^2 \sim M^2 =1$. We expect our setup to break down in this limit, and this is indeed what we observe.

As an example, consider the upper 1d bound of the y-axis in figure~\ref{fig:scalarplot}, $A^{0000}_{4,0}/q$. We computed the upper bound for $m=\{ 0.4,0.45,0.46,0.47,0.48,0.49,0.499,0.4999,0.5\}$. The result of the analysis is shown in table~\ref{tab:table4m2}. For the precise value $m=\frac12$ we are not able to put a bound on this observable.

\begin{table}[]
    \centering
    \begin{tabular}{|c|c|}
        \hline
        $m/M$ & Upper bound on $A^{0000}_{4,0}/q$ \\
        \hline
        0.4 & 7.219 \\\hline
        0.45 & 12.44 \\\hline
        0.46 & 14.97 \\\hline
        0.47 & 19.15 \\\hline
        0.48 & 27.50 \\\hline
        0.49 & 52.50 \\\hline
        0.499 & 502.5 \\\hline
        0.4999 & 5002 \\\hline
        0.5 & Unbounded \\\hline
    \end{tabular}
    \caption{Upper bound on $A^{0000}_{4,0}/q$ as $4m^2\to M^2$}
    \label{tab:table4m2}
\end{table}

\subsection{UV completions}
\label{sec:UVcompl}

In most cases where amplitudes saturating the positivity bounds are known, they arise from integrating out single particles at tree-level. Here we will compute the low-energy amplitudes from integrating out massive scalars and vectors, which are the only options which satisfy the Froissart bound.\footnote{Massive gravitons typically grow like $s^2$ and might therefore be considered \textit{marginal} with respect to the Froissart bound. See \cite{Henriksson:2021ymi} for a case where they were considered.}

For a parity-even scalar, we have two couplings. In the Lagrangian we could think of these as arising from the couplings $\phi A^\mu A_\mu$ and $\phi F^{\mu \nu} F_{\mu \nu}$. A pseudo-scalar $\tilde{\phi}$ has one coupling ($\tilde{\phi} F^{\mu \nu} \tilde F_{\mu \nu}$). There is a pseudo-vector with one possible coupling ($B_\mu A_\mu F^{\mu \nu}$) and a vector with one coupling ($\tilde B_\mu A_\mu \tilde F^{\mu \nu}$). Although these are all non-renormalisable couplings (therefore strictly speaking not UV-complete), they do satisfy the Regge bound and can therefore be employed as UV completions from an amplitude point of view.  

This counting of couplings matches and can be derived from the discussion in footnote~\ref{footnote:couplings}, which labels the states that can contribute to spin-0 and spin-1 partial waves in the parity-even and parity-odd sectors. We can directly use those given values of the matrix $\mathbf{c_{\ell, X}^{\lambda_i \lambda_j}}$, which determine $\rho^{\lambda_1 \lambda_2 \lambda_3, \lambda_4}_{\ell}(s)$ to compute the imaginary parts of the low-energy amplitudes. However we will give the full amplitudes in case they are of future interest.

\subsubsection{Scalar exchanges}

We consider first a real massive scalar with mass $M$. It can have two independent couplings:
\begin{equation}
    \mathcal L \supset \lambda^{(1)}_{\phi} F^{\mu\nu}F_{\mu\nu}\phi + \lambda^{(2)}_{\phi} A^\mu A_\mu \phi,
\end{equation}
where both $\lambda^{(i)}_{\phi}$ are real. We can also consider a pseudo-scalar $\tilde \phi$ with mass $\tilde M$, with the coupling
\begin{equation}
    \mathcal L \supset \tilde \lambda_{\phi} F^{\mu\nu}\tilde F_{\mu\nu}\tilde \phi.
\end{equation}

With these vertices, we find exchange amplitudes of the form
\begin{align}
    f(s|t,u)&=\frac{8\left(\lambda_\phi^{(1)}(s-2m^2)+\lambda_{\phi}^{(2)}\right)^2}{M^2-s} + 8\tilde\lambda_\phi^2\left(4m^2-3s+\frac{M^2(s-u)}{M^2-t}+\frac{M^2(s-t)}{M^2-u}\right),
    \\
    g(s|t,u)&=\frac{16\lambda_\phi^{(1)}\left(\lambda_\phi^{(1)}(s-2m^2)+\lambda_{\phi}^{(2)}\right)}{M^2-s}+16\tilde\lambda_\phi^2\left(-2+\frac{M^2}{M^2-t}+\frac{M^2}{M^2-u}\right),
    \\
    \tilde g(s,t,u)&=16\tilde\lambda_\phi^2\left(\frac{M^2}{M^2-u}-\frac{M^2+s-2m^2}{M^2-t}\right),
    \\
    h(s|t,u)&=16\left({\lambda_{\phi}^{(1)}}^2-\tilde\lambda_\phi^2\right)\left( \frac{1}{M^2-t}+ \frac{1}{M^2-u} \right),
    \\
    \tilde h(s,t,u)&=16\left({\lambda_{\phi}^{(1)}}^2+\tilde\lambda_\phi^2\right)\frac{s-t}{(M^2-s)(M^2-t)},
\end{align}
where we used the relation $s+t+u=4m^2$ in a non-exhaustive way to simplify the expressions.

\begin{figure}
    \centering
    \includegraphics[scale=0.65]{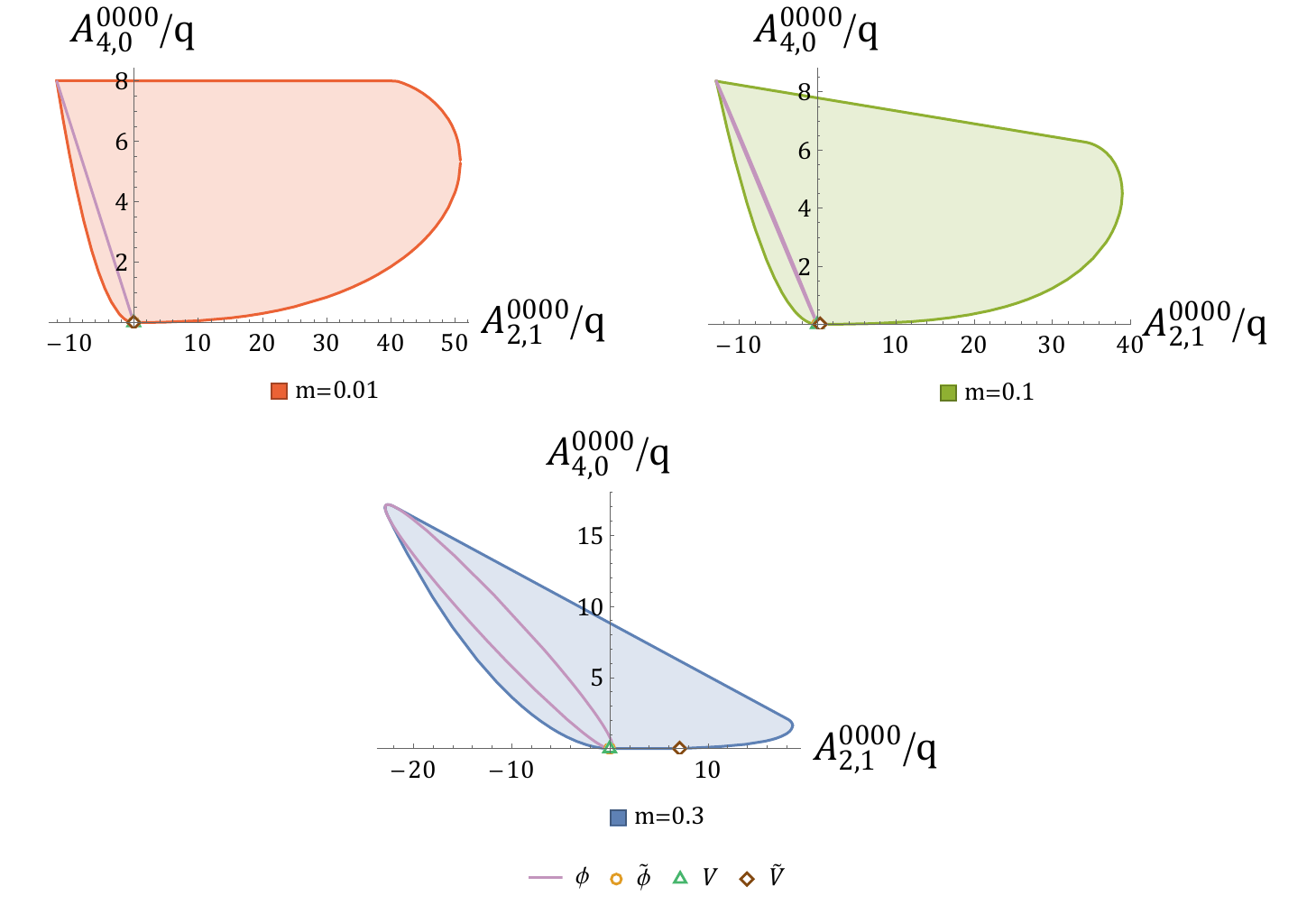}
    \caption{UV completions for the scalar-like plots.}
    \label{fig:UVscalarPlot}
\end{figure}

\begin{figure}
    \centering
    \includegraphics[scale=0.65]{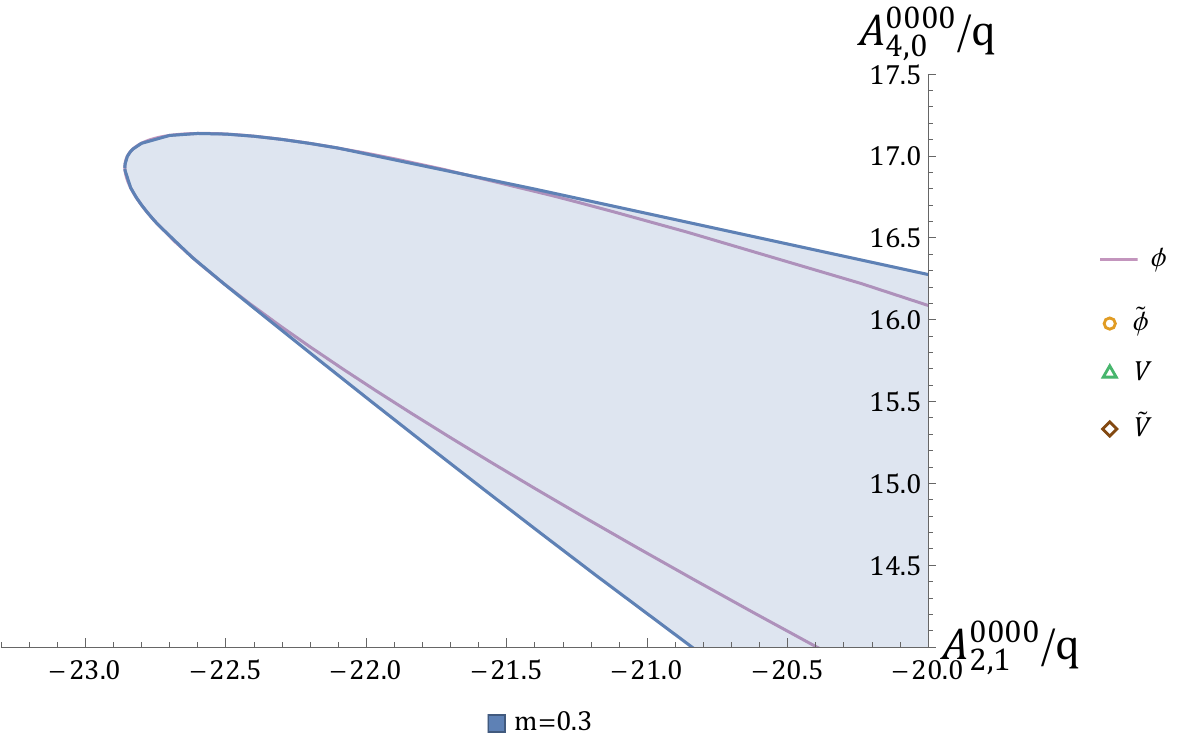}
    \caption{Close-up of the upper region of the scalar-like plot with $m=3/10$ in figure~\ref{fig:UVscalarPlot}.}
    \label{fig:UVscalarPlotZoom}
\end{figure}

In our plots, the scalar completion is represented by a closed curve that parametrizes the two couplings $\lambda_\phi^{(1,2)}$. Part of this curve always ends up on a kink, which gets ``smoothed out'' to fit the shape of the closed curve. In the longitudinal-helicity plots of figure~\ref{fig:UVscalarPlot}, we can focus on the upper left kink and compare with the massless limit. In that case, the spin-0 UV completion sits at the corner. Figure ~\ref{fig:UVscalarPlotZoom} is a zoomed version of the upper left kink for $m=3/10$. The curve nicely lies on the border of the region, which thus cannot shrink more than that.

\begin{figure}
    \centering
    \includegraphics[scale=0.61]{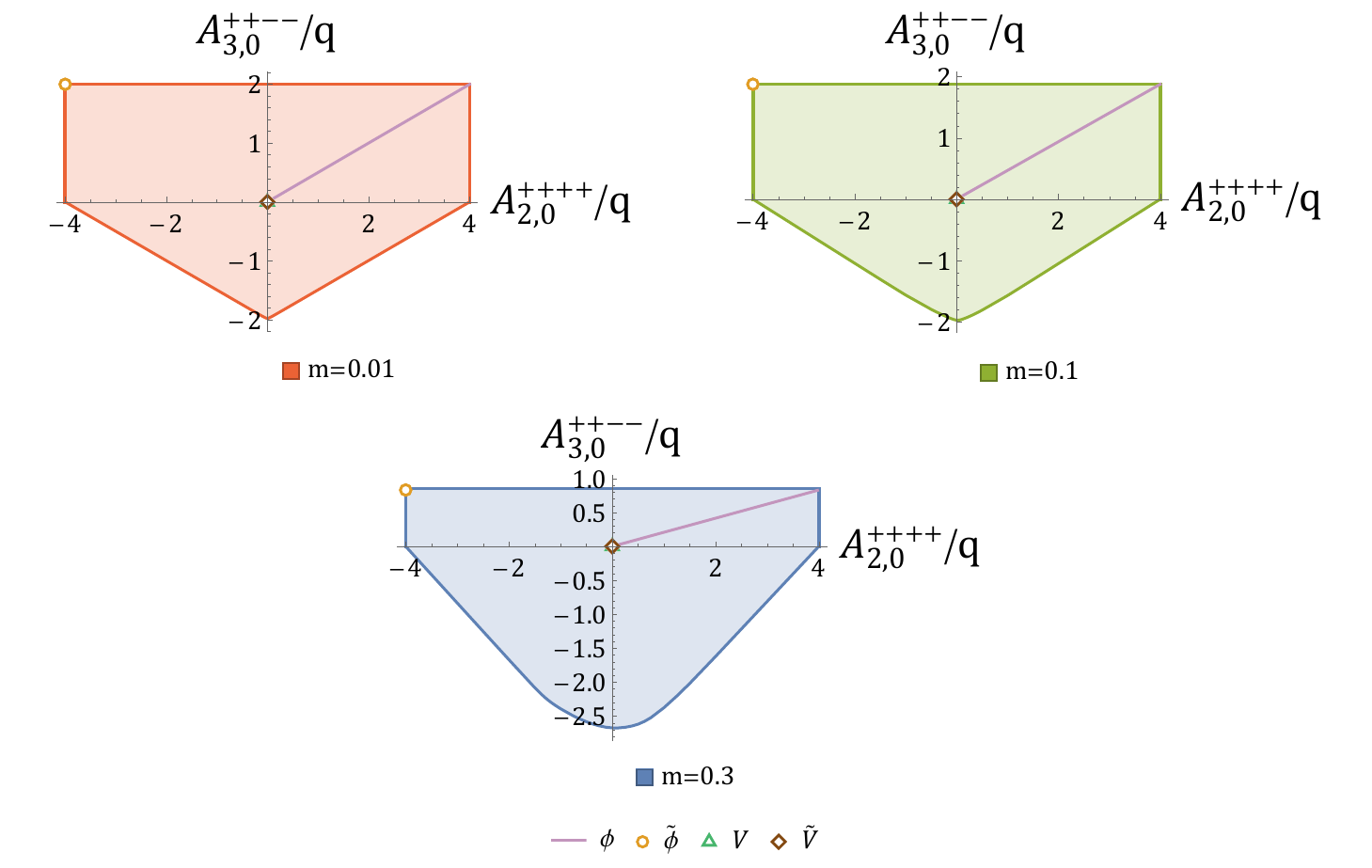}
    \caption{UV completions for the first photon-like plots.}
    \label{fig:UVphotonPlot1}
\end{figure}

\begin{figure}
    \centering
    \includegraphics[scale=0.61]{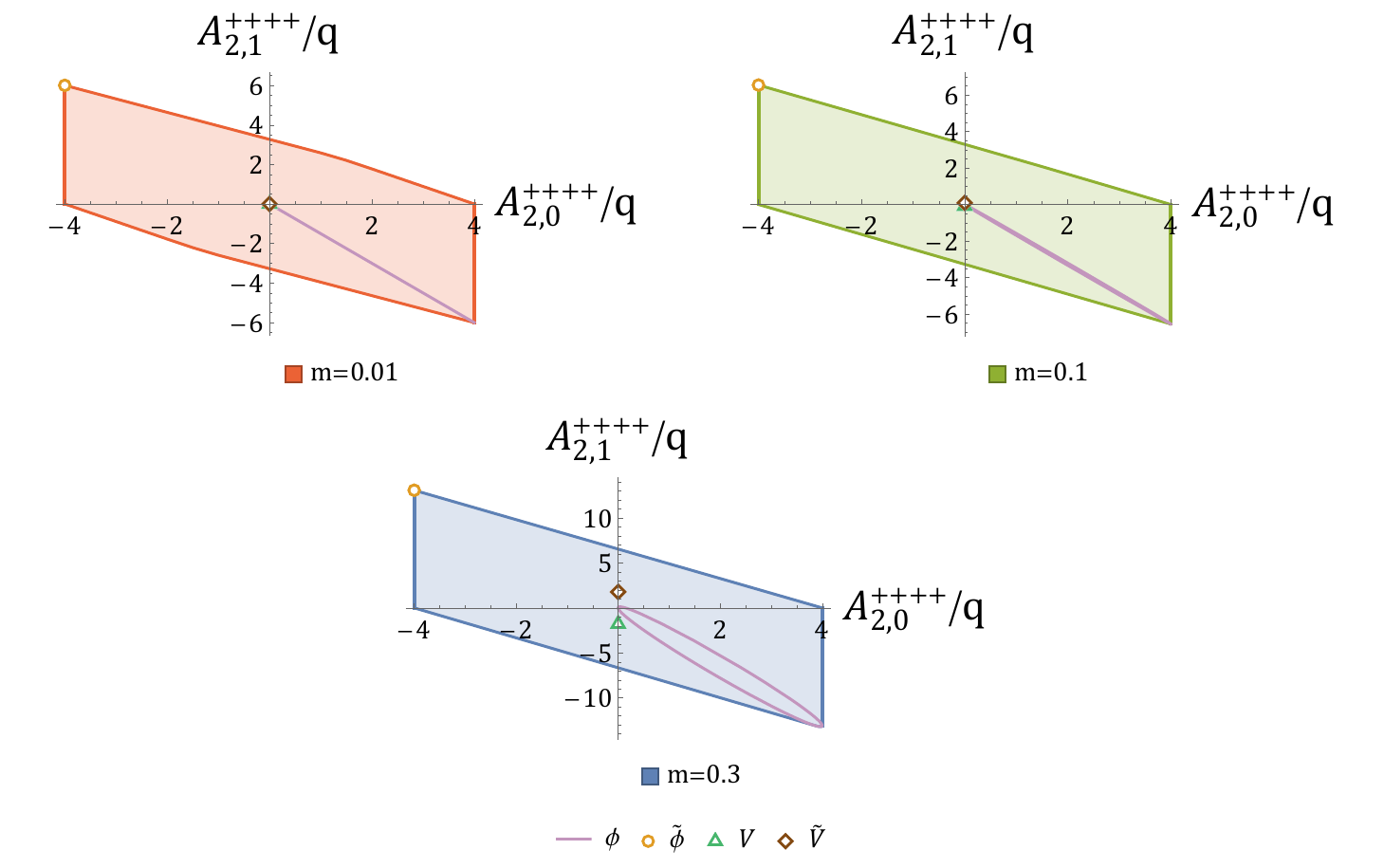}
    \caption{UV completions for the second photon-like plots.}
    \label{fig:UVphotonPlot2}
\end{figure}

Figures~\ref{fig:UVphotonPlot1} and~\ref{fig:UVphotonPlot2} show bounds on the transverse helicity amplitudes, populated with the UV completions. It is useful to compare with figure 12 of \cite{Henriksson:2022oeu} for the massless photon case. For all values of the mass, the pseudo-scalar completion (axion) stays in the upper left corner, while the scalar lives in the upper right corner and interpolates between the corner and the origin (in the photon case, the scalar completion is  a point ath the corner rather than a curve, since gauge invariance forbids the interaction $\phi A_\mu A^\mu$).

\begin{figure}
    \centering
    \includegraphics[scale=0.61]{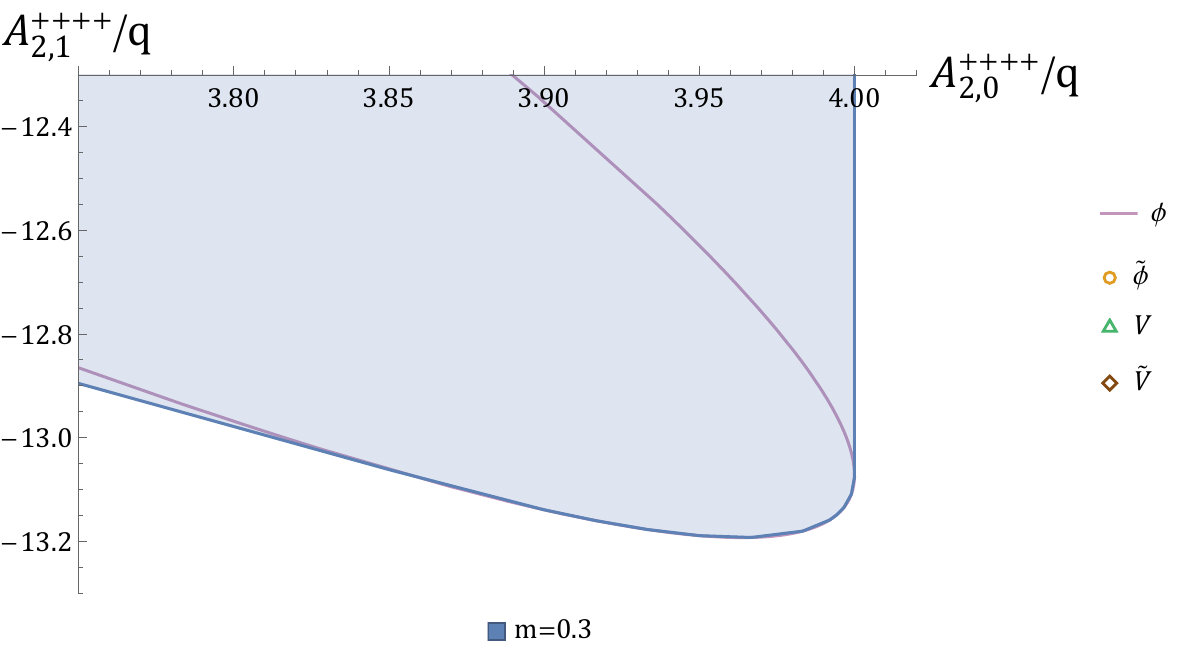}
    \caption{Close-up of the bottom right region of the second photon-like plot with $m=3/10$ in figure~\ref{fig:UVphotonPlot2}.}
    \label{fig:UVphotonPlot2Zoom}
\end{figure}

\begin{figure}
    \centering
    \includegraphics[scale=0.61]{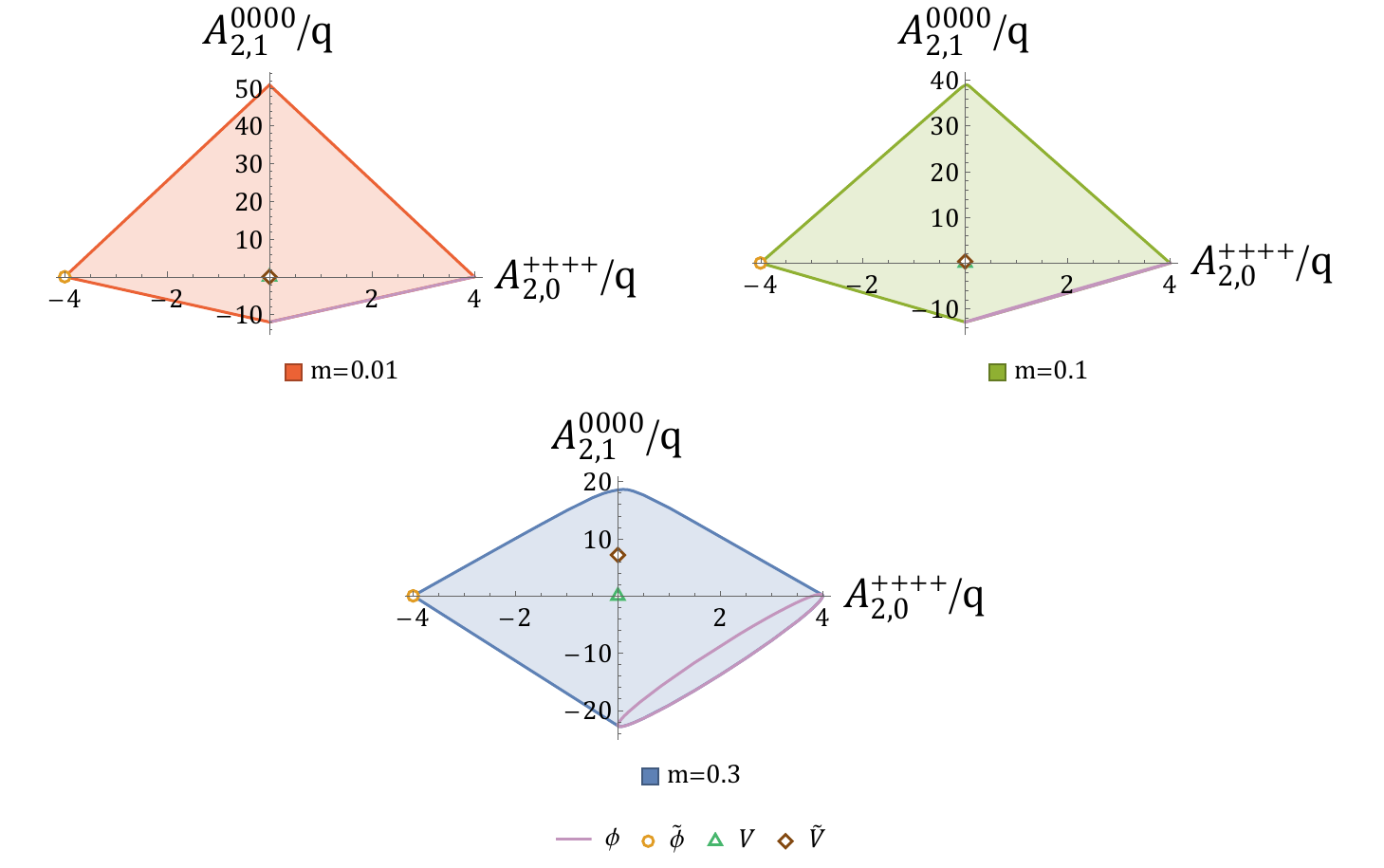}
    \caption{UV completions for the first mixed plots.}
    \label{fig:UVmixedPlot1}
\end{figure}

\begin{figure}
    \centering
    \includegraphics[scale=0.61]{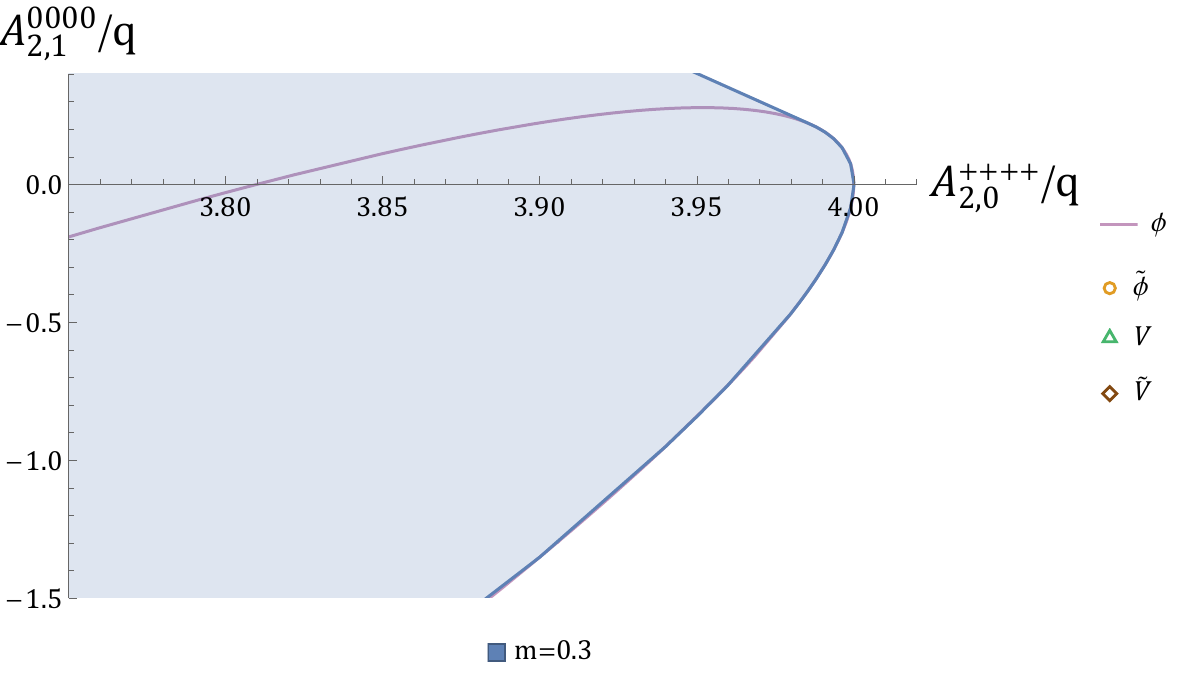}
    \caption{Close-up of the right kink of the first mixed plot with $m=3/10$ in figure~\ref{fig:UVmixedPlot1}. A similar thing happens for the kink at the bottom.}
    \label{fig:UVmixedPlot1Zoom}
\end{figure}

As promised, let us focus on figure~\ref{fig:UVphotonPlot2} for $m=3/10$. Near the bottom right we can see that the lower bound curves up precisely to allow the scalar completion to lie inside the allowed region. This is shown in figure~\ref{fig:UVphotonPlot2Zoom}. The same happens with the two kinks at the bottom and right in figure~\ref{fig:UVmixedPlot1}, again for $m=3/10$. A zoomed version of the right kink is showed in figure~\ref{fig:UVmixedPlot1Zoom}.

\subsubsection{Vector exchanges}

For a vector $B_\mu$ of mass $M$, which couples through
\begin{equation}
    \mathcal L \supset  B_\mu A_\nu \tilde{F}^{\mu\nu}, 
\end{equation}
the amplitudes are given by
\begin{align}
    f(s|t,u)&= \frac{2}{M^2} (4m^2 -3 s),
    \\
    g(s|t,u)&=- \frac{8}{M^2},
    \\
    \tilde g(s,t,u)&= -\frac{2}{M^2} \frac{2s - 4m^2 + M^2}{M^2 - t},
    \\
    h(s|t,u)&= - \frac{4}{M^2} \left( \frac{1}{M^2-t}+\frac{1}{M^2-u} \right),
    \\
    \tilde h(s,t,u)&= \frac{4}{M^2} \frac{s-t}{(M^2-s)(M^2-t)} .
\end{align}

For a pseudo-vector $B'_\mu$ with mass $M$ which couples through
\begin{equation}
    \mathcal L \supset  B'_\mu A_\nu F^{\mu\nu}
\end{equation}
we have 
\begin{align}
    f(s|t,u)&= - \frac{2}{M^2} s,
    \\
    g(s|t,u)&=- \frac{4}{M^2},
    \\
    \tilde g(s,t,u)&= -\frac{2}{M^2 -t},
    \\
    h(s|t,u)&=\frac{4}{M^2} \left( \frac{1}{M^2-t}+\frac{1}{M^2-u} \right),
    \\
    \tilde h(s,t,u)&= \frac{4}{M^2} \frac{s-t}{(M^2-s)(M^2-t)} .
\end{align}

\begin{figure}
    \centering
    \includegraphics[scale=0.61]{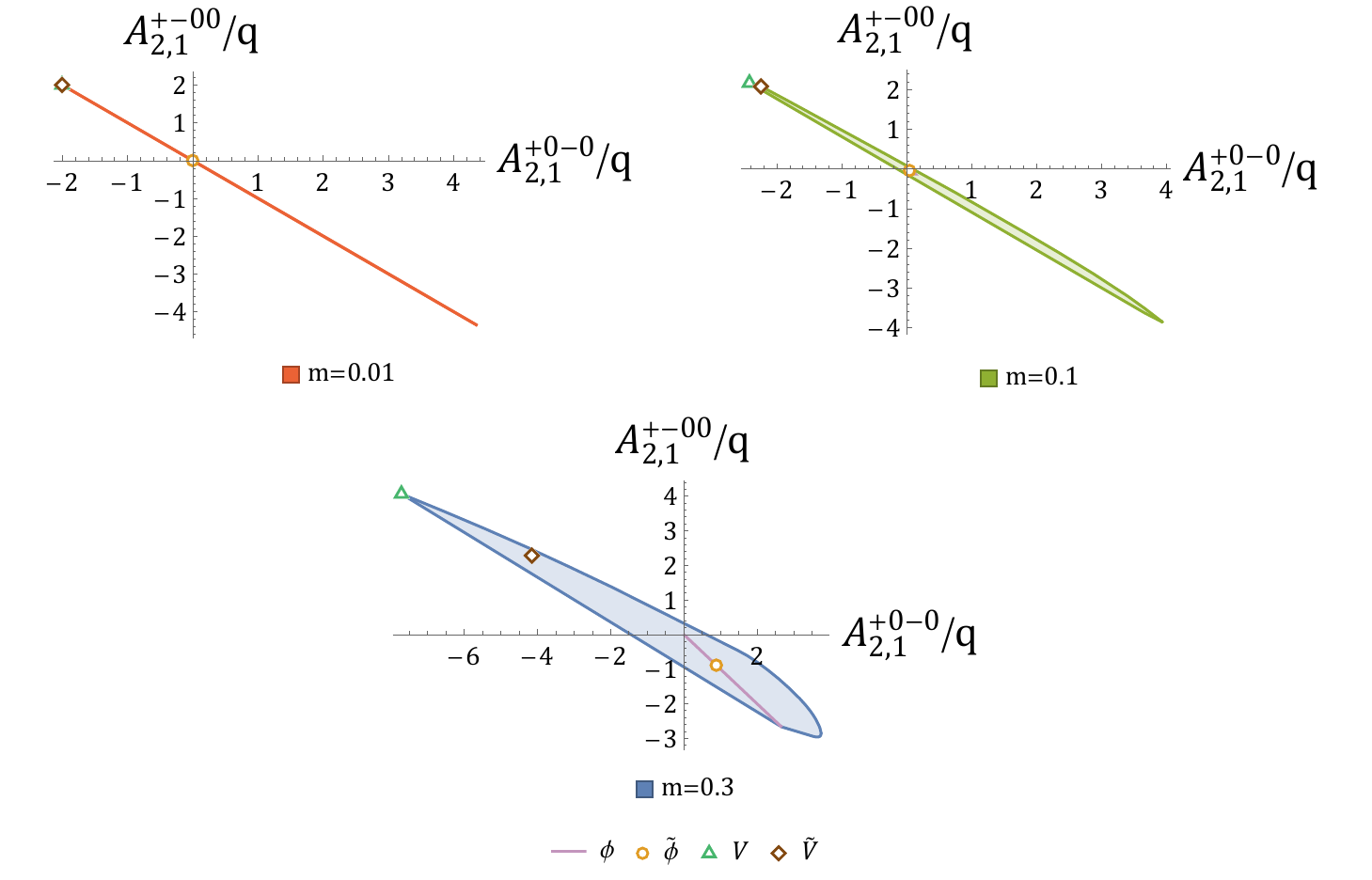}
    \caption{UV completions for the second mixed plots. The scalar completion curve ends at a kink.}
    \label{fig:UVmixedPlot2}
\end{figure}

In all our plots but one (figure~\ref{fig:UVmixedPlot2}), the vector and pseudo-vector have a peculiar feature: as $m\to0$, they approach the origin, where both observables vanish. One way to see this (focusing on the pseudo-vector completion for concreteness) goes through the St\"uckelberg representation of the massless limit, in which the massive vector is described by a photon and a derivatively coupled scalar. Then, such coupling $ B_\mu A_\nu F^{\mu\nu}$ is necessarily associated with the scalar component only  $B_\mu \partial_\nu\phi F^{\mu\nu}/m$, since spin-1 massive vectors $B_\mu$ cannot couple to two photons by the Landau-Yang theorem.
This implies that at small $m$ there can only be a coupling between
the longitudinal mode, the transverse modes, and the new vector $B$, if $\lambda_B/m$ does not vanish. This explains why the scattering amplitudes of two longitudinal and two transverse modes admit a non-vanishing vector/pseudo-vector completion in the massless limit, as can be seen in figure~\ref{fig:UVmixedPlot2}, while amplitudes with transverse only vanish.

\begin{figure}
    \centering
    \includegraphics[scale=0.61]{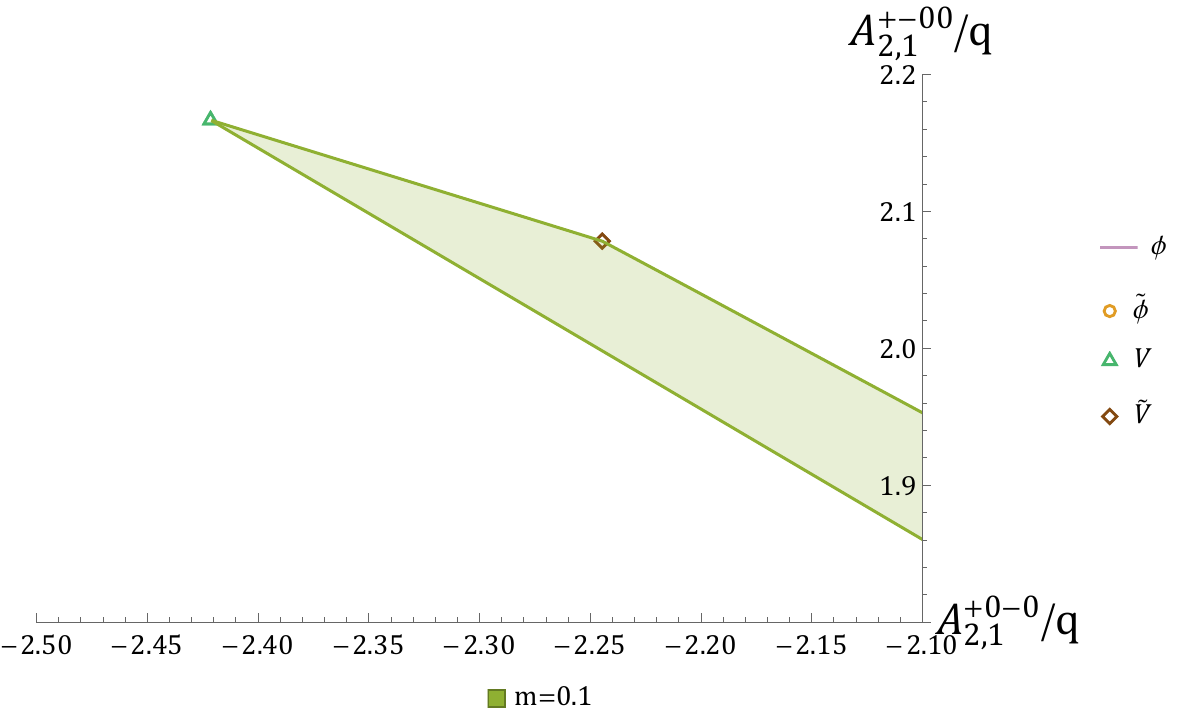}
    \caption{UV completions for the second mixed plots. Zoom on the kink for $m=1/10$.}
    \label{fig:UVmixedPlot2Zoom}
\end{figure}

For this special case of figure~\ref{fig:UVmixedPlot2}, the vector lives at the upper left kink of the plot for all three values of the mass. Notably, for $m=1/10$ a kink emerges, which is populated by the pseudo-vector completion: see~\ref{fig:UVmixedPlot2Zoom}. It seems like this kink merges with the pseudo-vector one in the massless limit, but further investigation is required to clarify this.

\section{Discussion}

The main outcome of this work has been to derive positivity bounds for massive vector particles. In doing this, however, we have addressed a number of issues that are generic to theories of massive higher spin:
 (a) to determine the most general set of 4pt amplitudes, which consist of general functions of Mandelstam invariants times contractions of momenta and external polarization, (b) to determine the action of crossing symmetry, which has a very complicated form on the helicity amplitudes but a much simpler form on the functions multiplying the structures, (c) to determine the constraints that unitarity imposes on the spectral densities of the helicity amplitudes, and (d) to use dispersion relations to derive sum rules and null constraints, which form the input to the by-now well understood numerical methods for finding optimal bounds on the EFT amplitudes.

The major difficulties stemmed from the fact that crossing symmetry, necessary to control the analytic structure of amplitudes in the complex energy plane, mixes every amplitude -- this is not the case, for instance, in scattering massless spin-one particles \cite{Henriksson:2021ymi, Henriksson:2022oeu}. 
We found that a mixed approach, partially based on
 helicity amplitudes $A^{\lambda_1 \lambda_2 \lambda_3 \lambda_4}$ (used to define observables), and in terms of the functions $f$, $g$, $\tilde{g}$, $h$ and $\tilde h$ with simple crossing properties (for deriving the null constraints), provides the most efficient path towards deriving positivity bounds.

 Positivity is inherently projective, and in this work we identified the appropriate ratios of observables that can be bounded in the context of massive spinning particles. Contrary to the massless case, where bounds could be expressed in terms of ratios between individual helicity amplitudes, here crossing symmetry implies that a minimum set of elastic amplitudes must appear at denominators. More precisely, generic bounds are expressed in terms of the sum of   transverse, longitudinal and mixed elastic combinations of amplitudes.

 As a  result we were able to identify the allowed regions in the space of EFT coefficients. 
 In the massless limit, we have shown how amplitude-observables of longitudinal helicities reproduce previous results on  spin-0 particles, while transverse ones match the results for photons.
We have also studied quantities that are unique to the massive case, such as the helicity non-conserving $+++0$ amplitude, and the sum of the $+-00$ and $+0-0$ ones. Bounds on these quantities become more and more stringent at small mass, and eventually they reduce the dimensionality of the parameter
space, as expected.

We compared our results to a set of possible ``partial UV completions'' -- partial in the sense that, although they do not provide a UV completion from the quantum field theoretical point of view, they furnish amplitudes which are consistent with the high-energy Regge behaviour.  These are the amplitudes that arise from integrating out scalars or vectors at tree-level.  
These completions allow us to understand important  features of our bounds and to track them as the mass is changed. 
A particularly interesting case is the longitudinal amplitude in figure \ref{fig:UVscalarPlot}, for which a kink in the massless limit morphs into a smooth curve at larger  values of $m$ and this is captured by a two-parameter family of scalar UV completions that arises in the massive case.

\paragraph{Future directions}
This work deals with the most important technical aspects of massive higher-spin amplitudes in a simple scenario of a single spin-1 massive state. In this way it paves the road to a number of interesting applications.

The weakly coupled EFT bootstrap developed in this paper is particularly well-suited to describing the interactions of the lightest spin-1 glueball in QCD in the limit of a large number of colors $N$. In this limit,  QCD is expected to be a  confining theory and the low-lying spectrum is dominated by stable, non-interacting glueballs and mesons. Exotic states like tetraquarks and molecules are notably absent, and hadron interactions are suppressed by powers of $1/\sqrt{N}$. The 't Hooft limit thus presents QCD as a theory of weakly-coupled hadrons, with interactions governed by a coupling that diminishes as $N$ approaches infinity. 

The study of QCD glueballs, even at large $N$, opens a pathway to understanding the low-energy dynamics of the strong nuclear force. Glueballs, composed solely of gluons, serve as unique probes into the non-perturbative regime of QCD. In order to do so, we need to include interactions with other low-energy states, such as scalar or spin-2 glueballs. The existence of a spin-2 resonance is expected to play a crucial role \cite{Albert:2023seb} and it will be important to explore its consequences and compare with lattice predictions for the glueballs' spectrum~\cite{Lucini:2010nv}.

Extending the analysis to include scattering of massive spin-2 (and possibly even higher spin) particles is an important step beyond the present work. This is particularly important in the context of QCD  and its higher-spin spectrum, as well as in massive gravity, where it was shown that the EFT has a very small region of validity \cite{Bellazzini:2023nqj}, \textit{i.e.} that the allowed regions of  coefficients  shrinks to zero if the graviton mass is much smaller than the cutoff scale. Ref.~\cite{Bellazzini:2023nqj} has not used the ``full'' set of positivity bounds, so it would be  interesting to improve that analysis to find sharp bounds on the allowed ratio $m / M$, relevant also to the QCD case.

Finally, it would be interesting to use these techniques in the context of  non-abelian massive gauge bosons, such as the $W$ and $Z$ bosons in the Standard Model (SM).\footnote{Ref.~\cite{Distler:2006if} and more recently \cite{Zhang:2018shp, Zhang:2020jyn} have used forward-limit bounds  to show that only a small percent of the parameter space that is usually considered in vector boson scattering is consistent with positivity.}
This could reveal potentially interesting UV completions to composite Higgs models, in which the Higgs boson only partially UV-completes $WW$ scattering.

\section*{Acknowledgments}

We would like to thank Jan Albert, Ilija Buri\'{c}, Simon Caron-Huot, Laura Engelbrecht, Denis Karateev, Stefanos Kousvos, Yue-Zhou Li, Scott Melville, Dario Sauro and Francesco Russo for a number of interesting and useful conversations about this work. This work has received funding from the European Research Council (ERC) under the European Union's Horizon 2020 research and innovation program (grant agreement no.~758903). B.M. is supported by NSERC (Canada). Partial funding also comes from the Mathematical Physics Laboratory of the CRM. The work of  S. R. and F. R. is supported by the Swiss National Science Foundation under grants no. 200021-205016 and PP00P2-206149.

\appendix
\addtocontents{toc}{\protect\setcounter{tocdepth}{1}}

\section{Notation and conventions}
\label{app:conventions}

In this paper, we define the amplitudes to be all-ingoing. In addition, we follow the conventions of Srednicki's QFT textbook \cite{Srednicki:2007qs} for the momenta and polarizations. This gives us: 

\begin{description}
\item[Metric] $\eta_{\mu\nu}=\mathrm{diag}(-1,1,1,1)$.
\item[Mandelstam variables] $s=-(p_1+p_2)^2=4\omega^2$, $t=-(p_1+p_3)^2$, $u=-(p_1+p_4)^2$. Thus $\cos\theta=1+\frac{2t}{s-4m^2}$.
\item[Standard momenta] $p^\mu_1=(\omega,0,0,p)$, $p^\mu_2=(\omega,0,0,-p)$, $p_3^\mu=(-\omega,-p \sin\theta,0,-p\cos\theta)$, $p_4^\mu=(-\omega,p \sin\theta,0,p \cos\theta)$, $m^2=\omega^2-p^2$.
\item[Assumptions for $s$-channel kinematics] $t<0$, $s>4m^2$. Moreover, we assume that $u=4m^2-s-t<0$, so that $stu>0$.
\item[Polarizations] For particle $1$, we assume $(\epsilon_1^+)^\mu=\frac1{\sqrt2}(0,1,-i,0)$, $(\epsilon_1^-)^\mu=\frac1{\sqrt2}(0,1,i,0)$ and $(\epsilon_1^0)^\mu=\frac1{m}(p,0,0,\omega)$. 

For particle $2$, we assume $(\epsilon_2^+)^\mu=\frac1{\sqrt2}(0,-1,-i,0)$, $(\epsilon_2^-)^\mu=\frac1{\sqrt2}(0,-1,i,0)$ and $(\epsilon_2^0)^\mu=\frac1{m}(p,0,0,-\omega)$. 

For particle $3$, we assume $(\epsilon_3^+)^\mu=-\frac1{\sqrt2}(0,\cos\theta,-i,-\sin\theta)$, $(\epsilon_3^-)^\mu=-\frac1{\sqrt2}(0,\cos\theta,i,-\sin\theta)$ and $(\epsilon_3^0)^\mu=-\frac1{m}(p,\omega\sin\theta,0,\omega\cos\theta)$.

 For particle $4$, we assume $(\epsilon_4^+)^\mu=-\frac1{\sqrt2}(0,-\cos\theta,-i,\sin\theta)$, $(\epsilon_4^-)^\mu=-\frac1{\sqrt2}(0,-\cos\theta,i,\sin\theta)$ and $(\epsilon_4^0)^\mu=-\frac1{m}(p,-\omega\sin\theta,0,-\omega\cos\theta)$.
\end{description}

\subsection{Wigner $d$ functions}
\label{app:wigner}

We use the standard Wigner $d$ functions, which for instance are implemented in Mathematica
\begin{equation}
    d_{\lambda_{12},\lambda_{43}}^\ell(\theta) = \texttt{WignerD[\{$\ell$,$\lambda_{43}$,$\lambda_{12}$\},$\theta$]},
\end{equation}
where $\lambda_{ij}=\lambda_i-\lambda_j$. For practical purposes, we find it convenient to use the closed-form expression in \cite{Caron-Huot:2022ugt},
\begin{align}\label{eq:Wignerdfunction}
    d_{\lambda_{12},\lambda_{43}}^\ell(\arccos x) &=\frac1{\Gamma(\lambda_{43}-\lambda_{12}+1)} \sqrt{\frac{\Gamma(\ell+\lambda_{43}+1)\Gamma(\ell-\lambda_{12}+1)}{\Gamma(\ell-\lambda_{43}+1)\Gamma(\ell+\lambda_{12}+1)}} 
    \left(\frac{1+x}2\right)^{\frac{\lambda_{43}+\lambda_{12}}2}
    \nonumber
    \\&\quad \times
    \left(\frac{1-x}2\right)^{\frac{\lambda_{43}-\lambda_{12}}2}
    {_2F_1}\left(\lambda_{43}-\ell,\ell+\lambda_{12}+1,\lambda_{43}-\lambda_{12}+1,\frac{1-x}2\right),
\end{align}
valid for $\lambda_{43}\geq \lambda_{12}$ and $\lambda_{43}+\lambda_{12}\geq 0$, together with the identities
\begin{align}
    d^\ell_{h,h'}(\arccos x)=(-1)^{h'-h}d^\ell_{h',h}(\arccos x)=d^\ell_{-h',-h}(\arccos x).
\end{align}

\section{Details on the amplitude parametrization}
\label{app:amplitudedetails}

Here we present a more full derivation of the exact parametrization used in the bulk of the paper.

\subsection{List of structures in general dimensions}
\label{app:structures}

Recall that we have defined $(\epsilon_i \epsilon_j) \equiv (\epsilon(k_i))^\mu  (\epsilon(k_j))_\mu$ and $(\epsilon_i k_j) \equiv (\epsilon(k_i))^\mu (k_j)\mu$. Then we can first write the basis of 19 structures, which provide a complete basis for $d>4$.

\begin{align}\label{eq:tildeepslist}
    \tilde{e}_1 \quad &= \quad (\epsilon_1\epsilon_2) \  (\epsilon_3 \epsilon_4) \, , \\
    \tilde{e}_2 \quad &= \quad (\epsilon_1 \epsilon_3) \ (\epsilon_2 \epsilon_4) \, , \\
    \tilde{e}_3 \quad &= \quad (\epsilon_1 \epsilon_4) \ (\epsilon_2 \epsilon_3) \, , \\ \\
    \tilde{e}_4  \quad &= \quad (\epsilon_1 \epsilon_{2}) \  (\epsilon_3 k_{4}) \ (\epsilon_4 k_{3}) + (\epsilon_3 \epsilon_{4}) \  (\epsilon_1 k_{2}) \ (\epsilon_2 k_{1}) \, , \\
    \tilde{e}_5  \quad &= \quad (\epsilon_1 \epsilon_{3}) \  (\epsilon_2 k_{4}) \ (\epsilon_4 k_{2}) + (\epsilon_2 \epsilon_{4}) \  (\epsilon_1 k_{3}) \ (\epsilon_3 k_{1}) \, , \\
    \tilde{e}_{6}  \quad &= \quad (\epsilon_1 \epsilon_{4}) \  (\epsilon_2 k_{3}) \ (\epsilon_3 k_{2}) + (\epsilon_2 \epsilon_{3}) \  (\epsilon_1 k_{4}) \ (\epsilon_4 k_{1}) \, , \\
    \tilde{e}_7  \quad &= \quad (\epsilon_1 \epsilon_{2}) \  (\epsilon_3 k_{1}) \ (\epsilon_4 k_{2}) + (\epsilon_3 \epsilon_{4}) \  (\epsilon_1 k_{3}) \ (\epsilon_2 k_{4}) \, , \\
    \tilde{e}_8  \quad &= \quad (\epsilon_1 \epsilon_{2}) \  (\epsilon_3 k_{2}) \ (\epsilon_4 k_{1}) + (\epsilon_3 \epsilon_{4}) \  (\epsilon_2 k_{3}) \ (\epsilon_1 k_{4}) \, , \\
    \tilde{e}_9  \quad &= \quad (\epsilon_1 \epsilon_{3}) \  (\epsilon_2 k_{1}) \ (\epsilon_4 k_{3}) + (\epsilon_2 \epsilon_{4}) \  (\epsilon_1 k_{2}) \ (\epsilon_3 k_{4}) \, , \\
    \tilde{e}_{10}  \quad &= \quad (\epsilon_1 \epsilon_{3} ) \  (\epsilon_2 k_{3}) \ (\epsilon_4 k_{1}) + (\epsilon_2 \epsilon_{4}) \  (\epsilon_1 k_{4}) \ (\epsilon_3 k_{2} ) \, , \\
    \tilde{e}_{11}  \quad &= \quad (\epsilon_1 \epsilon_{4}) \  (\epsilon_2 k_{1}) \ (\epsilon_3 k_{4}) + (\epsilon_2 \epsilon_{3}) \  (\epsilon_1 k_{2}) \ (\epsilon_4 k_{3}) \, , \\
    \tilde{e}_{12}  \quad &= \quad (\epsilon_1 \epsilon_{4} )\  (\epsilon_2 k_{4}) \ (\epsilon_3 k_{1}) + (\epsilon_2 \epsilon_{3}) \  (\epsilon_1 k_{3}) \ (\epsilon_4 k_{2}) \, , \\ \\
    \tilde{e}_{13}  \quad &= \quad (\epsilon_1 k_2) \ (\epsilon_2 k_1) \ (\epsilon_3 k_4) \ (\epsilon_4 k_3) \, , \\ 
    \tilde{e}_{14}  \quad &= \quad (\epsilon_1 k_3) \ (\epsilon_2 k_4) \ (\epsilon_3 k_1) \ (\epsilon_4 k_2) \, , \\ 
    \tilde{e}_{15}  \quad &= \quad (\epsilon_1 k_4) \ (\epsilon_2 k_3) \ (\epsilon_3 k_2) \ (\epsilon_4 k_1) \, , \\ 
    \tilde{e}_{16}  \quad &= \quad (\epsilon_1 k_2) \ (\epsilon_2 k_1) \ (\epsilon_3 k_1) \ (\epsilon_4 k_2) + (\epsilon_1 k_3) \ (\epsilon_2 k_4) \ (\epsilon_3 k_4) \ (\epsilon_4 k_3) \, , \\ 
    \tilde{e}_{17}  \quad &= \quad (\epsilon_1 k_3) \ (\epsilon_2 k_1) \ (\epsilon_3 k_4) \ (\epsilon_4 k_2 )+ (\epsilon_1 k_2) \ (\epsilon_2 k_4) \ (\epsilon_3 k_1) \ (\epsilon_4 k_3) \, , \\ 
    \tilde{e}_{18}  \quad &= \quad (\epsilon_1 k_2) \ (\epsilon_2 k_4) \ (\epsilon_3 k_4) \ (\epsilon_4 k_2) + (\epsilon_1 k_3) \ (\epsilon_2 k_1) \ (\epsilon_3 k_1) \ (\epsilon_4 k_3 )\, , \\ 
    \tilde{e}_{19}  \quad &= \quad (\epsilon_1 k_2) \ ( \epsilon_2 k_1) \ (\epsilon_3 k_2) \ (\epsilon_4 k_1) + (\epsilon_1 k_4) \ (\epsilon_2 k_3) \ (\epsilon_3 k_4) \ (\epsilon_4 k_3)  \, . \\ 
\end{align}

This provides a basis for boson-exchange and time-reversal symmetric amplitudes. In 4d, this basis is over-complete. This is because in 4d, only 4 vectors can be linearly independent -- thus there are relations between the polarization vectors and momenta that lead to additional structure constraints. One way to implement these constraints is to define 5-vectors $v_a$, $v_b$, $v_c$, $v_d$, and $v_e$, with $v  = (v_1, v_2, v_3, v_4, 0)$. Then it is clear that 
\begin{align}
    \varepsilon_{ABCDE} v_a^A v_b^B v_c^C v_d^D v_e^E = 0
\end{align}
because only 4 of the $v$s are linearly independent. If we apply this strategy to our kinematic 4-vectors by embedding them into 5d, we find three constraints:
\begin{align}
    \varepsilon_{ABCDE} \,  \epsilon_1^{A} \epsilon_2^{B} k_s^{C} k_t^{D} k_u^{E}  \, \varepsilon_{TWXYZ} \,  \epsilon_3^{T} \epsilon_4^{W} k_s^{X} k_t^{Y} k_u^{Z} \ = \ 0 \, , \\
    \, \varepsilon_{ABCDE} \,  \epsilon_1^{A} \epsilon_3^{B} k_s^{C} k_t^{D} k_u^{E}   \, \varepsilon_{TWXYZ} \,  \epsilon_2^{T} \epsilon_4^{W} k_s^{X} k_t^{Y} k_u^{Z} \ = \ 0 \, , \\
    \, \varepsilon_{ABCDE} \,  \epsilon_1^{A} \epsilon_4^{B} k_s^{C} k_t^{D} k_u^{E}  \, \varepsilon_{TWXYZ} \,  \epsilon_2^{T} \epsilon_3^{W} k_s^{X} k_t^{Y} k_u^{Z} \ = \ 0  \, .
\end{align}
Where $k_s = k_1 + k_2$, $k_t = k_1 + k_3$, and $k_u = k_1 + k_4$. This expression is easily simplified using identities for a product of epsilon tensors. These constraints are linearly dependent and lead to two constraints among the tensor structures. In the end, we find the following basis of structures, which comprise $e_i$ and which are used throughout the paper:
\begin{align}
\begin{split}
    {e}_1 \quad &= \quad (\epsilon_1\epsilon_2) \  (\epsilon_3 \epsilon_4) \, , \\
    {e}_2 \quad &= \quad (\epsilon_1 \epsilon_3) \ (\epsilon_2 \epsilon_4) \, , \\
    {e}_3 \quad &= \quad (\epsilon_1 \epsilon_4) \ (\epsilon_2 \epsilon_3) \, , \\ \\
    {e}_4  \quad &= \quad (\epsilon_1 \epsilon_{2}) \  (\epsilon_3 k_{4}) \ (\epsilon_4 k_{3}) + (\epsilon_3 \epsilon_{4}) \  (\epsilon_1 k_{2}) \ (\epsilon_2 k_{1}) \, , \\
    {e}_5  \quad &= \quad (\epsilon_1 \epsilon_{3}) \  (\epsilon_2 k_{4}) \ (\epsilon_4 k_{2}) + (\epsilon_2 \epsilon_{4}) \  (\epsilon_1 k_{3}) \ (\epsilon_3 k_{1}) \, , \\
    {e}_{6}  \quad &= \quad (\epsilon_1 \epsilon_{4}) \  (\epsilon_2 k_{3}) \ (\epsilon_3 k_{2}) + (\epsilon_2 \epsilon_{3}) \  (\epsilon_1 k_{4}) \ (\epsilon_4 k_{1}) \, , \\
    {e}_7  \quad &= \quad (\epsilon_1 \epsilon_{2}) \  (\epsilon_3 k_{1}) \ (\epsilon_4 k_{2}) + (\epsilon_3 \epsilon_{4}) \  (\epsilon_1 k_{3}) \ (\epsilon_2 k_{4}) \, , \\
    {e}_8  \quad &= \quad (\epsilon_1 \epsilon_{2}) \  (\epsilon_3 k_{2}) \ (\epsilon_4 k_{1}) + (\epsilon_3 \epsilon_{4}) \  (\epsilon_2 k_{3}) \ (\epsilon_1 k_{4}) \, , \\
    {e}_9  \quad &= \quad (\epsilon_1 \epsilon_{3}) \  (\epsilon_2 k_{1}) \ (\epsilon_4 k_{3}) + (\epsilon_2 \epsilon_{4}) \  (\epsilon_1 k_{2}) \ (\epsilon_3 k_{4}) \, , \\
    {e}_{10}  \quad &= \quad (\epsilon_1 \epsilon_{3} ) \  (\epsilon_2 k_{3}) \ (\epsilon_4 k_{1}) + (\epsilon_2 \epsilon_{4}) \  (\epsilon_1 k_{4}) \ (\epsilon_3 k_{2} ) \, , \\
    {e}_{11}  \quad &= \quad (\epsilon_1 \epsilon_{4}) \  (\epsilon_2 k_{1}) \ (\epsilon_3 k_{4}) + (\epsilon_2 \epsilon_{3}) \  (\epsilon_1 k_{2}) \ (\epsilon_4 k_{3}) \, , \\
    {e}_{12}  \quad &= \quad (\epsilon_1 \epsilon_{4} )\  (\epsilon_2 k_{4}) \ (\epsilon_3 k_{1}) + (\epsilon_2 \epsilon_{3}) \  (\epsilon_1 k_{3}) \ (\epsilon_4 k_{2}) \, , \\ \\
    {e}_{13}  \quad &= \quad (\epsilon_1 k_3) \ (\epsilon_2 k_4) \ (\epsilon_3 k_2) \ (\epsilon_4 k_1) + (\epsilon_1 k_4) \ (\epsilon_2 k_3) \ (\epsilon_3 k_1) \ (\epsilon_4 k_2) \, , \\ 
    {e}_{14}  \quad &= \quad (\epsilon_1 k_2) \ (\epsilon_2 k_3) \ (\epsilon_3 k_4) \ (\epsilon_4 k_1 )+ (\epsilon_1 k_4) \ (\epsilon_2 k_1) \ (\epsilon_3 k_2) \ (\epsilon_4 k_3) \, , \\ 
    {e}_{15}  \quad &= \quad (\epsilon_1 k_2) \ (\epsilon_2 k_4) \ (\epsilon_3 k_1) \ (\epsilon_4 k_3) + (\epsilon_1 k_3) \ (\epsilon_2 k_1) \ (\epsilon_3 k_4) \ (\epsilon_4 k_2 )\, , \\ 
    {e}_{16}  \quad &= \quad (\epsilon_1 k_3) \ (\epsilon_2 k_4) \ (\epsilon_3 k_1) \ (\epsilon_4 k_2) \, , \\ 
    {e}_{17}  \quad &= \quad (\epsilon_1 k_4) \ (\epsilon_2 k_3) \ (\epsilon_3 k_2) \ (\epsilon_4 k_1) \, . \\ 
\end{split}
\end{align}

A crossing symmetric basis also includes ${e}_{18}  = (\epsilon_1 k_2) \ (\epsilon_2 k_1) \ (\epsilon_3 k_4) \ (\epsilon_4 k_3)$, but this may be eliminated with the linear relation
\begin{align}\label{eq:18to17}
    e_{13}+e_{14}+e_{15} = e_{16} + e_{17} + e_{18} \, ,
\end{align}
which follows from momentum conservation. From the crossing-invariant basis of 18 structures, it is more clear how the functions given in~\eqref{eq:functions} arise. 

\subsection{Basis of functions}
\label{app:functions}

The crossing properties of the structures determine the crossing properties of the functions. First consider the amplitude
\begin{align}
\begin{split}
    A_\text{low} \ &= \ f_1(s|t,u) e_1 + f_1(t|s,u) e_2 + f_1(u|s,t) e_3 \\
    & \quad + g_1(s|t,u) e_4 + g_1(t|s,u) e_5  + g_1(u|s,t) e_6 \\
    & \quad + g_2(s,t,u) e_7 + g_2(s,u,t) e_8 + g_2(t,s,u) e_9  \\
    & \quad \qquad + g_2(t,u,s) e_{10} + g_2(u,s,t) e_{11} + g_2(u,t,s) e_{12} \\
    & \quad + h_1(s|t,u) e_{13} + h_1(t|s,u) e_{14}  + h_1(u|s,t) e_{15} \\
    & \quad + h_2(t|s,u) e_{16} + h_2(u|s,t) e_{17} + h_2(s|t,u) e_{18} \, .
\end{split}
\end{align}
Upon replacing $e_{18}$ using equation~\eqref{eq:18to17}, we find:
\begin{align}\label{eq:Alowgeneral}
    A_\text{Low} (s,t) \ &= \ f_1(s|t,u) e_1 + f_1(t|s,u) e_2 + f_1(u|s,t) e_3 
    \nonumber
    \\
    & \quad + g_1(s|t,u) e_4 + g_1(t|s,u) e_5  + g_1(u|s,t) e_6 
    \nonumber\\
    & \quad + g_2(s,t,u) e_7 + g_2(s,u,t) e_8 + g_2(t,s,u) e_9  
    \nonumber\\
    & \quad + g_2(t,u,s) e_{10} + g_2(u,s,t) e_{11} + g_2(u,t,s) e_{12} 
    \nonumber\\
    & \quad + \big(h_1(s|t,u)+h_2(s|t,u)\big) e_{13} + \big(h_1(t|s,u)+h_2(s|t,u)\big) e_{14}  
    \nonumber\\
    & \quad+ \big(h_1(u|s,t)+h_2(s|t,u)\big) e_{15} + \big(h_2(t|s,u)-h_2(s|t,u)\big) e_{16}
    \nonumber\\
    &\quad+ \big(h_2(u|s,t)-h_2(s|t,u)\big) e_{17}.
\end{align}
To avoid redundant coefficients, and for convenience, we redefine the functions as
\begin{align}
\begin{split}
    f(s|t) \ &= \ f_1(s|t, u), \\
    g(s|t) \ &= \ g_1(s|t, u), \\
    \tilde{g}(s, t) \ &= \ g_2(u, s, t), \\
    h(s|t) \ &= \ h_1(s|t, u) + \frac{1}{2} \big(h_2(t|s, u) + h_2(u|s, t) \big),\\ 
    \tilde{h}(s,t) \ &= \frac12 \ \big(h_2(s|t,u) - h_2(t|s,u)\big).
\end{split}
\end{align}
So we have three partially symmetric functions ($f$, $g$, and $h$), a function with no symmetry properties ($\tilde{g}$) and an $s-t$ antisymmetric function ($\tilde{h}$). The general amplitudes in our theory are parametrized by the following coefficients:  
\begin{align}
    f(s|t) \ & = \ f_{0,0} + f_{1,0} (t + u) + f_{2,0} (t + u)^2 + f_{0,1} t u +   \ldots,  \\
    g(s|t) \ & = \ g_{0,0} + g_{1,0} (t + u) + g_{2,0} (t + u)^2 + g_{0,1} t u + \ldots,  \\
    \tilde g(s,t) \ & = \ \tilde{g}_{0,0} + \tilde{g}_{1,0} s + \tilde{g}_{0,1} t + \tilde{g}_{2,0} s^2 + \tilde{g}_{1,1} st + \ldots,  \\
    h(s|t) \ & = \ h_{0,0} + h_{1,0} (t + u) + h_{2,0} (t + u)^2 + h_{0,1} t u + \ldots, \\
    \tilde h(s,t) \ & = \ (s-t) \left( \tilde{h}_{0,0} + \tilde{h}_{1,0} (s+t) + \tilde{h}_{2,0} (s+t)^2 + \tilde{h}_{0,1} st +   \ldots \right).
\end{align}
The resulting parametrization is 
\begin{align}
    A_\text{Low} (s,t) \ &= \ f(s|t) e_1 + f(t|u) e_2 + f(u|s) e_3 
    \nonumber
    \\
    & \quad + g(s|t) e_4 + g(t|u) e_5  + g(u|s) e_6 
    \nonumber\\
    & \quad + \tilde g(s,t) e_7 + \tilde g(s,u) e_8 + \tilde g(t,s) e_9  
     + \tilde g(t,u) e_{10} + \tilde g(u,s) e_{11} + \tilde g(u,t) e_{12} 
    \nonumber\\
    & \quad + \big( h(s|t) + \tilde{h}(s,t) + \tilde{h}(s,u)\big) e_{13} +\big( h(t|u) + \tilde{h}(s,u)\big) e_{14} \nonumber \\
    & \qquad \quad + \big( h(u|s) + \tilde{h}(s,t)\big) e_{15} 
    \nonumber\\
    & \quad + 2\tilde{h}(t,s) e_{16} + 2 \tilde{h}(u,s) e_{17}.
    \label{eq:Alowgeneralnew}
\end{align}

\section{Generalized Proca theory}
\label{app:Proca-theory}
 The technology of this paper is rather abstract, so it may be helpful to illustrate using the Generalized Proca theory (or ``Proca theory,'' for short), described by the Lagrangian \cite{deRham:2020yet}:
\begin{align}
\begin{split}
    \label{eq:proca_lagrangian}
    \mathcal{L}^{GP} &=  -\frac{1}{4} F^2 - \frac{1}{2} m^2 A^2 + b_1 m^4 (A^2)^2 + b_2 m^2 A^2 F^{\mu \nu}F_{\mu \nu} \\
    & \qquad  + b_3 m^2 A^2 \left[ \left( \partial \cdot A \right)^2 - \partial_\rho A_\sigma \partial^\sigma A^\rho \right] + b_4 m^2 F^{\mu \nu} F^\rho{}_\nu A_\mu A_\rho \\
    & \qquad \quad + b_5 F^{\mu \nu} F_{\nu \rho} F^{\rho \sigma} F_{\sigma \mu} + b_6 (F^{\mu \nu} F_{\mu \nu})^2 + b_7 \tilde F^{\mu \nu} \tilde F^{\rho \sigma} \partial_\rho A_\mu \partial_\sigma A_\nu .
\end{split}
\end{align}

In general, massive vector theories are constructed by allowing the vector to have derivative self-interactions, with the additional requirement that only three degrees of freedom propagate, or equivalently, that there are no ghosts.

As a starting point for massive spin-one theories we can take a simple deformation of the Maxwell action plus a mass term $A_\mu A^\mu$, where the mass term explicitly breaks the $U(1)$ gauge-invariance. This theory does not have a smooth $m \to 0$ because the massive theory has three degrees of freedom while the massless theory has only two. This is not a pathology per se but might represent an aesthetic problem if we expect that physics varies smoothly as we change the parameters of the theory. The St\"uckelberg trick cures this problem by identifying a new theory which is dynamically equivalent to the original, but which has a different massless limit. This new theory is obtained by the replacement
$  A_\mu \ \to \ A_\mu + \partial_\mu \phi$.
This transformation, plus rescaling $\phi \to \phi / m$ to ensure that $\phi$ has a canonical kinetic term, transforms the action to 
\begin{align}
    S_{\text{S}} \ = \ \int d^4x \left[ -\frac{1}{4} F_{\mu \nu} F^{\mu \nu} - \frac{1}{2} m^2 A_\mu A^\mu - m \partial_\mu \phi A^\mu - \frac{1}{2}  \partial_\mu \phi \partial^\mu \phi \right] \, .
\end{align}
This theory has a gauge symmetry $A \to A + d \lambda$, $\phi \to \phi - m \lambda$. Its massless limit is now smooth: we see that as $m \to 0$, we recover a decoupled massless vector and massless scalar.

Constructing more interesting massive spin-one theories requires adding self-interactions. It is well-known that higher-derivative interactions can cause ghostly degrees of freedom to enter the theory. So a basic question is: what interactions ensure the propagation of only three degrees of freedom? 

The requirement that only three degrees of freedom propagate is equivalent to the condition that one of the components of $A$ appears in the Lagrangian as a Lagrange multiplier enforcing a constraint. This, in turn, is satisfied if the Hessian of the Lagrangian $\ \frac{\delta^2 \mathcal{L}}{\delta \dot{A}^\mu \dot{A}^\nu}$ has a null eigenvector, which roughly implies that one linear combination of the components of $A$ appears with only a single time derivative and is thus non-dynamical.\footnote{This condition was first discussed in \cite{deRham:2011rn} in the context of massive gravity; see also \cite{deRham:2014zqa} and \cite{Heisenberg:2018vsk} for reviews discussing constraints and the Hessian matrix.}

The ``Generalized Proca theory,'' or simply ``Proca theory,'' discovered in \cite{Tasinato:2014eka, Heisenberg:2014rta} and further clarified in \cite{Allys:2015sht, BeltranJimenez:2016rff, Allys:2016jaq, Heisenberg:2018vsk}, is one of the few massive spin one theories which have (a) second-order equations of motion and (b) a non-dynamical time component. A different solution, inspired by the decoupling limit of massive gravity in AdS \cite{DeRham:2018axr}, was given in \cite{deRham:2020yet}.

The Proca action has no infinite sum of coefficients like in \eqref{eq:ffun}--\eqref{eq:tildehfun}. This notably contradicts the philosophy of this paper, which is to allow for the most general EFT. However this will make it a nice example because the finite number of coefficients will make it much easier to derive sum rules and null constraints for the theory. In what follows, we will give some examples of how to obtain bounds in the Proca theory, which should be useful for the reader trying to understand the formalism of the previous sections.

\subsection{Amplitudes in the Proca theory}
\label{app:procaamplitudes}
The amplitude arising from~\eqref{eq:proca_lagrangian} is 
\begin{align}
\begin{split}
         A_{\text{EFT}} \ &= \  8 b_1 m^4 (e_1 + e_2 + e_3)  \\
        & \quad + 8 b_2 m^2 \left( (s - 2m^2) e_1  + (t - 2m^2) e_2 + (u - 2m^2) e_3 + e_4 + e_5 + e_{6} \right) \\
        & \quad +4 b_3 m^2 ( e_4 + e_5 + e_{6} ) \\
        & \quad - 2 b_4 m^2 (s e_1 + t e_2 + u e_3 + 2 e_4 + 2 e_5 + 2 e_6 +  e_7 + e_8 + e_9 + e_{10} + e_{11} + e_{12} ) \\
        & \quad + b_5 \Big( (s^2 + (t-u)^2) e_1 + (t^2 + (s-u)^2) e_2 + (u^2 + (s-t)^2) e_3 + 4 s e_4 + 4 t e_5  \\
        & \quad \qquad \qquad \qquad + 4 u e_6 + 8(u-m^2) e_7 +  8(t-m^2) e_8 + 8(u-m^2) e_{9} +  8(s-m^2) e_{10} \\
        & \quad \qquad \qquad \qquad + 8(t-m^2) e_{11} +  8(s-m^2) e_{12} + 8 e_{13} + 8 e_{14} + 8 e_{15} \Big) \\
        & \quad + 8 b_6  \Big( (s - 2m^2)^2 e_1  + (t - 2m^2)^2 e_2 + (u - 2m^2)^2 e_3 + 2(s-2m^2) e_4 \\
        & \quad \qquad \qquad \qquad + 2(t-2m^2)e_5 + 2(u-2m^2)e_{6} + 4 e_{13} + 4 e_{14} + 4 e_{15}\Big)  \\
        & \quad + 4 b_7 \Big( (t+u) e_4 + (s + u) e_5 + (s + t) e_6 + (u-t) e_7 + (t-u) e_8 \\
        & \quad \qquad \qquad \qquad + (u-s) e_9 + (s-u) e_{10} + (t-s) e_{11} + (s-t) e_{12} \Big).
    \end{split}
    \label{eq:EFT-structure-amplitude}
\end{align}
From this amplitude, we can determine the functions that multiply the structures:
\begin{align}
    f(s|t) \ &= \ 8 m^4 (b_1 + 2 b_2 - b_4 + 2 b_5 + 4 b_6) + 2 m^2 (-4 b_2 + b_4 - 4 b_5 - 16 b_6) (t + u) \nonumber \\\label{eq:procaTermsSec2no1}
    & \qquad + (2 b_5 + 8 b_6) (t + u)^2 - 4 b_5 t u, \\
    g(s| t) \ &= \ 4m^2(2 b_2 + b_3 - b_4 + 4 b_5 + 8 b_6) - 4(b_5 + 4 b_6 - b_7)(t + u), \\
    \tilde g(s,t) \ &= \ -2m^2 (b_4 - 12 b_5 - 8 b_7) - (8 b_5 + 4 b_7) s - (8 b_5 + 8 b_7)t,  \\
    h(s| t) \ &= \ 8 b_5 + 32 b_6, \\
    \tilde h (s,t) \ &= \ 0.    
    \label{eq:procaTermsSec2no5}
\end{align}
or, in terms of the coefficients appearing in the functions, 
\begin{align}
\begin{split}
    f_0 \ &= \ 8 m^4 (b_1 + 2 b_2 - b_4 + 2 b_5 + 4 b_6), \\
    f_{1,0} \ &= \ 2m^2 (-4 b_2 + b_4 - 4 b_5 - 16 b_6), \\
    f_{2,0} \ &= \  2 b_5 + 8 b_6, \\
    f_{0,1} \ &= \ -4 b_5, \\
    g_0 \ &= \ 4m^2 (2 b_2 + b_3 - b_4 + 4 b_5 + 8 b_6), \\
    g_{1,0} \ &= \ -4(b_5 + 4 b_6 - b_7), \\
    \tilde{g}_0 \ &= \ -2 m^2 (b_4 -12 b_5 - 8 b_7), \\
    \tilde{g}_{1,0} \ &= \  -8 b_5 -4 b_7, \\
    \tilde{g}_{0,1} \ &= \ -8 b_5 - 8 b_7, \\
    h_0 \ &= \ 8 b_5 + 32 b_6. 
    \end{split}
\end{align}

\subsection{Forward limit sum rules for the Proca theory}
\label{sec:Proca-ruls-and-bounds}

In what follows we will write down the explicit form of the sum rules for the Proca theory, whose Lagrangian is given in \eqref{eq:proca_lagrangian}. Our goal is to find bounds on the coefficients $b_i$'s. In the process, we will also compare with the known results in the massless limit.

Recall that the ``master equation'' for computing sum rules is
\begin{equation}
	\begin{split}
		&\left( \res_{s=0}+\res_{s=4m^2} \right)
  \left[\frac{\tilde A^I_{\Low}(s,t)}{s^{1+k}}\right] =\sum_{\ell}16(2\ell+1)\int_{M^2}^\infty ds \, \mu(s) \Bigg[ \rho_{\ell}^I (s)\,  \frac{\tilde d^{\ell,I}(\theta)}{s^k}  \\
		& 
  \qquad \qquad 
  -\sum_J {\big(\tilde E(u(s),t)C\tilde E^{-1}(s,t) \big)^I}_J 
  \frac{s \, \rho_\ell^J(s) \tilde d^{\ell,J}(\theta)}{(4m^2-s-t)^{1+k}} \Bigg] ,
	\end{split}
\end{equation}
where we defined
\begin{equation}
	\tilde d^{\ell,I}(\theta) =\begin{cases}
		\frac{m s}{\sqrt{-t s (s+t-4m^2)}} d^{\ell,I}(\theta) \quad \mathrm{if} \quad I=2,4,9,10,11, \\
		d^{\ell,I}(\theta) \quad \mathrm{otherwise,}
	\end{cases}
\end{equation}
and
\begin{equation}
    d^{\ell}_{\lambda_{12}\lambda_{43}}(\theta) \rightarrow d^{\ell,I}(\theta), \quad \rho^{\lambda_1 \lambda_2 \lambda_3 \lambda_4}_\ell (s) \rightarrow \rho^I_\ell (s).
\end{equation}

Now we will assume that the low-energy is the generalized Proca theory. Then, using $k=2$ and $t=0$, we can find the sum rules for the coefficients of the theory. In 6 cases out of 17, the dispersion relations yield $0=0$, because the Wigner-$d$ functions in the right-hand side vanish at $t = 0$. The sum rules which we will examine will be
\begin{align}\label{eq:GPsumrules}
        4(b_5 +4b_6) &=\sum_{\ell}16(2\ell+1)\int_{M^2}^\infty ds \, \mu(s) \Bigg[ \left(\frac{s}{\left(s-4 m^2\right)^3}+\frac{1}{s^2}\right) \rho_{\ell}^1(s) \Bigg], \\
        4b_1 &=\sum_{\ell}16(2\ell+1)\int_{M^2}^\infty ds \, \mu(s) \Bigg[\left(\frac{s}{\left(s-4 m^2\right)^3}+\frac{1}{s^2}\right) \rho_{\ell}^5(s) \Bigg], \\
        2b_2 -b_3 +\frac{b_4}{2} &=\sum_{\ell}16(2\ell+1)\int_{M^2}^\infty ds \, \mu(s) \Bigg[-\frac{s \rho_{\ell}^8(s)}{\left(s -4 m^2\right)^3}+\frac{\rho_{\ell}^6(s)}{s^2} \Bigg], \\
        6b_5 +8b_6 &=\sum_{\ell}16(2\ell+1)\int_{M^2}^\infty ds \, \mu(s) \Bigg[\frac{s \rho_{\ell}^{12}(s)}{\left(s-4 m^2\right)^3}+\frac{\rho_{\ell}^{14}(s)}{s^2} \Bigg], \\
        b_4 &=\sum_{\ell}16(2\ell+1)\int_{M^2}^\infty ds \, \mu(s) \Bigg[\left(\frac{s}{\left(s-4 m^2\right)^3}+\frac{1}{s^2}\right) \rho_{\ell}^{16}(s) \Bigg].
\end{align}
Additionally, we have 
\begin{align}
    \label{eq:proca:b7}
            -\frac{b_7}{\sqrt{2}} &=\sum_{\ell}16(2\ell+1)\int_{M^2}^\infty ds \, \mu(s) \Bigg[ -\frac{\sqrt{2} m^2 \left( \rho_{\ell}^1(s)- \rho_{\ell}^6(s) + \rho_{\ell}^8(s) \right)}{\left(s -4 m^2\right)^3}
        \\\nonumber
        &\hspace{6cm} -\frac{2 \sqrt{\ell (\ell+1)} m \left(8 m^4-4 m^2 s+s^2\right) \rho_{\ell}^2(s)}{s^{3/2} \left(s-4 m^2\right)^3} \Bigg], \\\label{eq:proca:b3}
        -\sqrt{2}b_3 &=\sum_{\ell}16(2\ell+1)\int_{M^2}^\infty ds \, \mu(s) \Bigg[\frac{\sqrt{2} m^2 \left( \rho_{\ell}^5(s)- \rho_{\ell}^6(s) + \rho_{\ell}^8(s) + \rho_{\ell}^{16}(s) \right)}{\left(s -4 m^2\right)^3} \\\nonumber
        &\hspace{6cm} +\frac{2 \sqrt{\ell(\ell+1)} m \left(8 m^4-4 m^2 s+s^2\right) \rho_{\ell}^4(s)}{s^{3/2} \left(s-4 m^2\right)^3} \Bigg] \, ,
\end{align}
which come from amplitudes for which the sum of external polarizations is odd. They are proportional to $m^2$, which vanishes in the massless limit. These sum rules have an explicit dependence in $\ell$. This is because the definition for $\tilde{A}$ \eqref{eq:Atildedef} implies that the $t\to0$
limit contains first derivatives of the amplitude in $t$. Therefore, these are not forward quantities.

\paragraph{Massless limit}

Let us compare with the massless limit $m\rightarrow 0$, where the massive vector becomes a massless vector (\emph{i.e.} a photon) plus a real scalar field, and the two do not interact. In other words, the scalar decouples from the vector and becomes free. Therefore, to compare with the photon case we must not look at those helicity amplitudes that involve $\lambda_i =0$. As $m\rightarrow 0$, \eqref{eq:GPsumrules} greatly simplifies to
\begin{equation}
    \begin{split}
        2(2b_5 +8b_6) &=\sum_{\ell}16(2\ell+1)\int_{M^2}^\infty ds \, \mu(s) \Bigg[\frac{2\rho^1_\ell(s)}{s^2} \Bigg], \\
        4b_1 &=\sum_{\ell}16(2\ell+1)\int_{M^2}^\infty ds \, \mu(s) \Bigg[\frac{2\rho^5_\ell(s)}{s^2}\Bigg], \\
        2b_2 -b_3 +\frac{b_4}{2} &=\sum_{\ell}16(2\ell+1)\int_{M^2}^\infty ds \, \mu(s) \Bigg[ \frac{\rho^6_\ell(s)}{s^2} -\frac{\rho^8_\ell(s)}{s^2} \Bigg], \\
        6b_5 +8b_6 &=\sum_{\ell}16(2\ell+1)\int_{M^2}^\infty ds \, \mu(s) \Bigg[\frac{\rho^{14}_\ell(s)}{s^2} +\frac{\rho^{12}_\ell(s)}{s^2} \Bigg], \\
        b_4 &=\sum_{\ell}16(2\ell+1)\int_{M^2}^\infty ds \, \mu(s) \Bigg[\frac{2\rho^{16}_\ell(s)}{s^2} \Bigg].
    \end{split}
\end{equation}

The relevant sum rules for the transverse modes (see \eqref{eq:helamplorder}) are the two involving $\rho^1_\ell (s)$, $\rho^{12}_\ell (s)$ and $\rho^{14}_\ell (s)$. By comparing the spectral densities with \cite{Henriksson:2021ymi, Henriksson:2022oeu}\footnote{To connect with the notation in \cite{Henriksson:2021ymi}, note that $\rho^1_{\mathrm{photon}} \leftrightarrow \rho^1$, $\rho^2_{\mathrm{photon}} \leftrightarrow \rho^{12}$, $\rho^3_{\mathrm{photon}} \leftrightarrow \rho^{14}$, $\rho^4_{\mathrm{photon}} \leftrightarrow \rho^{13}$, $\rho^5_{\mathrm{photon}} \leftrightarrow \rho^3$.}, we can see that
\begin{equation}
    \begin{split}
        f_2 &\equiv 2b_5 +8b_6, \\
        g_2 &\equiv 6b_5 +8b_6,
    \end{split}
\end{equation}
where $g_2$ and $f_2$ appear in the photon scattering $A^{++--}_{\mathrm{photon}}=g_2 s^2 +\dots$ and $A^{++++}_{\mathrm{photon}}=f_2(s^2+t^2+u^2)+\dots$ respectively. Indeed, we see that the combinations $g_2 +f_2$, $g_2 -f_2$ yield
\begin{equation}
    \begin{split}
        g_2 +f_2 &=\sum_{\ell}16(2\ell+1)\int_{M^2}^\infty ds \, \mu(s) \frac1{s^2} \left[ \rho^{12}_\ell(s) +\rho^{14}_\ell(s) +\rho^1_\ell(s) \right] , \\
        g_2 -f_2 &=\sum_{\ell}16(2\ell+1)\int_{M^2}^\infty ds \, \mu(s) \frac1{s^2} \left[ \rho^{12}_\ell(s) +\rho^{14}_\ell(s) -\rho^{1}_\ell(s) \right] .
    \end{split}
\end{equation}
In fact, both combinations of densities $ \rho^{12}_\ell(s) +\rho^{14}_\ell(s) \pm \rho^{1}_\ell(s) $ are positive, and we find the result of \cite{Henriksson:2021ymi} that $g_2\pm f_2\geqslant 0$.

\subsection{Full matrix sum rules}

Let us return to the massive case. The matrices that correspond to \eqref{eq:GPsumrules} are
\begin{align}
    \nonumber
    \mathbf V^{ee}_{2 (2 b_5+8 b_6)}(\ell,s) &=\left(\begin{smallmatrix}
1&0&0&0\\ 0&0&0&0\\ 0&0&0&0\\ 0&0&0&0
\end{smallmatrix}\right)\left( \frac{s}{(s-4m^2)^3}+\frac1{s^2} \right),\\
\nonumber\mathbf V^{oe}_{2 (2 b_5+8 b_6)}(\ell,s) &=\begin{pmatrix}
-1&0\\ 0&0
\end{pmatrix}\left( \frac{s}{(s-4m^2)^3} +\frac1{s^2} \right), \\
    \mathbf V^{eo}_{2 (2 b_5+8 b_6)}(\ell,s) &=\begin{pmatrix}
0&0\\ 0&0
\end{pmatrix},&\mathbf V^{oo}_{2 (2 b_5+8 b_6)}(\ell,s) &=\big(\,0\,\big),
\end{align}

\begin{align}
    \nonumber
    \mathbf V^{ee}_{4 b_1}(\ell,s) &=\left(\begin{smallmatrix}
0&0&0&0\\ 0&0&0&0\\ 0&0&0&0\\ 0&0&0&1
\end{smallmatrix}\right)\left( \frac{s}{(s-4m^2)^3} +\frac1{s^2} \right), \quad & \mathbf V^{oe}_{4 b_1}(\ell,s) &=\begin{pmatrix}
0&0\\ 0&0
\end{pmatrix}, 
\\
    \mathbf V^{eo}_{4 b_1}(\ell,s) &=\begin{pmatrix}
0&0\\ 0&0
\end{pmatrix}, \quad & \mathbf V^{oo}_{4b_1}(\ell,s) &=\big(\,0\,\big),
\end{align}

\begin{align}
    \nonumber
    \mathbf V^{ee}_{2b_2 -b_3 +\frac12 b_4}(\ell,s) &=\left(\begin{smallmatrix}
0&0&0&\frac1{2s^2}\\ 0&0&0&0\\ 0&0&-\frac{s}{(s-4 m^2)^3}&0\\ \frac1{2s^2}&0&0&0
\end{smallmatrix}\right), \quad & \mathbf V^{oe}_{2b_2 -b_3 +\frac12 b_4}(\ell,s) &=\begin{pmatrix}
0&0\\ 0&\frac{s}{(s-4m^2)^3}
\end{pmatrix}, 
\\
    \mathbf V^{eo}_{2b_2 -b_3 +\frac12 b_4}(\ell,s) &=\begin{pmatrix}
0&0\\ 0&-\frac{s}{(s-4m^2)^3}
\end{pmatrix}, \quad & \mathbf V^{oo}_{2b_2 -b_3 +\frac12 b_4}(\ell,s) &=\bigg(\,\frac{s}{(s-4m^2)^3}\,\bigg),
\end{align}

\begin{align}
    \nonumber
    \mathbf V^{ee}_{6b_5 +8b_6}(\ell,s) &=\left(\begin{smallmatrix}
\frac{s}{(s-4m^2)^3}&0&0&0\\ 0&\frac1{s^2}&0&0\\ 0&0&0&0\\ 0&0&0&0
\end{smallmatrix}\right), \quad & \mathbf V^{oe}_{6b_5 +8b_6}(\ell,s) &=\begin{pmatrix}
1&0\\ 0&0
\end{pmatrix}\frac{s}{(s-4m^2)^3}, 
\\
    \mathbf V^{eo}_{6b_5 +8b_6}(\ell,s) &=\begin{pmatrix}
1&0\\ 0&0
\end{pmatrix}\frac1{s^2}, \quad & \mathbf V^{oo}_{6b_5 +8b_6}(\ell,s) &=\big(\,0\,\big),
\end{align}
\begin{align}
    \nonumber
    \mathbf V^{ee}_{b_4}(\ell,s) &=\left(\begin{smallmatrix}
0&0&0&0\\ 0&0&0&0\\ 0&0&\frac{s}{(s-4m^2)^3} +\frac1{s^2}&0\\ 0&0&0&0
\end{smallmatrix}\right), \quad & \mathbf V^{oe}_{b_4}(\ell,s) &=\begin{pmatrix}
0&0\\ 0&1
\end{pmatrix} \left(\frac{s}{(s-4m^2)^3} +\frac1{s^2}\right), 
\\
    \mathbf V^{eo}_{b_4}(\ell,s) &=\begin{pmatrix}
0&0\\ 0&1
\end{pmatrix} \left(\frac{s}{(s-4m^2)^3} +\frac1{s^2} \right), \quad & \mathbf V^{oo}_{b_4}(\ell,s) &=\bigg(\,\frac{s}{(s-4m^2)^3} +\frac1{s^2}\,\bigg).
\end{align}

\paragraph{Null constraints} In the forward limit, not including \eqref{eq:proca:b7} and \eqref{eq:proca:b3} (the sum rules for $b_3$ and $b_7$), the only null constraints are
\begin{equation}
    \begin{split}
        0 &=\sum_{\ell}16(2\ell+1)\int_{M^2}^\infty ds \, \mu(s) \Bigg[ \left( \frac{s}{(s-4m^2)^3} -\frac1{s^2} \right) \rho_{\ell}^{6}(s) +\left( \frac{s}{(s-4m^2)^3} -\frac1{s^2} \right) \rho_{\ell}^{8}(s) \Bigg], \\
        0 &=\sum_{\ell}16(2\ell+1)\int_{M^2}^\infty ds \, \mu(s) \Bigg[ \left( \frac{s}{(s-4m^2)^3} -\frac1{s^2} \right)\rho_{\ell}^{12}(s) 
        +\left( \frac1{s^2} -\frac{s}{(s-4m^2)^3} \right) \rho_{\ell}^{14}(s) \Bigg].
    \end{split}
\end{equation}
These null constraints do not appear when we take the subtraction point to be $s=2m^2$ instead of $s=0$, as in section \ref{sec:forward-limit-bounds}. Instead, the sum rules coming from $\rho^6$ and $\rho^8$ (resp. $\rho^{12}$ and $\rho^{14}$) are duplicates of each other.

\subsection{A positive coefficient in the Proca theory}\label{sec:proca}

 Consider the amplitude
\begin{equation}
    \tilde A^{12}=A^{++--}\,,
\end{equation}
which in our conventions represents the process $++\to++$.
Using the parametrizations \eqref{eq:procaTermsSec2no1}--\eqref{eq:procaTermsSec2no5}, we find that the sum rule $\mathcal I_{k}^{12}(t)$ of \eqref{eq:mastersumrule} for $k=2$ evaluates to
\begin{equation}
    \mathcal I_{2}^{12}(t)_\Low=\oint\frac{ds}{2\pi i}\frac{\tilde A_{\Low}^{12}(s,t)}{s^3}=6b_5+8b_6 =: g_2,
\end{equation}
which for the specific action \eqref{eq:proca_lagrangian} is independent of $t$. 
In the massless limit $m\to0$, this constant agrees with the $g_2$ coefficient appearing in the photon scattering amplitude $A^{++--}_{\mathrm{photon}}=g_2s^2+\ldots$ in the parametrization of \cite{Henriksson:2021ymi}.

Now consider the high-energy density for $t=0$.
The contribution from the right-hand cut involves the partial wave density $c^{++}_{\ell,X}(s)$. It therefore directly follows that the contributions from right-hand cut are,
\begin{align}
    \mathbf V^{ee}_{g_2}(\ell,s)|_{\text{rh cut}}&=\left(\begin{smallmatrix}
1&0&0&0\\ 0&0&0&0\\ 0&0&0&0\\ 0&0&0&0
\end{smallmatrix}\right)\frac{d_{0,0}^{\ell}(0)}{s^2}, 
&
    \mathbf V^{oe}_{g_2}(\ell,s)|_{\text{rh cut}}&=\begin{pmatrix}
1&0\\ 0&0
\end{pmatrix}\frac{d_{0,0}^{\ell}(0)}{s^2}, 
\\
    \mathbf V^{eo}_{g_2}(\ell,s)|_{\text{rh cut}}&=\begin{pmatrix}
0&0\\ 0&0
\end{pmatrix},
&
    \mathbf V^{oo}_{g_2}(\ell,s)|_{\text{rh cut}}&=\big(\,0\,\big)\,.
\end{align}
Here we used that $t=0$ corresponds to the scattering angle $\theta=0$. In this limit, $d^\ell_{0,0}(\theta)=1$ for any spin $\ell$.

For the left-hand cut, we use crossing as described in section~\ref{sec:crossing}.
We find
\begin{align}
    \mathbf V^{ee}_{g_2}(\ell,s)|_{\text{lh cut}}&=\left(\begin{smallmatrix}
0 & 0 & 0 & 0 \\
 0 & 1 & 0 & 0 \\
 0 & 0 & 0 & 0 \\
 0 & 0 & 0 & 0
\end{smallmatrix}\right)\frac{s\, d^\ell_{2,2}(0)}{(s-4m^2)^3}, 
&
    \mathbf V^{oe}_{g_2}(\ell,s)|_{\text{lh cut}}&=\begin{pmatrix}
0&0\\0&0
\end{pmatrix}, 
\\
    \mathbf V^{eo}_{g_2}(\ell,s)|_{\text{lh cut}}&=\begin{pmatrix}
1&0\\0&0
\end{pmatrix}\frac{s\, d^\ell_{2,2}(0)}{(s-4m^2)^3},
&
    \mathbf V^{oo}_{g_2}(\ell,s)|_{\text{lh cut}}&=\big(\,0 \,\big).
\end{align}
Putting the pieces together, we have
\begin{align}
    \mathbf V^{ee}_{g_2}(\ell,s)&=\begin{pmatrix}
\frac{1}{s^2} & 0 & 0 & 0 \\
 0 & \frac{s}{(s-4 m^2)^3} & 0 & 0 \\
 0 & 0 & 0 & 0 \\
 0 & 0 & 0 & 0
\end{pmatrix}, 
&
    \mathbf V^{oe}_{g_2}(\ell,s)&=\begin{pmatrix}
\frac{1}{s^2} & 0 \\
 0 & 0 
\end{pmatrix}, 
\\
    \mathbf V^{eo}_{g_2}(\ell,s)&=\begin{pmatrix}
\frac{s}{(s-4 m^2)^3} & 0 \\
 0 & 0
\end{pmatrix},
&
    \mathbf V^{oo}_{g_2}(\ell,s)&=\big(\,0 \,\big),
\end{align}
where we used that $d^\ell_{0,0}(0)=d^\ell_{2,2}(0)=1$

By direct inspection, the matrices appearing in the sum rules for $g_2$ are positive semi-definite because $s\geqslant M^2 >4m^2$, so
\begin{equation}
   \mathbf V^{\pm\pm}_{g_2}\succcurlyeq0\,.
\end{equation}
We can therefore immediately conclude that
\begin{equation}
    g_2\geqslant 0\,,
\end{equation}
i.e. the same as the well-known positivity condition for a massless vector (photon) \cite{Cheung:2014ega,Henriksson:2021ymi}.
This is in agreement with section \ref{sec:forward-limit-bounds}, where we show the equivalent bound for the general case.

\section{Dealing with square roots}
\label{sec:roots}

In \eqref{eq:Atildedef} we introduce an improved amplitude to remove some $\sqrt{stu}$ factors due to the external polarizations. The Wigner $d$ functions \eqref{eq:Wignerdfunction} are proportional to $\sqrt{tu}$ whenever there is an odd number of zero helicity $\lambda_i =0$. So, due to crossing, we have that in general every non-forward sum rule is proportional to $\sqrt{s}$. Because SDPB only works with polynomials, we need to take care of these square roots. Let us explain what we do here.

As we stressed, $\sqrt{s}$ only appears whenever a spectral density has an odd number of zero helicities. There are five of these:
\begin{align}
    \rho_\ell^2, \quad \rho_\ell^9, &\quad \rho_\ell^{10}\,, \quad \rho_\ell^{11}\,, \qquad \text{one 0 helicity}\,,\\
    &\rho_\ell^4, \qquad \qquad  \qquad \, \text{three 0 helicities}\,.
\end{align}
From here, we can also see how these square roots affect the matrices after we write the spectral densities in terms of $c^{\lambda_i\lambda_j}_{X,\ell}(s)$. To get rid of these factors, it suffices to multiply the matrix $\mathbf V$ by a suitable diagonal matrix of the form $(1,\dots,\sqrt{s},\dots,1)$ from the left and the right. This is in practice equivalent to rescaling the vectors $c^{\lambda_i\lambda_j}_{X,\ell}(s)$. In turn, this can be seen as a (positive) rescaling of the spectral densities. We thus use the following prescription:
\begin{equation}
    \rho_\ell^I \rightarrow \left( \frac{\sqrt{s}}{m} \right)^{\#\text{zeroes in }I} \rho_\ell^I\,.
\end{equation}

\section*{}

\bibliography{cite.bib}

\providecommand{\href}[2]{#2}\begingroup\raggedright\begin{thebibliography}{10}

\bibitem{Pham:1985cr}
T.N.~Pham and T.N.~Truong, \emph{{Evaluation of the derivative quartic terms of
  the meson chiral Lagrangian from forward dispersion relation}},
  \href{https://doi.org/10.1103/PhysRevD.31.3027}{\emph{Phys. Rev. D}
  {\bfseries 31} (1985) 3027}.

\bibitem{Pennington:1994kc}
M.R.~Pennington and J.~Portoles, \emph{{The Chiral Lagrangian parameters,
  $\bar\ell_1$, $\bar\ell_2$, are determined by the $\rho$-resonance}},
  \href{https://doi.org/10.1016/0370-2693(94)01551-M}{\emph{Phys. Lett. B}
  {\bfseries 344} (1995) 399}
  [\href{https://arxiv.org/abs/hep-ph/9409426}{{\ttfamily hep-ph/9409426}}].

\bibitem{Ananthanarayan:1994hf}
B.~Ananthanarayan, D.~Toublan and G.~Wanders, \emph{{Consistency of the chiral
  pion pion scattering amplitudes with axiomatic constraints}},
  \href{https://doi.org/10.1103/PhysRevD.51.1093}{\emph{Phys. Rev. D}
  {\bfseries 51} (1995) 1093}
  [\href{https://arxiv.org/abs/hep-ph/9410302}{{\ttfamily hep-ph/9410302}}].

\bibitem{Comellas:1995hq}
J.~Comellas, J.I.~Latorre and J.~Taron, \emph{{Constraints on chiral
  perturbation theory parameters from QCD inequalities}},
  \href{https://doi.org/10.1016/0370-2693(95)01110-C}{\emph{Phys. Lett. B}
  {\bfseries 360} (1995) 109}
  [\href{https://arxiv.org/abs/hep-ph/9507258}{{\ttfamily hep-ph/9507258}}].

\bibitem{Dita:1998mh}
P.~Dita, \emph{{Positivity constraints on chiral perturbation theory pion pion
  scattering amplitudes}},
  \href{https://doi.org/10.1103/PhysRevD.59.094007}{\emph{Phys. Rev. D}
  {\bfseries 59} (1999) 094007}
  [\href{https://arxiv.org/abs/hep-ph/9809568}{{\ttfamily hep-ph/9809568}}].

\bibitem{Adams:2006sv}
A.~Adams, N.~Arkani-Hamed, S.~Dubovsky, A.~Nicolis and R.~Rattazzi,
  \emph{{Causality, analyticity and an IR obstruction to UV completion}},
  \href{https://doi.org/10.1088/1126-6708/2006/10/014}{\emph{JHEP} {\bfseries
  10} (2006) 014} [\href{https://arxiv.org/abs/hep-th/0602178}{{\ttfamily
  hep-th/0602178}}].

\bibitem{Rattazzi:2008pe}
R.~Rattazzi, V.S.~Rychkov, E.~Tonni and A.~Vichi, \emph{{Bounding scalar
  operator dimensions in 4D CFT}},
  \href{https://doi.org/10.1088/1126-6708/2008/12/031}{\emph{JHEP} {\bfseries
  12} (2008) 031} [\href{https://arxiv.org/abs/0807.0004}{{\ttfamily
  0807.0004}}].

\bibitem{Bellazzini:2020cot}
B.~Bellazzini, J.~Elias~Mir\'o, R.~Rattazzi, M.~Riembau and F.~Riva,
  \emph{{Positive moments for scattering amplitudes}},
  \href{https://doi.org/10.1103/PhysRevD.104.036006}{\emph{Phys. Rev. D}
  {\bfseries 104} (2021) 036006}
  [\href{https://arxiv.org/abs/2011.00037}{{\ttfamily 2011.00037}}].

\bibitem{Tolley:2020gtv}
A.J.~Tolley, Z.-Y.~Wang and S.-Y.~Zhou, \emph{{New positivity bounds from full
  crossing symmetry}},
  \href{https://doi.org/10.1007/JHEP05(2021)255}{\emph{JHEP} {\bfseries 05}
  (2021) 255} [\href{https://arxiv.org/abs/2011.02400}{{\ttfamily
  2011.02400}}].

\bibitem{Caron-Huot:2020cmc}
S.~Caron-Huot and V.~Van~Duong, \emph{{Extremal effective field theories}},
  \href{https://doi.org/10.1007/JHEP05(2021)280}{\emph{JHEP} {\bfseries 05}
  (2021) 280} [\href{https://arxiv.org/abs/2011.02957}{{\ttfamily
  2011.02957}}].

\bibitem{Sinha:2020win}
A.~Sinha and A.~Zahed, \emph{{Crossing symmetric dispersion relations in
  quantum field theories}},
  \href{https://doi.org/10.1103/PhysRevLett.126.181601}{\emph{Phys. Rev. Lett.}
  {\bfseries 126} (2021) 181601}
  [\href{https://arxiv.org/abs/2012.04877}{{\ttfamily 2012.04877}}].

\bibitem{Arkani-Hamed:2020blm}
N.~Arkani-Hamed, T.-C.~Huang and Y.-T.~Huang, \emph{{The EFT-hedron}},
  \href{https://doi.org/10.1007/JHEP05(2021)259}{\emph{JHEP} {\bfseries 05}
  (2021) 259} [\href{https://arxiv.org/abs/2012.15849}{{\ttfamily
  2012.15849}}].

\bibitem{Manohar:2008tc}
A.V.~Manohar and V.~Mateu, \emph{{Dispersion relation bounds for pi pi
  scattering}}, \href{https://doi.org/10.1103/PhysRevD.77.094019}{\emph{Phys.
  Rev. D} {\bfseries 77} (2008) 094019}
  [\href{https://arxiv.org/abs/0801.3222}{{\ttfamily 0801.3222}}].

\bibitem{Mateu:2008gv}
V.~Mateu, \emph{{Universal bounds for $SU(3)$ low energy constants}},
  \href{https://doi.org/10.1103/PhysRevD.77.094020}{\emph{Phys. Rev. D}
  {\bfseries 77} (2008) 094020}
  [\href{https://arxiv.org/abs/0801.3627}{{\ttfamily 0801.3627}}].

\bibitem{Nicolis:2009qm}
A.~Nicolis, R.~Rattazzi and E.~Trincherini, \emph{{Energy's and amplitudes'
  positivity}}, \href{https://doi.org/10.1007/JHEP05(2010)095}{\emph{JHEP}
  {\bfseries 05} (2010) 095} [\href{https://arxiv.org/abs/0912.4258}{{\ttfamily
  0912.4258}}]. [Erratum: {\em JHEP} {\bf 11} (2011) 128].

\bibitem{Baumann:2015nta}
D.~Baumann, D.~Green, H.~Lee and R.A.~Porto, \emph{{Signs of analyticity in
  single-field inflation}},
  \href{https://doi.org/10.1103/PhysRevD.93.023523}{\emph{Phys. Rev. D}
  {\bfseries 93} (2016) 023523}
  [\href{https://arxiv.org/abs/1502.07304}{{\ttfamily 1502.07304}}].

\bibitem{Bellazzini:2015cra}
B.~Bellazzini, C.~Cheung and G.N.~Remmen, \emph{{Quantum gravity constraints
  from unitarity and analyticity}},
  \href{https://doi.org/10.1103/PhysRevD.93.064076}{\emph{Phys. Rev. D}
  {\bfseries 93} (2016) 064076}
  [\href{https://arxiv.org/abs/1509.00851}{{\ttfamily 1509.00851}}].

\bibitem{Bellazzini:2016xrt}
B.~Bellazzini, \emph{{Softness and amplitudes\textquoteright{} positivity for
  spinning particles}},
  \href{https://doi.org/10.1007/JHEP02(2017)034}{\emph{JHEP} {\bfseries 02}
  (2017) 034} [\href{https://arxiv.org/abs/1605.06111}{{\ttfamily
  1605.06111}}].

\bibitem{Cheung:2016yqr}
C.~Cheung and G.N.~Remmen, \emph{{Positive signs in massive gravity}},
  \href{https://doi.org/10.1007/JHEP04(2016)002}{\emph{JHEP} {\bfseries 04}
  (2016) 002} [\href{https://arxiv.org/abs/1601.04068}{{\ttfamily
  1601.04068}}].

\bibitem{Bonifacio:2016wcb}
J.~Bonifacio, K.~Hinterbichler and R.A.~Rosen, \emph{{Positivity constraints
  for pseudolinear massive spin-2 and vector Galileons}},
  \href{https://doi.org/10.1103/PhysRevD.94.104001}{\emph{Phys. Rev. D}
  {\bfseries 94} (2016) 104001}
  [\href{https://arxiv.org/abs/1607.06084}{{\ttfamily 1607.06084}}].

\bibitem{Cheung:2016wjt}
C.~Cheung and G.N.~Remmen, \emph{{Positivity of curvature-squared corrections
  in gravity}},
  \href{https://doi.org/10.1103/PhysRevLett.118.051601}{\emph{Phys. Rev. Lett.}
  {\bfseries 118} (2017) 051601}
  [\href{https://arxiv.org/abs/1608.02942}{{\ttfamily 1608.02942}}].

\bibitem{deRham:2017avq}
C.~de~Rham, S.~Melville, A.J.~Tolley and S.-Y.~Zhou, \emph{{Positivity bounds
  for scalar field theories}},
  \href{https://doi.org/10.1103/PhysRevD.96.081702}{\emph{Phys. Rev. D}
  {\bfseries 96} (2017) 081702}
  [\href{https://arxiv.org/abs/1702.06134}{{\ttfamily 1702.06134}}].

\bibitem{Bellazzini:2017fep}
B.~Bellazzini, F.~Riva, J.~Serra and F.~Sgarlata, \emph{{Beyond positivity
  bounds and the fate of massive gravity}},
  \href{https://doi.org/10.1103/PhysRevLett.120.161101}{\emph{Phys. Rev. Lett.}
  {\bfseries 120} (2018) 161101}
  [\href{https://arxiv.org/abs/1710.02539}{{\ttfamily 1710.02539}}].

\bibitem{deRham:2017zjm}
C.~de~Rham, S.~Melville, A.J.~Tolley and S.-Y.~Zhou, \emph{{UV complete me:
  Positivity Bounds for Particles with Spin}},
  \href{https://doi.org/10.1007/JHEP03(2018)011}{\emph{JHEP} {\bfseries 03}
  (2018) 011} [\href{https://arxiv.org/abs/1706.02712}{{\ttfamily
  1706.02712}}].

\bibitem{deRham:2017imi}
C.~de~Rham, S.~Melville, A.J.~Tolley and S.-Y.~Zhou, \emph{{Massive Galileon
  positivity bounds}},
  \href{https://doi.org/10.1007/JHEP09(2017)072}{\emph{JHEP} {\bfseries 09}
  (2017) 072} [\href{https://arxiv.org/abs/1702.08577}{{\ttfamily
  1702.08577}}].

\bibitem{Hinterbichler:2017qyt}
K.~Hinterbichler, A.~Joyce and R.A.~Rosen, \emph{{Massive spin-2 scattering and
  asymptotic superluminality}},
  \href{https://doi.org/10.1007/JHEP03(2018)051}{\emph{JHEP} {\bfseries 03}
  (2018) 051} [\href{https://arxiv.org/abs/1708.05716}{{\ttfamily
  1708.05716}}].

\bibitem{Bonifacio:2017nnt}
J.~Bonifacio, K.~Hinterbichler, A.~Joyce and R.A.~Rosen, \emph{{Massive and
  massless spin-2 scattering and asymptotic superluminality}},
  \href{https://doi.org/10.1007/JHEP06(2018)075}{\emph{JHEP} {\bfseries 06}
  (2018) 075} [\href{https://arxiv.org/abs/1712.10020}{{\ttfamily
  1712.10020}}].

\bibitem{Bellazzini:2017bkb}
B.~Bellazzini, F.~Riva, J.~Serra and F.~Sgarlata, \emph{{The other effective
  fermion compositeness}},
  \href{https://doi.org/10.1007/JHEP11(2017)020}{\emph{JHEP} {\bfseries 11}
  (2017) 020} [\href{https://arxiv.org/abs/1706.03070}{{\ttfamily
  1706.03070}}].

\bibitem{Bonifacio:2018vzv}
J.~Bonifacio and K.~Hinterbichler, \emph{{Bounds on amplitudes in effective
  theories with massive spinning particles}},
  \href{https://doi.org/10.1103/PhysRevD.98.045003}{\emph{Phys. Rev. D}
  {\bfseries 98} (2018) 045003}
  [\href{https://arxiv.org/abs/1804.08686}{{\ttfamily 1804.08686}}].

\bibitem{deRham:2018qqo}
C.~de~Rham, S.~Melville, A.J.~Tolley and S.-Y.~Zhou, \emph{{Positivity bounds
  for massive spin-1 and spin-2 fields}},
  \href{https://doi.org/10.1007/JHEP03(2019)182}{\emph{JHEP} {\bfseries 03}
  (2019) 182} [\href{https://arxiv.org/abs/1804.10624}{{\ttfamily
  1804.10624}}].

\bibitem{Zhang:2018shp}
C.~Zhang and S.-Y.~Zhou, \emph{{Positivity bounds on vector boson scattering at
  the LHC}}, \href{https://doi.org/10.1103/PhysRevD.100.095003}{\emph{Phys.
  Rev. D} {\bfseries 100} (2019) 095003}
  [\href{https://arxiv.org/abs/1808.00010}{{\ttfamily 1808.00010}}].

\bibitem{Bellazzini:2018paj}
B.~Bellazzini and F.~Riva, \emph{{New phenomenological and theoretical
  perspective on anomalous ZZ and Z\ensuremath{\gamma} processes}},
  \href{https://doi.org/10.1103/PhysRevD.98.095021}{\emph{Phys. Rev. D}
  {\bfseries 98} (2018) 095021}
  [\href{https://arxiv.org/abs/1806.09640}{{\ttfamily 1806.09640}}].

\bibitem{Bellazzini:2019xts}
B.~Bellazzini, M.~Lewandowski and J.~Serra, \emph{{Positivity of amplitudes,
  weak gravity conjecture, and modified gravity}},
  \href{https://doi.org/10.1103/PhysRevLett.123.251103}{\emph{Phys. Rev. Lett.}
  {\bfseries 123} (2019) 251103}
  [\href{https://arxiv.org/abs/1902.03250}{{\ttfamily 1902.03250}}].

\bibitem{Melville:2019wyy}
S.~Melville and J.~Noller, \emph{{Positivity in the sky: Constraining dark
  energy and modified gravity from the UV}},
  \href{https://doi.org/10.1103/PhysRevD.101.021502}{\emph{Phys. Rev. D}
  {\bfseries 101} (2020) 021502}
  [\href{https://arxiv.org/abs/1904.05874}{{\ttfamily 1904.05874}}]. [Erratum:
  {\em Phys. Rev. D} {\bf 102} (2020) 049902].

\bibitem{deRham:2019ctd}
C.~de~Rham and A.J.~Tolley, \emph{{Speed of gravity}},
  \href{https://doi.org/10.1103/PhysRevD.101.063518}{\emph{Phys. Rev. D}
  {\bfseries 101} (2020) 063518}
  [\href{https://arxiv.org/abs/1909.00881}{{\ttfamily 1909.00881}}].

\bibitem{Alberte:2019xfh}
L.~Alberte, C.~de~Rham, A.~Momeni, J.~Rumbutis and A.J.~Tolley,
  \emph{{Positivity constraints on interacting spin-2 fields}},
  \href{https://doi.org/10.1007/JHEP03(2020)097}{\emph{JHEP} {\bfseries 03}
  (2020) 097} [\href{https://arxiv.org/abs/1910.11799}{{\ttfamily
  1910.11799}}].

\bibitem{Alberte:2019zhd}
L.~Alberte, C.~de~Rham, A.~Momeni, J.~Rumbutis and A.J.~Tolley,
  \emph{{Positivity constraints on interacting pseudo-linear spin-2 fields}},
  \href{https://doi.org/10.1007/JHEP07(2020)121}{\emph{JHEP} {\bfseries 07}
  (2020) 121} [\href{https://arxiv.org/abs/1912.10018}{{\ttfamily
  1912.10018}}].

\bibitem{Bi:2019phv}
Q.~Bi, C.~Zhang and S.-Y.~Zhou, \emph{{Positivity constraints on aQGC: carving
  out the physical parameter space}},
  \href{https://doi.org/10.1007/JHEP06(2019)137}{\emph{JHEP} {\bfseries 06}
  (2019) 137} [\href{https://arxiv.org/abs/1902.08977}{{\ttfamily
  1902.08977}}].

\bibitem{Remmen:2019cyz}
G.N.~Remmen and N.L.~Rodd, \emph{{Consistency of the Standard Model Effective
  Field Theory}}, \href{https://doi.org/10.1007/JHEP12(2019)032}{\emph{JHEP}
  {\bfseries 12} (2019) 032}
  [\href{https://arxiv.org/abs/1908.09845}{{\ttfamily 1908.09845}}].

\bibitem{Ye:2019oxx}
G.~Ye and Y.-S.~Piao, \emph{{Positivity in the effective field theory of
  cosmological perturbations}},
  \href{https://doi.org/10.1140/epjc/s10052-020-7973-z}{\emph{Eur. Phys. J. C}
  {\bfseries 80} (2020) 421}
  [\href{https://arxiv.org/abs/1908.08644}{{\ttfamily 1908.08644}}].

\bibitem{Herrero-Valea:2019hde}
M.~Herrero-Valea, I.~Timiryasov and A.~Tokareva, \emph{{To positivity and
  beyond, where Higgs-dilaton inflation has never gone before}},
  \href{https://doi.org/10.1088/1475-7516/2019/11/042}{\emph{JCAP} {\bfseries
  11} (2019) 042} [\href{https://arxiv.org/abs/1905.08816}{{\ttfamily
  1905.08816}}].

\bibitem{Zhang:2020jyn}
C.~Zhang and S.-Y.~Zhou, \emph{{Convex geometry perspective on the (Standard
  Model) Effective Field Theory Space}},
  \href{https://doi.org/10.1103/PhysRevLett.125.201601}{\emph{Phys. Rev. Lett.}
  {\bfseries 125} (2020) 201601}
  [\href{https://arxiv.org/abs/2005.03047}{{\ttfamily 2005.03047}}].

\bibitem{Trott:2020ebl}
T.~Trott, \emph{{Causality, unitarity and symmetry in effective field theory}},
  \href{https://doi.org/10.1007/JHEP07(2021)143}{\emph{JHEP} {\bfseries 07}
  (2021) 143} [\href{https://arxiv.org/abs/2011.10058}{{\ttfamily
  2011.10058}}].

\bibitem{Zhang:2021eeo}
C.~Zhang, \emph{{SMEFTs living on the edge: determining the UV theories from
  positivity and extremality}},
  \href{https://doi.org/10.1007/JHEP12(2022)096}{\emph{JHEP} {\bfseries 12}
  (2022) 096} [\href{https://arxiv.org/abs/2112.11665}{{\ttfamily
  2112.11665}}].

\bibitem{Wang:2020jxr}
Y.-J.~Wang, F.-K.~Guo, C.~Zhang and S.-Y.~Zhou, \emph{{Generalized positivity
  bounds on chiral perturbation theory}},
  \href{https://doi.org/10.1007/JHEP07(2020)214}{\emph{JHEP} {\bfseries 07}
  (2020) 214} [\href{https://arxiv.org/abs/2004.03992}{{\ttfamily
  2004.03992}}].

\bibitem{Li:2021lpe}
X.~Li, H.~Xu, C.~Yang, C.~Zhang and S.-Y.~Zhou, \emph{{Positivity in multifield
  effective field theories}},
  \href{https://doi.org/10.1103/PhysRevLett.127.121601}{\emph{Phys. Rev. Lett.}
  {\bfseries 127} (2021) 121601}
  [\href{https://arxiv.org/abs/2101.01191}{{\ttfamily 2101.01191}}].

\bibitem{Du:2021byy}
Z.-Z.~Du, C.~Zhang and S.-Y.~Zhou, \emph{{Triple crossing positivity bounds for
  multi-field theories}},
  \href{https://doi.org/10.1007/JHEP12(2021)115}{\emph{JHEP} {\bfseries 12}
  (2021) 115} [\href{https://arxiv.org/abs/2111.01169}{{\ttfamily
  2111.01169}}].

\bibitem{Davighi:2021osh}
J.~Davighi, S.~Melville and T.~You, \emph{{Natural selection rules: new
  positivity bounds for massive spinning particles}},
  \href{https://doi.org/10.1007/JHEP02(2022)167}{\emph{JHEP} {\bfseries 02}
  (2022) 167} [\href{https://arxiv.org/abs/2108.06334}{{\ttfamily
  2108.06334}}].

\bibitem{Chowdhury:2021ynh}
S.D.~Chowdhury, K.~Ghosh, P.~Haldar, P.~Raman and A.~Sinha, \emph{{Crossing
  symmetric spinning S-matrix bootstrap: EFT bounds}},
  \href{https://arxiv.org/abs/2112.11755}{{\ttfamily 2112.11755}}.

\bibitem{Henriksson:2021ymi}
J.~Henriksson, B.~McPeak, F.~Russo and A.~Vichi, \emph{{Rigorous bounds on
  light-by-light scattering}},
  \href{https://doi.org/10.1007/JHEP06(2022)158}{\emph{JHEP} {\bfseries 06}
  (2022) 158} [\href{https://arxiv.org/abs/2107.13009}{{\ttfamily
  2107.13009}}].

\bibitem{Caron-Huot:2021enk}
S.~Caron-Huot, D.~Mazac, L.~Rastelli and D.~Simmons-Duffin, \emph{{AdS bulk
  locality from sharp CFT bounds}},
  \href{https://doi.org/10.1007/JHEP11(2021)164}{\emph{JHEP} {\bfseries 11}
  (2021) 164} [\href{https://arxiv.org/abs/2106.10274}{{\ttfamily
  2106.10274}}].

\bibitem{Caron-Huot:2022jli}
S.~Caron-Huot, Y.-Z.~Li, J.~Parra-Martinez and D.~Simmons-Duffin,
  \emph{{Graviton partial waves and causality in higher dimensions}},
  \href{https://doi.org/10.1103/PhysRevD.108.026007}{\emph{Phys. Rev. D}
  {\bfseries 108} (2023) 026007}
  [\href{https://arxiv.org/abs/2205.01495}{{\ttfamily 2205.01495}}].

\bibitem{Caron-Huot:2022ugt}
S.~Caron-Huot, Y.-Z.~Li, J.~Parra-Martinez and D.~Simmons-Duffin,
  \emph{{Causality constraints on corrections to Einstein gravity}},
  \href{https://doi.org/10.1007/JHEP05(2023)122}{\emph{JHEP} {\bfseries 05}
  (2023) 122} [\href{https://arxiv.org/abs/2201.06602}{{\ttfamily
  2201.06602}}].

\bibitem{Bern:2021ppb}
Z.~Bern, D.~Kosmopoulos and A.~Zhiboedov, \emph{{Gravitational effective field
  theory islands, low-spin dominance, and the four-graviton amplitude}},
  \href{https://doi.org/10.1088/1751-8121/ac0e51}{\emph{J. Phys. A} {\bfseries
  54} (2021) 344002} [\href{https://arxiv.org/abs/2103.12728}{{\ttfamily
  2103.12728}}].

\bibitem{Henriksson:2022oeu}
J.~Henriksson, B.~McPeak, F.~Russo and A.~Vichi, \emph{{Bounding violations of
  the weak gravity conjecture}},
  \href{https://doi.org/10.1007/JHEP08(2022)184}{\emph{JHEP} {\bfseries 08}
  (2022) 184} [\href{https://arxiv.org/abs/2203.08164}{{\ttfamily
  2203.08164}}].

\bibitem{Fernandez:2022kzi}
C.~Fernandez, A.~Pomarol, F.~Riva and F.~Sciotti, \emph{{Cornering
  large-N$_{c}$ QCD with positivity bounds}},
  \href{https://doi.org/10.1007/JHEP06(2023)094}{\emph{JHEP} {\bfseries 06}
  (2023) 094} [\href{https://arxiv.org/abs/2211.12488}{{\ttfamily
  2211.12488}}].

\bibitem{Albert:2023jtd}
J.~Albert and L.~Rastelli, \emph{{Bootstrapping Pions at Large $N$. Part II:
  Background Gauge Fields and the Chiral Anomaly}},
  \href{https://arxiv.org/abs/2307.01246}{{\ttfamily 2307.01246}}.

\bibitem{Bellazzini:2023nqj}
B.~Bellazzini, G.~Isabella, S.~Ricossa and F.~Riva, \emph{{Massive gravity is
  not positive}},
  \href{https://doi.org/10.1103/PhysRevD.109.024051}{\emph{Phys. Rev. D}
  {\bfseries 109} (2024) 024051}
  [\href{https://arxiv.org/abs/2304.02550}{{\ttfamily 2304.02550}}].

\bibitem{Mcpeak:2023wmq}
B.~McPeak, M.~Venuti and A.~Vichi, \emph{{Adding subtractions: comparing the
  impact of different Regge behaviors}},
  \href{https://arxiv.org/abs/2310.06888}{{\ttfamily 2310.06888}}.

\bibitem{Cheung:2014ega}
C.~Cheung and G.N.~Remmen, \emph{{Infrared consistency and the weak gravity
  conjecture}}, \href{https://doi.org/10.1007/JHEP12(2014)087}{\emph{JHEP}
  {\bfseries 12} (2014) 087} [\href{https://arxiv.org/abs/1407.7865}{{\ttfamily
  1407.7865}}].

\bibitem{CarrilloGonzalez:2022fwg}
M.~Carrillo~Gonzalez, C.~de~Rham, V.~Pozsgay and A.J.~Tolley, \emph{{Causal
  effective field theories}},
  \href{https://doi.org/10.1103/PhysRevD.106.105018}{\emph{Phys. Rev. D}
  {\bfseries 106} (2022) 105018}
  [\href{https://arxiv.org/abs/2207.03491}{{\ttfamily 2207.03491}}].

\bibitem{CarrilloGonzalez:2023cbf}
M.~Carrillo~Gonz\'alez, C.~de~Rham, S.~Jaitly, V.~Pozsgay and A.~Tokareva,
  \emph{{Positivity-causality competition: a road to ultimate EFT consistency
  constraints}},  \href{https://arxiv.org/abs/2307.04784}{{\ttfamily
  2307.04784}}.

\bibitem{CarrilloGonzalez:2023rmc}
M.~Carrillo~Gonz\'alez, \emph{{Bounds on EFT's in an expanding Universe}},
  \href{https://arxiv.org/abs/2312.07651}{{\ttfamily 2312.07651}}.

\bibitem{Cremonini:2023epg}
S.~Cremonini, B.~McPeak and Y.~Tang, \emph{{Electric shocks: bounding
  Einstein-Maxwell theory with time delays on boosted RN backgrounds}},
  \href{https://arxiv.org/abs/2312.17328}{{\ttfamily 2312.17328}}.

\bibitem{Chowdhury:2019kaq}
S.D.~Chowdhury, A.~Gadde, T.~Gopalka, I.~Halder, L.~Janagal and S.~Minwalla,
  \emph{{Classifying and constraining local four photon and four graviton
  S-matrices}}, \href{https://doi.org/10.1007/JHEP02(2020)114}{\emph{JHEP}
  {\bfseries 02} (2020) 114}
  [\href{https://arxiv.org/abs/1910.14392}{{\ttfamily 1910.14392}}].

\bibitem{Chowdhury:2023fwb}
S.D.~Chowdhury, V.~Kumar, S.~Kundu and A.~Rahaman, \emph{{Regge constraints on
  local four-point scattering amplitudes of massive particles with spin}},
  \href{https://arxiv.org/abs/2311.17015}{{\ttfamily 2311.17015}}.

\bibitem{Lucini:2001ej}
B.~Lucini and M.~Teper, \emph{{$SU(N)$ gauge theories in four-dimensions:
  Exploring the approach to $N = \infty$}},
  \href{https://doi.org/10.1088/1126-6708/2001/06/050}{\emph{JHEP} {\bfseries
  06} (2001) 050} [\href{https://arxiv.org/abs/hep-lat/0103027}{{\ttfamily
  hep-lat/0103027}}].

\bibitem{Lucini:2004my}
B.~Lucini, M.~Teper and U.~Wenger, \emph{{Glueballs and k-strings in SU(N)
  gauge theories: Calculations with improved operators}},
  \href{https://doi.org/10.1088/1126-6708/2004/06/012}{\emph{JHEP} {\bfseries
  06} (2004) 012} [\href{https://arxiv.org/abs/hep-lat/0404008}{{\ttfamily
  hep-lat/0404008}}].

\bibitem{Lucini:2010nv}
B.~Lucini, A.~Rago and E.~Rinaldi, \emph{{Glueball masses in the large N
  limit}}, \href{https://doi.org/10.1007/JHEP08(2010)119}{\emph{JHEP}
  {\bfseries 08} (2010) 119} [\href{https://arxiv.org/abs/1007.3879}{{\ttfamily
  1007.3879}}].

\bibitem{Athenodorou:2021qvs}
A.~Athenodorou and M.~Teper, \emph{{$SU(N)$ gauge theories in 3+1 dimensions:
  glueball spectrum, string tensions and topology}},
  \href{https://doi.org/10.1007/JHEP12(2021)082}{\emph{JHEP} {\bfseries 12}
  (2021) 082} [\href{https://arxiv.org/abs/2106.00364}{{\ttfamily
  2106.00364}}].

\bibitem{deRham:2022sdl}
C.~de~Rham, L.~Engelbrecht, L.~Heisenberg and A.~L\"uscher, \emph{{Positivity
  bounds in vector theories}},
  \href{https://doi.org/10.1007/JHEP12(2022)086}{\emph{JHEP} {\bfseries 12}
  (2022) 086} [\href{https://arxiv.org/abs/2208.12631}{{\ttfamily
  2208.12631}}].

\bibitem{Bellazzini:2021oaj}
B.~Bellazzini, M.~Riembau and F.~Riva, \emph{{IR side of positivity bounds}},
  \href{https://doi.org/10.1103/PhysRevD.106.105008}{\emph{Phys. Rev. D}
  {\bfseries 106} (2022) 105008}
  [\href{https://arxiv.org/abs/2112.12561}{{\ttfamily 2112.12561}}].

\bibitem{Bern:2010ue}
Z.~Bern, J.J.M.~Carrasco and H.~Johansson, \emph{{Perturbative Quantum Gravity
  as a Double Copy of Gauge Theory}},
  \href{https://doi.org/10.1103/PhysRevLett.105.061602}{\emph{Phys. Rev. Lett.}
  {\bfseries 105} (2010) 061602}
  [\href{https://arxiv.org/abs/1004.0476}{{\ttfamily 1004.0476}}].

\bibitem{Hebbar:2020ukp}
A.~Hebbar, D.~Karateev and J.~Penedones, \emph{{Spinning S-matrix bootstrap in
  4d}}, \href{https://doi.org/10.1007/JHEP01(2022)060}{\emph{JHEP} {\bfseries
  01} (2022) 060} [\href{https://arxiv.org/abs/2011.11708}{{\ttfamily
  2011.11708}}].

\bibitem{Correia:2020xtr}
M.~Correia, A.~Sever and A.~Zhiboedov, \emph{{An analytical toolkit for the
  S-matrix bootstrap}},
  \href{https://doi.org/10.1007/JHEP03(2021)013}{\emph{JHEP} {\bfseries 03}
  (2021) 013} [\href{https://arxiv.org/abs/2006.08221}{{\ttfamily
  2006.08221}}].

\bibitem{Buric:2023ykg}
I.~Buric, F.~Russo and A.~Vichi, \emph{{Spinning partial waves for scattering
  amplitudes in d dimensions}},
  \href{https://doi.org/10.1007/JHEP10(2023)090}{\emph{JHEP} {\bfseries 10}
  (2023) 090} [\href{https://arxiv.org/abs/2305.18523}{{\ttfamily
  2305.18523}}].

\bibitem{Simmons-Duffin:2015qma}
D.~Simmons-Duffin, \emph{{A semidefinite program solver for the conformal
  bootstrap}}, \href{https://doi.org/10.1007/JHEP06(2015)174}{\emph{JHEP}
  {\bfseries 06} (2015) 174}
  [\href{https://arxiv.org/abs/1502.02033}{{\ttfamily 1502.02033}}].

\bibitem{Caron-Huot:2021rmr}
S.~Caron-Huot, D.~Mazac, L.~Rastelli and D.~Simmons-Duffin, \emph{{Sharp
  boundaries for the Swampland}},
  \href{https://doi.org/10.1007/jhep07(2021)110}{\emph{JHEP} {\bfseries 07}
  (2021) 110} [\href{https://arxiv.org/abs/2102.08951}{{\ttfamily
  2102.08951}}].

\bibitem{Chiang:2021ziz}
L.-Y.~Chiang, Y.-t.~Huang, W.~Li, L.~Rodina and H.-C.~Weng, \emph{{Into the
  EFThedron and UV constraints from IR consistency}},
  \href{https://doi.org/10.1007/JHEP03(2022)063}{\emph{JHEP} {\bfseries 03}
  (2022) 063} [\href{https://arxiv.org/abs/2105.02862}{{\ttfamily
  2105.02862}}].

\bibitem{Albert:2023seb}
J.~Albert, J.~Henriksson, L.~Rastelli and A.~Vichi, \emph{{Bootstrapping mesons
  at large $N$: Regge trajectory from spin-two maximization}},
  \href{https://arxiv.org/abs/2312.15013}{{\ttfamily 2312.15013}}.

\bibitem{Distler:2006if}
J.~Distler, B.~Grinstein, R.A.~Porto and I.Z.~Rothstein, \emph{{Falsifying
  models of new physics via WW scattering}},
  \href{https://doi.org/10.1103/PhysRevLett.98.041601}{\emph{Phys. Rev. Lett.}
  {\bfseries 98} (2007) 041601}
  [\href{https://arxiv.org/abs/hep-ph/0604255}{{\ttfamily hep-ph/0604255}}].

\bibitem{Srednicki:2007qs}
M.~Srednicki, \emph{{Quantum field theory}}, Cambridge University Press (2007).

\bibitem{deRham:2020yet}
C.~de~Rham and V.~Pozsgay, \emph{{New class of Proca interactions}},
  \href{https://doi.org/10.1103/PhysRevD.102.083508}{\emph{Phys. Rev. D}
  {\bfseries 102} (2020) 083508}
  [\href{https://arxiv.org/abs/2003.13773}{{\ttfamily 2003.13773}}].

\bibitem{deRham:2011rn}
C.~de~Rham, G.~Gabadadze and A.J.~Tolley, \emph{{Ghost free Massive Gravity in
  the St\"uckelberg language}},
  \href{https://doi.org/10.1016/j.physletb.2012.03.081}{\emph{Phys. Lett. B}
  {\bfseries 711} (2012) 190}
  [\href{https://arxiv.org/abs/1107.3820}{{\ttfamily 1107.3820}}].

\bibitem{deRham:2014zqa}
C.~de~Rham, \emph{{Massive gravity}},
  \href{https://doi.org/10.12942/lrr-2014-7}{\emph{Living Rev. Rel.} {\bfseries
  17} (2014) 7} [\href{https://arxiv.org/abs/1401.4173}{{\ttfamily
  1401.4173}}].

\bibitem{Heisenberg:2018vsk}
L.~Heisenberg, \emph{{A systematic approach to generalisations of General
  Relativity and their cosmological implications}},
  \href{https://doi.org/10.1016/j.physrep.2018.11.006}{\emph{Phys. Rept.}
  {\bfseries 796} (2019) 1} [\href{https://arxiv.org/abs/1807.01725}{{\ttfamily
  1807.01725}}].

\bibitem{Tasinato:2014eka}
G.~Tasinato, \emph{{Cosmic acceleration from Abelian symmetry breaking}},
  \href{https://doi.org/10.1007/JHEP04(2014)067}{\emph{JHEP} {\bfseries 04}
  (2014) 067} [\href{https://arxiv.org/abs/1402.6450}{{\ttfamily 1402.6450}}].

\bibitem{Heisenberg:2014rta}
L.~Heisenberg, \emph{{Generalization of the Proca action}},
  \href{https://doi.org/10.1088/1475-7516/2014/05/015}{\emph{JCAP} {\bfseries
  05} (2014) 015} [\href{https://arxiv.org/abs/1402.7026}{{\ttfamily
  1402.7026}}].

\bibitem{Allys:2015sht}
E.~Allys, P.~Peter and Y.~Rodriguez, \emph{{Generalized Proca action for an
  Abelian vector field}},
  \href{https://doi.org/10.1088/1475-7516/2016/02/004}{\emph{JCAP} {\bfseries
  02} (2016) 004} [\href{https://arxiv.org/abs/1511.03101}{{\ttfamily
  1511.03101}}].

\bibitem{BeltranJimenez:2016rff}
J.~Beltran~Jimenez and L.~Heisenberg, \emph{{Derivative self-interactions for a
  massive vector field}},
  \href{https://doi.org/10.1016/j.physletb.2016.04.017}{\emph{Phys. Lett. B}
  {\bfseries 757} (2016) 405}
  [\href{https://arxiv.org/abs/1602.03410}{{\ttfamily 1602.03410}}].

\bibitem{Allys:2016jaq}
E.~Allys, J.P.~Beltran~Almeida, P.~Peter and Y.~Rodr\'\i{}guez, \emph{{On the
  4D generalized Proca action for an Abelian vector field}},
  \href{https://doi.org/10.1088/1475-7516/2016/09/026}{\emph{JCAP} {\bfseries
  09} (2016) 026} [\href{https://arxiv.org/abs/1605.08355}{{\ttfamily
  1605.08355}}].

\bibitem{DeRham:2018axr}
C.~De~Rham, K.~Hinterbichler and L.A.~Johnson, \emph{{On the (A)dS Decoupling
  Limits of Massive Gravity}},
  \href{https://doi.org/10.1007/JHEP09(2018)154}{\emph{JHEP} {\bfseries 09}
  (2018) 154} [\href{https://arxiv.org/abs/1807.08754}{{\ttfamily
  1807.08754}}].

\end{thebibliography}\endgroup

\bibliographystyle{JHEP.bst}

\end{document}